\documentclass[12pt]{article}
\usepackage{tikz}
\usepackage{graphicx}
\usepackage{mathrsfs}
\usepackage[utf8]{inputenc}
\usepackage{longtable}
\usepackage{float}
\usepackage[english]{babel}
\usepackage[dvipsnames]{xcolor}
\usepackage{tcolorbox}
\usepackage{colortbl}
\usepackage{mdframed}
\usepackage{fancyvrb}
\usepackage{adjustbox}
\usepackage{multirow}
\usepackage{rotating}
\usepackage{threeparttable,tablefootnote,booktabs,stackengine}
\usepackage{etoolbox}
\usepackage{placeins}
\usepackage{bbm}
\usepackage{pdflscape}
\usepackage{vcell}
\usepackage{subcaption}
\usepackage{tabularx,ragged2e,enumitem,array,makecell}
\usepackage[margin=1in]{geometry}
\usepackage{amsmath,amsthm,amssymb,scrextend,amsfonts}
\usepackage[justification=centering]{caption}
\usepackage[toc,page]{appendix}
\usepackage{setspace}
\usepackage{lipsum}
\usepackage{footmisc}
\usepackage{spverbatim}
\usepackage{float}
\usepackage{graphicx}
\usepackage{titling}
\usepackage[dvipsnames]{xcolor}
\usepackage{xurl} 
\usepackage{lmodern}
\usepackage{anyfontsize}
\usepackage{verbatim}
\usepackage{url}
\usepackage{placeins}
\usepackage{titletoc}
\usepackage{bbding}
\usepackage[
    colorlinks=true,
    hypertexnames=false,
    linkcolor=MidnightBlue,
    filecolor=MidnightBlue,
    citecolor=MidnightBlue,
    urlcolor=MidnightBlue
]{hyperref}
\usepackage{cleveref}
\usepackage{siunitx}
\usepackage[pagestyles]{titlesec}
\usepackage{natbib}
 \bibpunct[, ]{(}{)}{,}{a}{}{,}%
\usepackage{bibunits}
\usepackage{pdfpages}

\defaultbibliographystyle{template/informs2014}
\defaultbibliography{main}

\AtBeginEnvironment{tablenotes}{\setstretch{1}\scriptsize}

\usetikzlibrary{arrows.meta,positioning,shapes.geometric,calc}

\definecolor{darkblue}{RGB}{0,0,120}
\xdefinecolor{ut-blue-dark}{RGB}{65,90,140}
\xdefinecolor{ut-blue}{RGB}{0,105,170}
\xdefinecolor{ut-blue-l}{RGB}{127,180,212}
\xdefinecolor{ut-blue-light}{RGB}{80,170,200}
\xdefinecolor{ut-grey-mid}{RGB}{143,143,149}
\xdefinecolor{ut-grey-light}{RGB}{208,207,210}
\xdefinecolor{ut-grey-llight}{RGB}{231,230,231}
\xdefinecolor{ut-blue}{RGB}{0,105,170}

\setlist[itemize]{leftmargin=*,nosep,topsep=0pt,partopsep=0pt}

\newcolumntype{Y}{>{\RaggedRight\arraybackslash}X}
\newcolumntype{L}{>{\raggedright\arraybackslash}X}
\newcolumntype{R}{>{\raggedleft\arraybackslash}p{1.5cm}}

\newcolumntype{P}[1]{>{\RaggedRight\arraybackslash}p{#1}}

\renewcommand{\arraystretch}{1.18}

\DeclareCaptionLabelFormat{andtable}{#1~#2  \&  \tablename~\thetable}

\newcolumntype{x}[1]{%
>{\centering\hspace{0pt}}p{#1}}%

\author{}

\newcommand{\fixfigspacing}{%
  \setlength{\baselineskip}{\the\fontdimen6\font}%
}

\providecommand{\FIGURE}[3]{%
    \caption{#2}
    \centering
    #1
    \ifthenelse{\isempty{#3}}
        {}
        {%
        \par\noindent\raggedright\footnotesize\singlespacing {#3}
        }
}

\providecommand{\noteslabel}[1]{%
  \StrCount{#1}{.}[\mycount] 
  \ifthenelse{\mycount = 1}{Note}{Notes}%
}
\providecommand{\TABLE}[3]{%
\caption{#1}
\centering #2\par
\ifthenelse{\isempty{#3}}%
    {}  
    {%
    \begin{tablenotes}[flushleft]
    \scriptsize%
    \item \hspace{1.4mm} {#3}
    \end{tablenotes}
    }
}

\usepackage{natbib}
 \bibpunct[, ]{(}{)}{,}{a}{}{,}%

\makeatletter
\newcommand\blfootnote[1]{%
  \begingroup
  \renewcommand\thefootnote{}%
  \renewcommand\Hy@raisedlink[1]{}%
  \footnotetext{\noindent #1}%
  \endgroup
}
\makeatother

\titleformat{\section}
  {\normalfont\bfseries \Large}{\thesection}{10pt}{}
\newpagestyle{mystyle}{
  \sethead[][\thesection.\enspace\sectiontitle][]{}{\thesection~\sectiontitle}{}
\setfoot{}{\thepage}{}}

\titleformat{\subsection}
  {\normalfont\bfseries \large}{\thesubsection}{10pt}{}
\newpagestyle{mystyle2}{
  \sethead[][\thesubsection.\enspace\subsectiontitle][]{}{\thesubsection~\subsectiontitle}{}
\setfoot{}{\thepage}{}}

\begin{document}
 
\date{\today}
\newgeometry{top=1cm, bottom=2cm, left=2cm, right=2cm}

\begin{titlepage}

\singlespacing

\title{Does TikTok Promote or Cannibalize Music Streaming? Estimands and Identification with Heavy-Tailed Outcomes}

\author{
  Daniel Winkler,
  Christian Hotz-Behofsits,
  Nils Wlömert,\\
  Dominik Papies,
  J\={u}ra Liaukonyt\.{e}}

\maketitle
\vspace{-0.8cm}

\blfootnote{\noindent The authors thank seminar and conference participants at Tilburg University, UNSW, Monash University, Bocconi University, University of Groningen, WU Vienna, Marketing Science (Sydney), the 5th International Pricing Symposium (Munich), and Marketing Analytics Symposium Sydney for helpful comments. The authors also thank Andrey Simonov and Matt McGranaghan for helpful discussions and suggestions.}

\noindent 
\thispagestyle{empty}

\vspace{0.1cm}

\begin{abstract}
			\singlespacing

\noindent We study how TikTok affects demand for music on paid streaming platforms. We use Universal Music Group's (UMG) global withdrawal of its catalog from TikTok as a quasi-natural experiment. Recent work using this setting reaches mixed conclusions about whether TikTok promotes or cannibalizes streaming demand. We show that these findings can be reconciled by making the estimand explicit: with heavy-tailed exposure and outcomes, common difference-in-differences (DiD) implementations in levels, logs, and Poisson answer different economic questions. In our data, the top 10\% of songs account for 96\% of TikTok creations and 76\% of Spotify streams, which makes the distinction between the typical song and the economically consequential song central. We find that removing TikTok access lowers Spotify demand for UMG titles, with losses concentrated among viral songs and little economically meaningful change for the long tail. Because the viral head accounts for a disproportionate share of listening and revenue, these losses drive aggregate implications. A TikTok creator-side analysis shows that some activity reallocates toward non-UMG audio when UMG content is unavailable. This substitution is limited in magnitude but economically relevant for interpreting the treatment effect because streaming compensation depends on relative stream shares. Finally, using the 2025 U.S.\ TikTok outage, which affected all labels symmetrically and is not subject to the label-specific spillover concern as the UMG withdrawal, we find corroborating evidence that disruptions to TikTok access reduce monetized streaming. We also provide a practitioner companion that guides the choice of DiD estimands, estimators, and diagnostics in heavy-tailed outcome settings.

\end{abstract}

\vspace{0mm}\noindent

\end{titlepage}
\restoregeometry

\begin{bibunit}

\setstretch{2}
\onehalfspacing

\vspace{-8mm}

\section{Introduction}
\label{intro}

The digitization of music and the rise of social media have reshaped how music is consumed and monetized. Short-form video platforms have become a major gateway into listening on paid streaming services, and in many cases they also function as a music consumption environment in their own right. Viral trends on these platforms can propel previously unknown songs and revive older catalog titles, with downstream effects on Spotify streams, radio airtime, and chart performance \citep{maheshwari_tiktok_2023}. At the same time, a substantial share of users engage with platforms such as TikTok specifically to listen to and interact with music, which makes music central to the user experience \citep{IFPI2022engaging}. Licensing negotiations between platforms and rights holders increasingly hinge on a core economic question: does TikTok primarily \emph{promote} monetized listening by stimulating discovery, or does it \emph{cannibalize} monetized listening by absorbing attention that would otherwise translate into streams on revenue-sharing services?

Answering this question is challenging because music demand is highly concentrated. In our data, the top 10\% of songs ranked by Spotify streams account for 76\% of total streams, while the top 10\% ranked by TikTok creations account for 96\% of total creations, with the remaining long tail exhibiting low baseline activity. This pattern is typical for many digital markets, platforms, and media products \citep[e.g.,][]{rosen_economics_1981, salganik_experimental_2006,fleder_blockbuster_2009,johnson_emergence_2014,koenig_technical_2023,wlomert_interplay_2024}: low marginal costs expand the amount of content available, while search, recommendation, and social sharing concentrate attention on a relatively small set of winners.\footnote{Throughout the paper we use ``heavy-tailed'' and ``highly skewed'' interchangeably to describe outcomes that are highly concentrated in the right tail. Such outcomes are also commonly referred to as power-law outcomes when the upper tail is well approximated by a power-law distribution.} This concentration creates a central methodological challenge: the typical song and the economically consequential song are not the same. A specification that gives equal weight to each song primarily summarizes the long tail, while a specification that tracks total streams is driven by the viral head. Credible inference therefore requires not only plausibly exogenous variation in TikTok exposure, but also an empirical strategy that makes the relevant aggregation explicit: which estimand is being recovered, how treatment effects are aggregated across songs, and which part of the outcome distribution carries the economic interpretation.

This paper studies TikTok’s role in the music ecosystem using a sharp and plausibly exogenous change in exposure: Universal Music Group’s (UMG) global withdrawal of its catalog from TikTok in February 2024 following failed licensing negotiations \citep{universal_music_group_open_2024,ingham_tiktoks_2024}. The event removed a large share of commercially important content from the platform at once and therefore provides a rare opportunity to quantify how short-form video exposure translates into monetized listening.\footnote{The complement-versus-cannibalization question connects to a broader quasi-experimental literature on content removals and restorations across digital platforms, which finds both promotional and displacement effects depending on the platform, content type, and part of the demand distribution. Web Appendix \ref{app:literature} summarizes this evidence.}

Recent quasi-experimental work has begun to exploit the same natural experiment but reaches different headline conclusions. \citet{cheng_value_2024} use log-transformed outcomes as their primary specification and report modest positive effects on Spotify streaming for UMG titles that were used on TikTok prior to the dispute, which they interpret as cannibalization of paid streaming. \citet{bairathi_lambrecht_rao_2024}, in contrast, use levels and Poisson pseudo-maximum-likelihood (PPML) specifications and document declines in streams for affected songs, which they interpret as consistent with TikTok playing a promotional role. We reconcile these findings by showing that, with heavy-tailed outcomes and heterogeneous treatment effects, different outcome transformations target different estimands, so the resulting DiD estimates can differ across implementations and can even reverse sign. By ``reconcile,'' we mean that, \textit{within our data and research design}, we can reproduce the qualitative patterns emphasized in the prior studies when we adopt the corresponding estimands and associated transformations.

More generally, we emphasize that, with heavy-tailed positive outcomes, the choice between levels, logs, and Poisson specifications should not be treated as a routine robustness exercise: each specification corresponds to a different parallel-trends structure and places implicit weight on different parts of the outcome distribution \citep{log_gravity}.\footnote{This point is not specific to heavy-tailed outcome environments. It applies whenever outcomes are uneven and economic interpretation depends on how effects aggregate across units: different functional forms implicitly target different estimands and therefore answer different economic questions. The difference is especially consequential with heavy-tailed outcomes, which is why we focus on this case in the paper.} In particular, one of the key modeling decisions is whether additive trends in levels or proportional (multiplicative) dynamics are more credible in the absence of treatment \citep{mcconnell_cant_2024}. In our setting, pre-treatment trajectories are much closer to multiplicative growth than to additive change, because absolute changes scale strongly with baseline popularity while proportional changes are comparatively stable. This evidence motivates our focus on log-based and log-link DiD specifications and the interpretation of estimated effects as proportional changes in demand. 

At the same time, because streaming outcomes are heavily skewed, even multiplicative specifications can target different estimands. An unweighted log-linear OLS regression estimates the DiD in mean log streams, which corresponds to an equal-weighted log change across songs and is naturally interpreted as a “typical song” percent change. This estimand is most informative when proportional treatment effects are similar across the distribution, or when the research question concerns the typical song rather than aggregate streams. This is not the case in our setting. For the typical long-tail title, removing TikTok access produces no economically meaningful change in streaming demand. In contrast, highly viral titles experience meaningful streaming declines when TikTok exposure disappears, and this pattern is visible across specifications once we allow treatment effects to vary by baseline TikTok virality. Because the viral head accounts for a disproportionate share of total listening, the aggregate implication is clearly negative. For this reason, we emphasize estimands that are more directly informative about aggregate economic impacts. A weighted log-linear specification shifts the log-DiD toward songs that account for more baseline listening, which makes the implied proportional effect closer to a stream-weighted one. A PPML model with a log link specifies a multiplicative conditional mean in levels; aggregating fitted means across songs yields a model-implied proportional change in expected total streams for the treated group. In our setting, these two approaches deliver very similar results.

The paper makes several contributions. First, it provides new evidence on whether short-form video promotes or cannibalizes monetized consumption, and it reconciles the existing conflicting findings. Leveraging the UMG withdrawal, we find, consistent with \citet{bairathi_lambrecht_rao_2024}, that TikTok exposure complements monetized streaming: when UMG content becomes unavailable on TikTok, demand for UMG titles on Spotify declines. This decline is concentrated in the viral head. The typical long-tail track exhibits no economically meaningful change, while losses are concentrated among highly TikTok-exposed titles. Because listening is extremely concentrated among these songs, the viral head declines dominate aggregate implications, and the overall effect is clearly negative. Second, we show that part of the estimated treatment effect likely reflects cross-catalog reallocation. When UMG content becomes unavailable on TikTok, creators and listeners may shift attention toward non-UMG titles, which can increase control-group streams and widen the treated-control gap. In a standard DiD interpretation, this channel raises an interference concern. In our setting, however, it is also economically informative. Spotify royalties are allocated based on each rights holder's share of total monetized streams \citep{spotify_royalties}, so reallocation toward competing catalogs is part of the revenue consequence for the treated catalog. To assess whether the promotional effect persists when this reallocation channel is shut down, we complement the UMG analysis with a second, independent disruption to TikTok access: the 2025 U.S.\ TikTok outage, which affects all labels symmetrically. The outage analysis provides consistent evidence that TikTok exposure promotes monetized streaming. Finally, we contribute methodologically by clarifying how functional-form choices in DiD interact with heavy-tailed outcomes and heterogeneous effects. With heavy-tailed outcomes, common DiD implementations map to different estimands and therefore carry different economic interpretations. We show that making these estimands explicit is a prerequisite for credible inference in settings where a small set of units drives most economically relevant activity.

\section{Institutional Background and Data}
\label{background}
\noindent TikTok continues to be one of the fastest-growing online platforms, nearing 2 billion users worldwide and about 170 million users in the United States. Consequently, the \textit{New York Times} describes TikTok as ``one of the world’s most influential online outlets’’ \citep{sisario_universal_2024}. A distinctive feature of TikTok, relative to platforms such as YouTube, is that consumption is organized around very short videos, with a recommended length of 21–34 seconds \citep{stokel-walker_tiktok_2022}, and these videos are tightly coupled to music through user-generated creations. This architecture gives TikTok a plausible role in shaping music consumption outside the platform, but it also makes the relevant mechanism a priori ambiguous: TikTok can operate as a promotional channel, a substitution channel, or both, depending on how exposure translates into downstream listening.

In this section, we describe UMG's dispute with TikTok and how it generates a sharp disruption to TikTok-side exposure, which we use to study spillovers to Spotify. We also overview the data used in the analysis.

\subsection{UMG’s TikTok Withdrawal and Re-entry}
\label{subsec:umg_tiktok_dispute}

On January 30, 2024, Universal Music Group (UMG) announced the breakdown of licensing negotiations with TikTok, and the termination of their license agreement became effective January 31, 2024 \citep{sisario_universal_2024}. The dispute resulted in an abrupt change in TikTok-side availability for UMG content: UMG audio was blocked for use in new user-generated videos, existing videos containing UMG audio were muted, and UMG artists’ music videos were removed from the platform. While reminiscent of past content disputes between digital platforms and rights holders \citep{wlomert_interplay_2024}, the UMG--TikTok standoff was unusually consequential given the importance of UMG’s catalog in mainstream listening. 

\begin{figure}[ht!]
\centering
\caption{Weekly User-generated Creations for UMG Songs in the TikTok Top 100}
\includegraphics[width=0.9\textwidth]{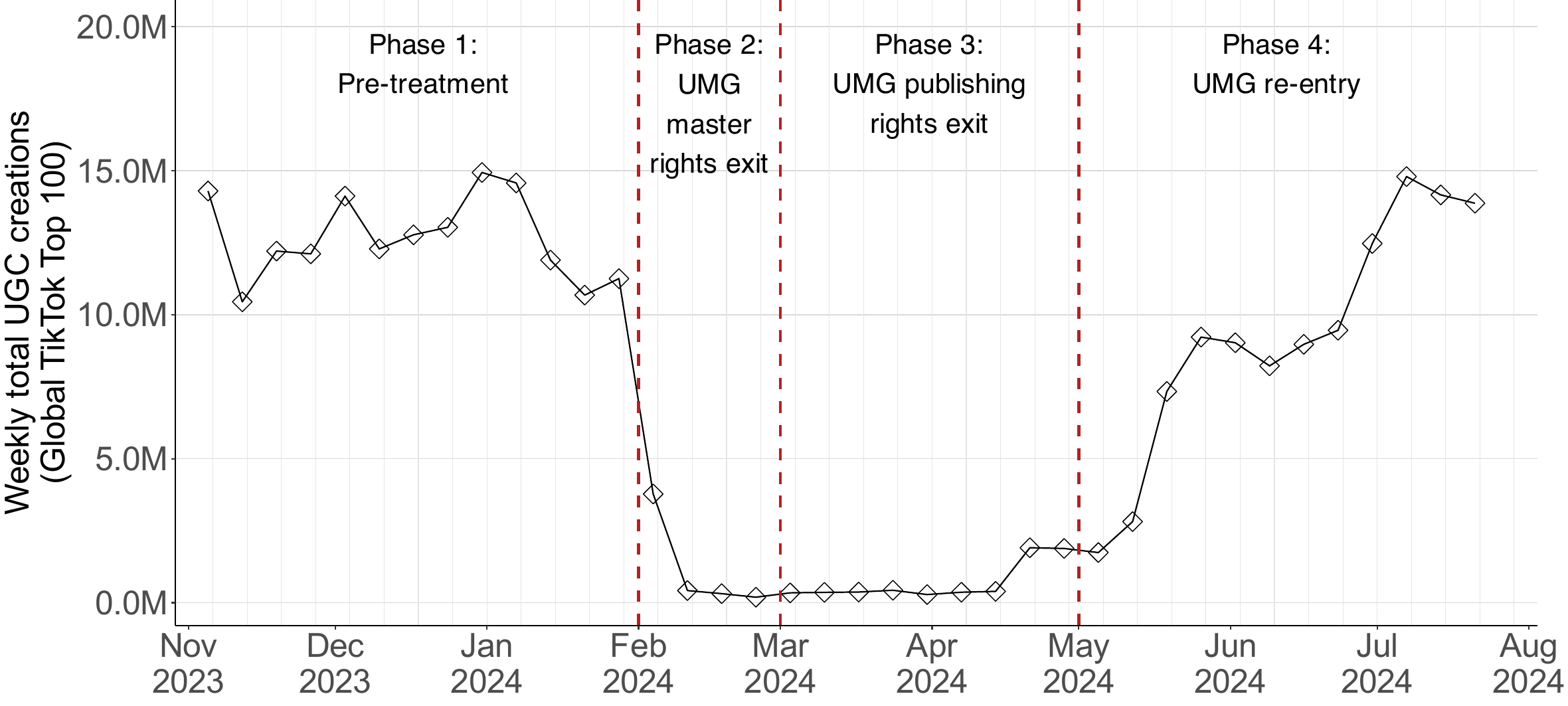}

\label{fig:tiktok_chart_global}

\vspace{0.4em}
\begin{minipage}{0.99\textwidth}
\scriptsize
\textit{Notes:} The figure shows the total number of user-generated TikTok videos that include songs distributed by UMG, aggregated across all UMG songs appearing in the TikTok Top 100 charts. Dashed vertical lines indicate key events: UMG’s master rights exit (Phase 2), publishing rights exit (Phase 3), and subsequent re-entry (Phase 4).
\end{minipage}
\vspace{-0.5cm}
\end{figure}

We organize the analysis around the dispute phases highlighted in \autoref{fig:tiktok_chart_global}. Phase 1 is the pre-treatment period, which runs from November 30, 2023 through January 31, 2024. The treatment begins on February 1, 2024, when UMG’s catalog becomes unavailable on TikTok following the breakdown in licensing negotiations, which we refer to as the master-rights exit (Phase 2). We then distinguish a subsequent escalation in early March 2024, when the dispute extends to publishing rights (Phase 3).\footnote{Master rights cover the specific recorded performance of a song, while publishing rights cover the underlying musical composition (lyrics and melody). These rights are distinct and often held by different parties---for example, a song’s master recording may be owned by one label (e.g., Sony or Warner), while its publishing rights are administered by another (e.g., UMG), and vice versa---so they can be licensed or withdrawn independently \citep{Witrand2025_master_publishing_rights}. Hence, Phase~3 (the publishing-rights exit) additionally affects songwriters and music publishers, and it also creates collateral damage for non-UMG recordings whose underlying compositions are administered by UMG's publishing arm. This means that a dispute initiated by a single label can spill over to affect artists and rights holders who are not party to the negotiation, which has implications for how collective bargaining in the music industry is structured.} Finally, we define Phase 4 as the re-entry period beginning in early May 2024, when UMG content is restored on TikTok. We mark these phase boundaries consistently throughout the paper, and in all figures with time-series outcomes we indicate the onset of Phase 2 and the subsequent Phase 3 and Phase 4 transitions with red vertical reference lines.

The dispute generates a discontinuous and directly measurable disruption to TikTok-side exposure. It shows up immediately in user-generated TikTok videos that use UMG songs, which we refer to throughout the paper as \textit{TikTok creations}. \autoref{fig:tiktok_chart_global} documents this first-stage pattern using weekly TikTok creations in the TikTok Top 100 that include UMG-distributed songs. The series drops sharply at the dispute onset, remains near zero during the exit phases, and then recovers after re-entry. This collapse in UMG-linked TikTok creations validates that the treatment is clearly visible in the data.

As mentioned above, the direction of spillovers from TikTok to paid streaming platforms is ex ante ambiguous, which is reflected both in the institutional debate and in the emerging empirical evidence. On one hand, TikTok may operate primarily as a discovery and promotion channel, which increases subsequent listening on licensed streaming platforms \citep{herstand_whos_2024, ingham_tiktoks_2024, yang_ugc_2024}. On the other hand, the scale of music playback embedded in short-form video, under compensation arrangements that differ from streaming, suggests a plausible substitution margin, particularly for younger listeners \citep{ingham_crises_2022, ingham_so_2022}. Consistent with this tension, \citet{cheng_value_2024} report small, positive average percentage effects on Spotify streaming for UMG titles that were used on TikTok prior to the dispute, which they interpret as \textit{substitution} toward paid streaming. \citet{bairathi_lambrecht_rao_2024}, in contrast, document an average decline in streams for affected titles, which is consistent with TikTok playing a \textit{promotional role}. We use this quasi-exogenous shock to measure the net spillover effect on Spotify and, at the same time, to reconcile these mixed findings by showing how alternative DiD implementations can deliver different conclusions in settings with heavy-tailed outcomes.

\subsection{Data}
\label{data}

We combine five data sources to measure (i) streaming outcomes, (ii) TikTok-side exposure and its disruption during the dispute, (iii) rights ownership needed to define treated and control songs, and (iv) Spotify playlist inclusion and playlist reach, which we use to study whether the withdrawal affects downstream playlist placement and exposure.

Soundcharts, a data aggregation platform targeted at the music industry, provides global Spotify stream counts and track-level TikTok availability, both in the pre period and during the dispute. We use these availability signals to determine which songs were removed from TikTok, which is critical for isolating the shock to TikTok-side exposure. Chartmetric, another music industry data platform, independently measures TikTok availability over the same horizon. We use Soundcharts and Chartmetric jointly to cross-validate availability and to fill in missing information in either source. We match master-recording ownership at the song level using data from GfK Entertainment (compiler of the official music charts in various countries), which allows us to identify UMG repertoire. We separately identify publishing rights using information from multiple international collecting societies\footnote{Publishing data are assembled from the French, Canadian, and German music rights collecting societies (i.e., SACEM, SOCAN, and GEMA) and supplemented with MusicBrainz. SACEM and SOCAN are queried for works with publisher names starting with ``Universal,'' while GEMA is accessed via ISRCs and common song-title prefixes. Tracks are linked to publishing records using a hierarchical procedure: ISRC matches where available, followed by exact performer–title matches and conservative fuzzy matching.}, and exclude from the control group non-UMG master recordings for which UMG holds publishing rights, ensuring that control songs are unaffected by the Phase-3 publishing-rights removal. Throughout the analysis related to the UMG–TikTok dispute, treated songs are UMG tracks that were used on TikTok in the pre period (that is, with strictly positive pre-treatment UGC creations). Finally, we collect country-level streaming data from Luminate for 10 countries for our analysis of the U.S.\ TikTok outage.

\begin{table}[ht!]
\caption{Pre-treatment Comparability Between Treated UMG and Matched Control Songs}
\label{tab:matchquality_songs}
\centering
\renewcommand{\arraystretch}{1}
\begin{adjustbox}{max width=\textwidth}
\begin{tabular}{lcccccc}
\toprule
\multicolumn{1}{c}{} & \multicolumn{2}{c}{Treated (UMG)} & \multicolumn{2}{c}{Control (Matched)} & \multicolumn{2}{c}{Differences (T--C)} \\
 & Mean & SD & Mean & SD & Mean diff. & SMD \\
\midrule

Spotify streams (weekly mean) 
& 142{,}545.36 & 659{,}382.16 
& 142{,}465.52 & 659{,}780.39 
& 79.84 & 0.000 \\

TikTok creations at treatment
& 12{,}655.54 & 154{,}713.71 
& 19{,}381.25 & 360{,}911.05 
& -6{,}725.70 & -0.017 \\

Spotify playlist followers at treatment
& 641{,}002.47 & 3{,}224{,}857.59 
& 636{,}879.69 & 3{,}324{,}117.97 
& 4{,}122.78 & 0.001 \\

Song age at treatment (weeks) 
& 929.18 & 855.86 
& 912.50 & 836.66 
& 16.68 & 0.014 \\

\addlinespace
N matched pairs 
& \multicolumn{2}{c}{53{,}753} 
& \multicolumn{2}{c}{53{,}753} 
& \multicolumn{2}{c}{} \\
\bottomrule
\end{tabular}
\end{adjustbox}

\vspace{0.1cm}

\noindent\parbox{\textwidth}{%
\setstretch{1}\scriptsize\textit{Notes}: The table reports pre-treatment balance for the matched sample used in the main analysis. Each treated UMG song is matched to one non-UMG control song; controls may be reused across treated songs. Details on the matching procedure are provided in the Web Appendix~\ref{app:matching_procedures}. The matched sample consists of 53{,}753 treated songs and 53{,}753 matched control observations, corresponding to 29{,}650 unique control songs. All variables are measured in the pre-treatment period unless noted otherwise. Mean differences are reported as treated (T) minus control (C); SMD denotes the standardized mean difference. Descriptive statistics for the full sample are provided in the Web Appendix~\ref{app:descriptives}.
}
\vspace{-0.5cm}
\end{table}

The Web Appendix \ref{app:sample} provides details on the sample construction. Our main sample consists of 53{,}753 treated songs and an equal number of 1:1 matched control titles.\footnote{Related studies examining the same event analyze treated samples of 35{,}837 songs in \citet{cheng_value_2024} and 21{,}526 songs in \citet{bairathi_lambrecht_rao_2024}, which reflects differences in data access and sampling frames.}

\paragraph{Treated songs.} We restrict the treated set to UMG songs with strictly positive pre-treatment TikTok creations. We define this indicator based on whether any videos available in the pre period use the song. This includes both videos created earlier but still available (i.e., not deleted) and newly created videos. This restriction targets the titles for which the dispute plausibly changes TikTok-side exposure, because only songs that were actually used in UGC prior to the withdrawal lose an existing channel. In contrast, for UMG songs with zero pre-treatment creations, the dispute does not meaningfully change TikTok exposure, which implies that any estimated post-dispute changes in their Spotify streams would be difficult to attribute to the withdrawal mechanism. We therefore use these zero-creation UMG tracks as a placebo group, where the prediction is that we should not detect a corresponding break in Spotify streaming around the dispute for songs that were not on TikTok in the first place.

\paragraph{Control songs.}
To construct a credible counterfactual for each treated song, we implement a 1:1 minimum-distance matching procedure. Starting from the universe of non-UMG tracks, we restrict the pool of potential controls to songs by different artists that fall within the same pre-treatment Spotify-popularity quartile (based on streams) and the same TikTok-virality decile. Virality is measured as a stock, defined as the cumulative number of TikTok user-generated videos a song has accrued by the end of the pre period; because scraping is irregular and we combine Soundcharts and Chartmetric, we use the largest stock count observed across scrapes and take the higher value when both sources are available for a song.\footnote{Soundcharts scrapes the TikTok creations for each song at irregular intervals. For the least popular songs with fewer than 100 creations in 15 days, they update the number every 30 days, while the most popular songs with over 10,000 creations get updated daily. Therefore, we define the virality measure over the entire pre-removal period.} Within this restricted pool, we select the control song that minimizes a weighted Euclidean distance to the treated song, where distances are computed based on (i) the nine-week pre-treatment Spotify streaming trajectory and (ii) standardized differences in song age and pre-treatment TikTok creations. To ensure close alignment in pre-treatment dynamics, we further exclude treated–control pairs for which the resulting match distance, normalized by the treated song’s average pre-treatment streaming level, exceeds the threshold of 0.1. Controls may be reused across treated songs; the resulting matched sample consists of 53,753 treated songs and 53,753 matched control observations, corresponding to 29,650 unique control songs.\footnote{Web Appendix~\ref{app:matching_procedures} reports robustness checks using alternative matching choices, including matching without replacement, exact matching on the pre-treatment TikTok-virality decile, and restricting the control pool to a single other major label. Because some control songs are reused across treated songs, we report two-way cluster-robust standard errors by song id and matched-pair id, which allows for arbitrary correlation within songs, including reused controls, and within matched pairs. Web Appendix~\ref{app:clustering_dimensions} reports inference under alternative clustering structures.} 

\autoref{tab:matchquality_songs} shows that treated and matched control songs are closely comparable across all pre-treatment characteristics, including TikTok creations, Spotify streaming, playlist exposure, and song age. Standardized mean differences for all variables fall well below the commonly used threshold of 0.1 \citep{austin_introduction_2011}, indicating a high degree of balance and suggesting that treated and control songs are comparable along these observable dimensions.

\section{Estimands, Estimators, and Economic Interpretation}
\label{sec:spec_diagnostics}

\subsection{Concentration in Outcomes: Viral Head and Long Tail}
\label{subsec:concentration_viral_head}

Both Spotify streams and TikTok creations in our sample exhibit the extreme concentration typical of heavy-tailed distributions in digital markets. We measure TikTok-side activity using creations, defined as the number of user-generated videos that incorporate a given song. Creations provide a meaningful measure of how actively a song is used in user-generated content on the platform. To measure TikTok exposure more directly, we would ideally also observe song-level TikTok views, but these data are not publicly available at broad scale. In a supplementary panel for a small set of songs, where we observe both creations and TikTok views, we find that the two measures are highly correlated ($r = 0.67$). We also find that views per creation increase with creation count, with a log-log elasticity of $0.89$. Thus, songs with more user-generated videos also tend to receive more views per video. This implies that views are even more concentrated than creations, which makes our creation-based measure conservative: it captures meaningful variation in TikTok activity, while likely understating the extent to which that activity is concentrated among the most popular songs. In this sense, creations provide a conservative basis for our methodological argument: the TikTok-side activity we observe is already highly concentrated, and view-based exposure would be even more dominated by the viral head.

\begin{figure}[!htbp] 
	\centering
	\caption{Concentration and Correlation Between TikTok Creations and Streams} 
	\label{fig:conc_and_corr}
    \begin{subfigure}{0.5\textwidth}
		\centering
		\caption{Concentration of Creations and Streams}
        \label{fig:conc_lorenz}
        \includegraphics[width=\textwidth]{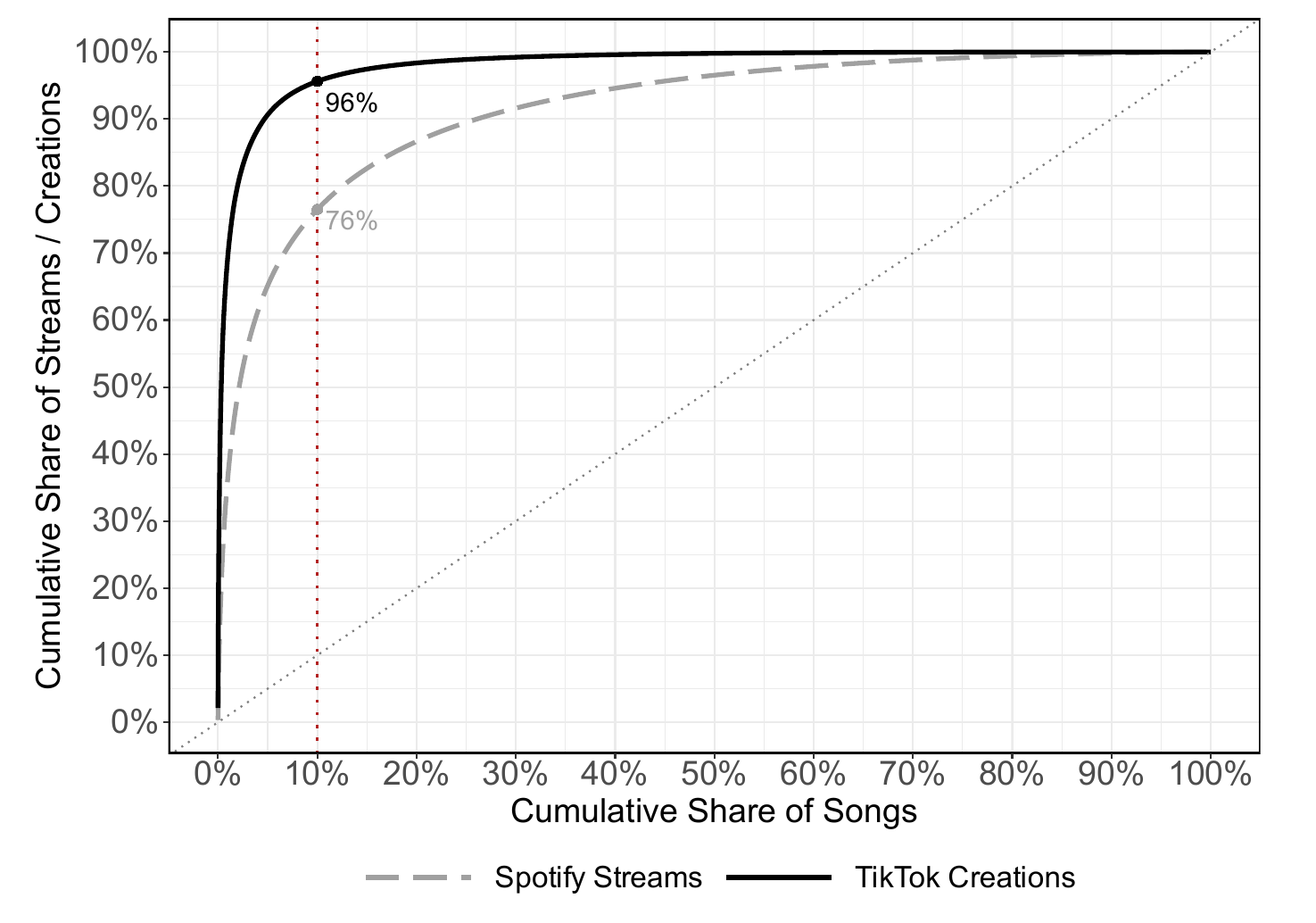}
	\end{subfigure}%
	\hfill
	\begin{subfigure}{0.5\textwidth}
		\centering
		\caption{Correlation Between Creations and Streams}
        \label{fig:concentration_side_by_side}
		\centering
		\includegraphics[width=1\textwidth, trim=0 0 0 0cm, clip]{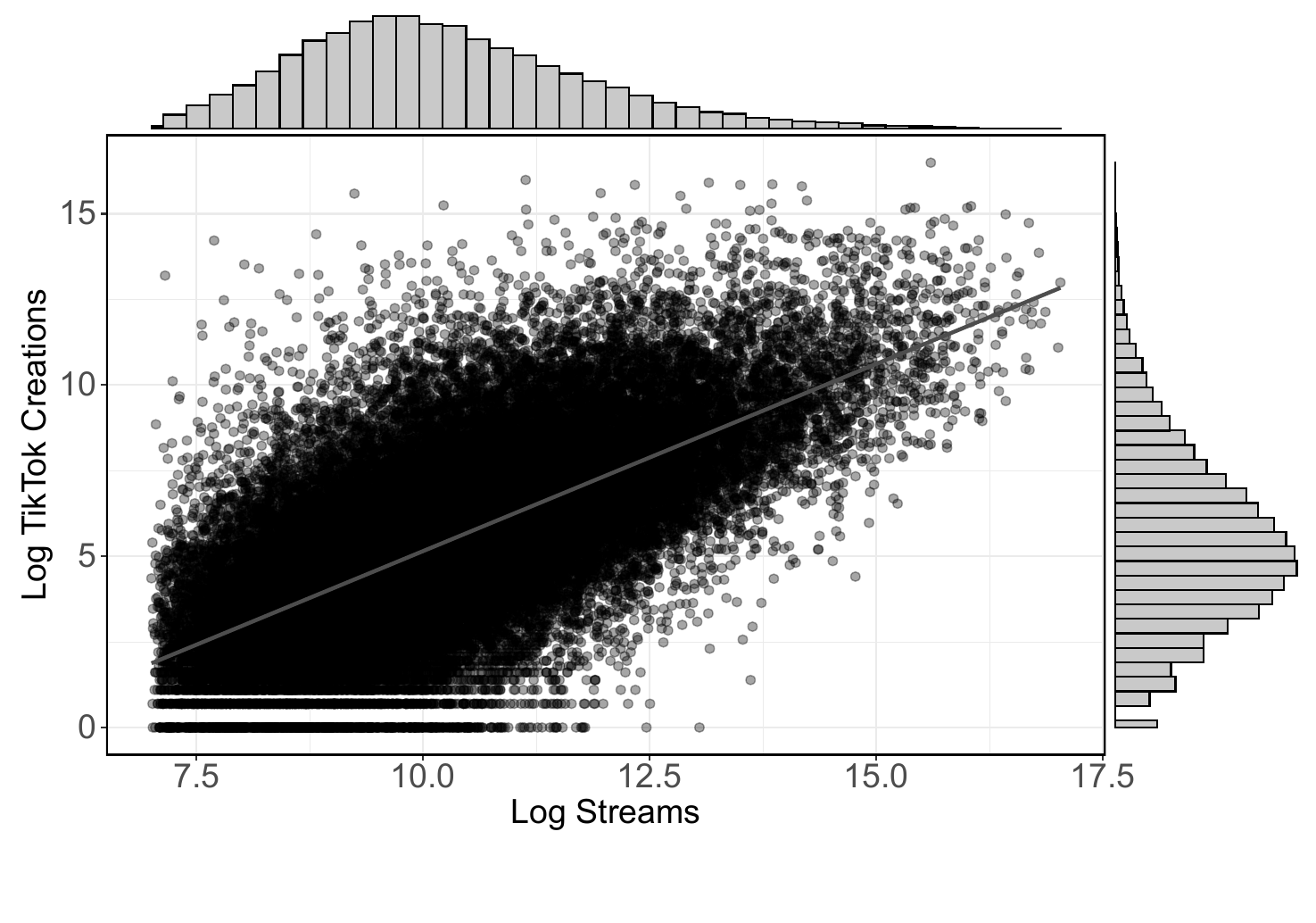}
	\end{subfigure}
	\raggedright\scriptsize
	\parbox{1\textwidth}{\textit{Notes}: Panel~(a) plots the cumulative share of TikTok creations and music streams against the cumulative share of songs, sorted in descending order of each outcome. The top 10\% of songs account for approximately 95.6\% of all TikTok creations and 76.5\% of all streams. Panel~(b) shows the association between log TikTok creations and log Spotify streams, with marginal histograms illustrating the distributions along each axis ($r = 0.65, p < 0.001$).}
    \vspace{-0.5cm}
\end{figure}

\begin{figure}[!htbp] 
	\centering
	\caption{Joint Distribution of Spotify Streams and TikTok Creations Across Deciles} 
	\label{fig:joint_dist}
	\begin{subfigure}{0.5\textwidth}
		\centering
		\caption{\% of Spotify Streams}
        \includegraphics[width=1\textwidth, trim=0 0 0 3cm, clip]{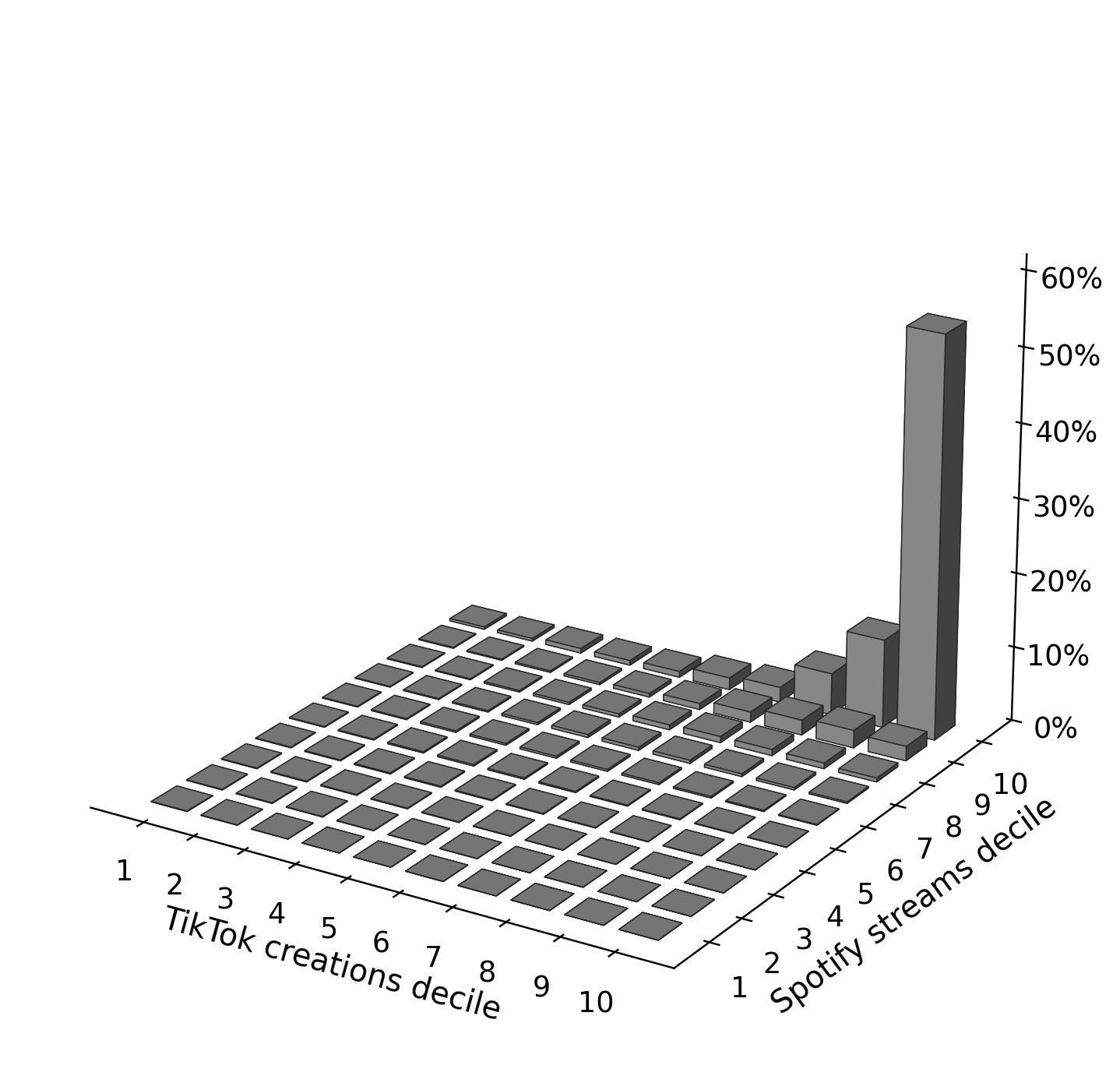}
	\end{subfigure}%
	\hfill
	\begin{subfigure}{0.5\textwidth}
		\centering
		\caption{\% of TikTok Creations}
		\centering
		\includegraphics[width=1\textwidth, trim=0 0 0 3cm, clip]{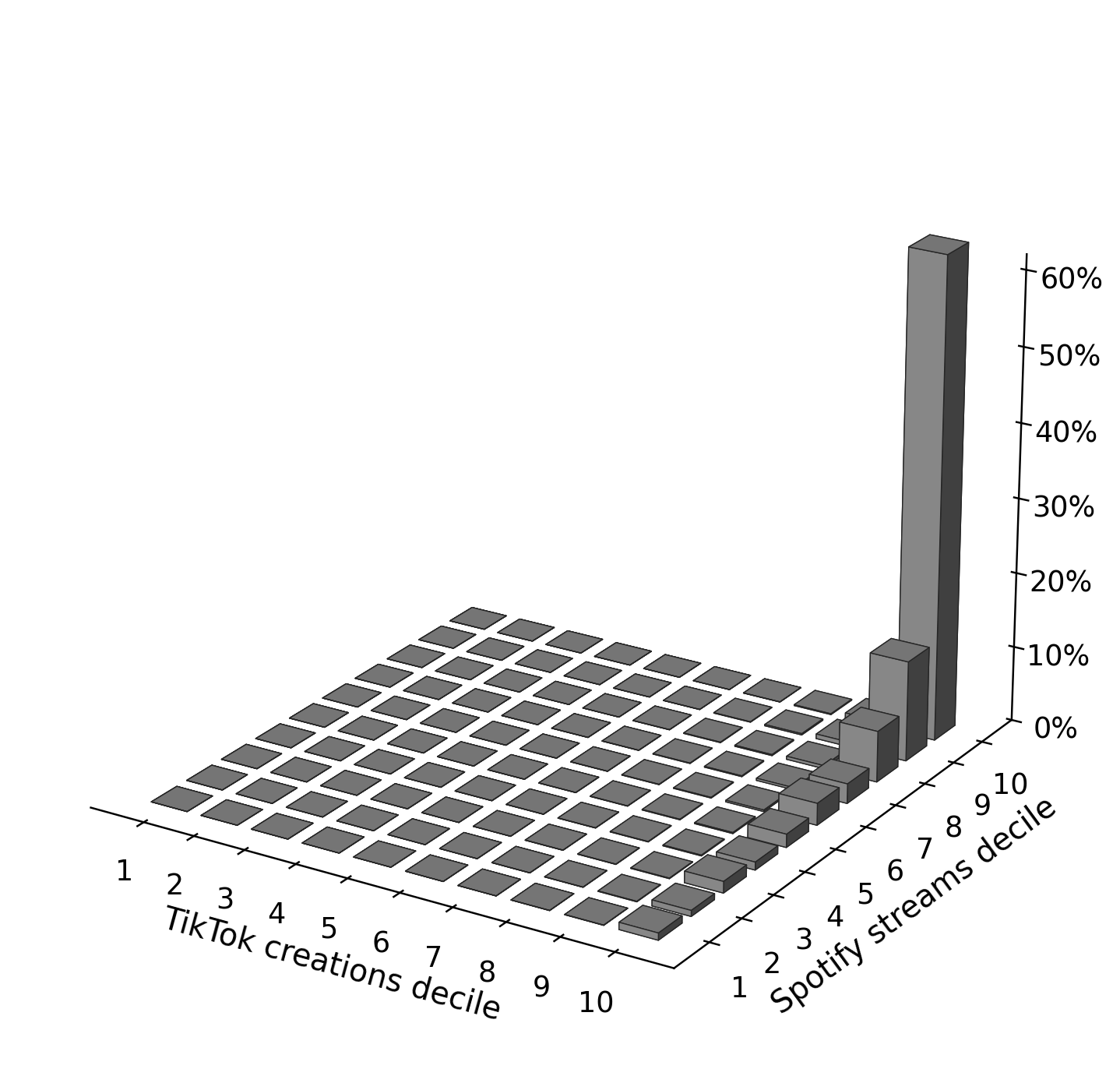}
	\end{subfigure}
	\raggedright\scriptsize
	\parbox{1\textwidth}{\textit{Notes}: The X-axis bins songs by deciles of TikTok creations, while the Y-axis bins the same songs by deciles of Spotify streams. Each bar corresponds to a cell in this 10$\times$10 grid. Panel (a) plots the share of total Spotify streams accounted for by songs in each TikTok--Spotify decile cell. Panel (b) plots the analogous share of total TikTok creations. Decile thresholds and within-decile means are available in the Web Appendix \ref{app:decile_cutoffs}.}
    \vspace{-0.5cm}
\end{figure}

\autoref{fig:conc_lorenz} plots Lorenz curves for TikTok creations and Spotify streams, and \autoref{fig:concentration_side_by_side} shows that the two measures are positively correlated ($r = 0.65$) but not identical. A nontrivial portion of TikTok-viral songs falls outside the top Spotify decile and vice versa. \autoref{fig:joint_dist} summarizes this joint structure by binning treated songs into a 10$\times$10 grid of TikTok-creation and Spotify-stream deciles. The top-right cell, songs simultaneously in the top decile of both creations and streams, accounts for the majority of all platform activity.

Since the withdrawal operates through TikTok-side availability, and the top TikTok creation decile accounts for approximately 96\% of all TikTok creations in the treated sample, we define the \textit{viral head} as the top 10\% of songs in the treated UMG sample ranked by pre-treatment TikTok creations. This group captures nearly all of the TikTok-side activity that can plausibly transmit to Spotify, while the remaining 90\% of songs, which we refer to as the \textit{long tail}, collectively account for only 4\% of pre-treatment UGC. This concentration has direct implications for our identification strategy: it motivates matching on TikTok-virality deciles and organizing the analysis by pre-treatment virality. It is also central to the economic interpretation of the estimates, because the typical song and the economically consequential song are not the same in this setting.

\subsection{Estimands and Estimators}
\label{subsec:why_consequential}

The distributional structure documented above makes the distinction between estimands and estimators first order. We use \emph{estimand} to refer to what we aim to learn about in the population, and \emph{estimator} to refer to the sample rule used to estimate it. In the discussion that follows, we use estimator and specification somewhat interchangeably as shorthand for the full empirical implementation. Technically, however, a \emph{specification} is broader: it combines the functional form, the maintained restriction on untreated evolution, the weighting scheme, and the fixed effects structure. With heavy-tailed outcomes, these choices jointly determine the economic interpretation of a DiD coefficient and the part of the outcome distribution that drives it. We begin with the canonical two-way fixed effects DiD in levels:
\begin{equation}
\label{eq:did_baseline_levels}
Y_{it}
=
\tau\,(\text{Treat}_i \times \text{Post}_t)
+\mu_i+\gamma_t+\varepsilon_{it},
\end{equation}
where $Y_{it}$ denotes Spotify streams for song $i$ in week $t$, $\mu_i$ and $\gamma_t$ are song and week fixed effects, $\text{Treat}_i$ indicates that song $i$ is treated (UMG title exposed to the withdrawal), and $\text{Post}_t$ indicates weeks after the withdrawal.\footnote{Results are unchanged when we replace song fixed effects with pair-by-song fixed effects, which allow the same control title to have a separate fixed effect in each matched pair where it appears.} Unless otherwise noted, our baseline DiD design defines the pre-period as Phase~1 (November 30, 2023 through January 31, 2024) and defines the post-period as the withdrawal window spanning Phases~2 and~3 (February 1, 2024 through April 25, 2024).

The same underlying treated--control contrast can be implemented with different functional forms and weighting choices. \emph{OLS in levels} summarizes changes in mean streams in absolute units and corresponds to parallel trends in absolute changes. \emph{Log-OLS} summarizes changes in mean log streams, naturally interpreted as equal-weighted proportional changes across songs. Previous research has established that log-OLS can be misleading about mean effects when log-variance changes around treatment because $\mathbb{E}[\log Y]$ does not move one-for-one with $\log \mathbb{E}[Y]$ (Jensen's inequality) \citep{log_gravity}. \emph{Weighted log-OLS} preserves the proportional interpretation but shifts the log-DiD toward songs that account for more baseline listening, using predetermined weights based on Phase~1 Spotify streams. Finally, \emph{Poisson pseudo-maximum-likelihood (PPML)} with a log link estimates a multiplicative conditional-mean model for nonnegative outcomes. It is widely used in gravity settings following \citet{log_gravity} and provides a natural way to implement nonlinear DiD specifications \citep{ciani_did_2019,wooldridge_simple_2023}. Rather than transforming the outcome and running OLS on $\log Y$, PPML works directly with the untransformed outcome $Y$ while assuming a multiplicative structure for the conditional mean. This means it targets the average effect on actual streams, not on log streams, so $\exp(\delta)-1$ is directly interpretable as a percent change in the conditional mean. Because PPML only requires the conditional mean to be correctly specified, it is robust to misspecification of the variance function, unlike log-OLS, whose coefficient can be biased for mean-stream effects when the variance of $\log Y$ changes around treatment.

In our setting, these choices are not interchangeable robustness checks: they target different estimands and impose different counterfactual trend restrictions. \autoref{tab:estimator_comparison} summarizes the key differences and properties of the different estimators. For applied guidance on selecting the DiD estimand and estimator under heavy-tailed outcomes, see the Practitioner's Companion. 

\begin{table}[ht!]
\renewcommand{\arraystretch}{1.15}
\footnotesize

\centering
\caption{A Guide to DiD Estimators with Heavy-tailed Outcomes}
\label{tab:estimator_comparison}

\begin{threeparttable}
    \begin{tabular}{P{1.7cm} P{3.8cm} P{4.6cm} P{4.6cm}}
    \toprule
    Estimator & Estimand, coefficient \& implicit weighting & Identifying restriction & When to use \\
    \midrule
    Levels OLS
    & DiD in mean: $\Delta\Delta\,\mathbb{E}[Y]$. Coefficient = absolute change in units. Large units dominate (via squared residuals).
    & \emph{Additive parallel trends}: untreated outcomes evolve in parallel on the level scale; treatment shifts outcomes by a fixed number of units.
    & When absolute, unit-denominated effects are the object of interest \emph{and} treated/control units are similar in baseline scale (or matched), so an additive counterfactual is credible. \\[2pt]
    Log OLS
    & DiD in mean log: $\Delta\Delta\,\mathbb{E}[\log Y]$. Coefficient $\approx$ proportional change for the typical unit. Equal weight per unit.
    & \emph{Multiplicative parallel trends} in the \emph{geometric} mean. If the coefficient is additionally interpreted as a proxy for $\Delta\Delta\,\log\mathbb{E}[Y]$, stable $\text{Var}(\log Y)$ across treatment $\times$ time cells is also required. Undefined at $Y=0$; $\log(1+Y)$ workaround yields unit-dependent ATEs that are not percentage effects.
    
    & When the typical-unit proportional effect is the target, $Y > 0$, units are comparable in scale, and $Var(\log Y)$ is stable across treatment $\times$ time cells.\\[2pt]
    Weighted log OLS
    & Weighted DiD in mean log: coefficient $\approx$ size-weighted proportional change. Weights proportional to pre-treatment outcome share (or other chosen weights). Viral head dominates.
    & Same as log OLS, applied to the reweighted population (e.g., pre-period shares). 
    & When aggregate (e.g., revenue- or platform-level) proportional effects are the target and a transparent, weight-driven log specification is desired. Useful sensitivity check for PPML when $Y>0$ and log-scale variance is stable.\\[2pt]
    PPML (log link)
    & DiD in the arithmetic mean: $\Delta\Delta\,\log\mathbb{E}[Y]$. Coefficient $\delta$; $\exp(\delta)-1$ = proportional change in $\mathbb{E}[Y]$. Weight proportional to predicted level.
    & \emph{Multiplicative parallel trends} in the \emph{arithmetic} mean: $\mathbb{E}[Y]$ scales by $\exp(\delta)$ under treatment. Only the conditional mean must be correctly specified; robust to variance misspecification. Natively handles $Y=0$, avoiding the unit-dependent ATEs from $\log(1+Y)$.
    & \emph{Recommended default for the population-total \% estimand under heavy-tailed outcomes}, especially with zeros, treatment-induced changes in dispersion, or aggregate proportional effects as the target.\\
    \bottomrule
    \end{tabular}%
    \begin{tablenotes}[para,flushleft]
    \scriptsize 
        \emph{Notes:} The table summarizes four common DiD implementations for non-negative outcomes. ``Estimand'' is the population object the coefficient targets, with implicit weighting describing which units disproportionately influence the pooled estimate. ``Identifying restriction'' refers to what is required for the \emph{coefficient} to be unbiased, not merely to the validity of standard errors. PPML and weighted log OLS often deliver similar conclusions; PPML is preferred when log-scale variance is unstable or zeros are present. See \citet{log_gravity} for log OLS vs.\ PPML, \citet{SolonHaiderWooldridge2015} on regression weights, and \citet{ChenRoth2024} on the unit-dependence of ATEs for common transformations.
    \end{tablenotes}
\end{threeparttable}
\vspace{-0.5cm}
\end{table}

\subsection{Different Estimators Yield Different Conclusions}
\label{subsec:understanding_divergence}

We begin by documenting how standard specifications of the DiD design can yield sharply different estimates in our setting.

\begin{table}[ht!]
    \centering
  {\singlespacing
    \renewcommand{\arraystretch}{1.0}
    \setlength{\tabcolsep}{5pt}
    
    \caption{DiD Estimates by Specification: Overall vs.\ Split by Baseline Virality}
    \small
    \label{tab:did_pooled_vs_split_specs}
    
    \begin{threeparttable}
    \begin{tabular}{lcccc}
    \toprule
     & Levels OLS & Log OLS & Weighted log OLS & PPML \\
    \midrule
    \multicolumn{5}{l}{\emph{Panel A. Overall effect}} \\
    $\text{Treat}_i \times \text{Post}_t$
    & $-4{,}661.2^{***}$ 
    & $0.0063^{***}$ 
    & $-0.0286^{***}$ 
    & $-0.0310^{***}$ \\
     & $(1{,}075.2)$ & $(0.0011)$ & $(0.0057)$ & $(0.0063)$ \\
    \addlinespace
    Song FE & Yes & Yes & Yes & Yes \\
    Week FE & Yes & Yes & Yes & Yes \\
    Observations & \multicolumn{4}{c}{2{,}365{,}132} \\
    Avg. Pre-streams (treated) & \multicolumn{4}{c}{142{,}545.4} \\
    $R^2$ / Pseudo $R^2$ & 0.9820 & 0.9946 & 0.9945 & 0.9846 \\
    \midrule
    \multicolumn{5}{l}{\emph{Panel B. Bottom 9 deciles vs.\ top decile}} \\
    $\text{Treat}_i \times \text{Post}_t$ (Bottom 9 deciles)
    & $-214.9$
    & $0.0085^{***}$
    & $0.0014$
    & $0.0036$ \\
     & $(315.3)$ & $(0.0011)$ & $(0.0038)$ & $(0.0043)$ \\
    $\text{Treat}_i \times \text{Post}_t$ (Top decile)
    & $-43{,}789.3^{***}$
    & $-0.0129^{***}$
    & $-0.0488^{***}$
    & $-0.0541^{***}$ \\
     & $(9{,}372.6)$ & $(0.0032)$ & $(0.0086)$ & $(0.0094)$ \\
    \addlinespace
    Song FE & Yes & Yes & Yes & Yes \\
    Week$\times$Group FE & Yes & Yes & Yes & Yes \\
    Observations & \multicolumn{4}{c}{2{,}365{,}132} \\
    Avg. Pre-streams (Bottom 9 d.) & \multicolumn{4}{c}{64{,}002.2} \\
    Avg. Pre-streams (Top d.) & \multicolumn{4}{c}{833{,}725.3} \\
    $R^2$ / Pseudo $R^2$ & 0.9821 & 0.9946 & 0.9945 & 0.9849 \\
    \bottomrule
    \end{tabular}
    \begin{tablenotes}[para,flushleft]
    \scriptsize
    \emph{Notes:} Panel A reports pooled DiD estimates from four specifications: OLS in levels, OLS in logs, weighted OLS in logs, and Poisson pseudo-maximum-likelihood (PPML) with a log link. Panel B allows the treatment effect to differ between songs in the bottom nine baseline TikTok-virality deciles and songs in the top decile. For the levels specification, given the reported pre-treatment streams for treated songs and scaling the levels coefficients by these baselines implies effects of $-3.27\%$ in Panel A, $-0.34\%$ for the bottom nine deciles in Panel B, and $-5.25\%$ for the top decile in Panel B. The weighted log-OLS specification weights observations by each song's pre-treatment stream share. All specifications report Huber–White sandwich standard errors, two-way clustered by song and matched pair. Details on the matching procedure are provided in the Web Appendix~\ref{app:matching_procedures}. Significance levels: $^{*}p<0.05$, $^{**}p<0.01$, $^{***}p<0.001$.
    \end{tablenotes}
    \end{threeparttable}
   }
   \vspace{-0.5cm}
\end{table}

Results in Panel A of \autoref{tab:did_pooled_vs_split_specs} illustrate the core empirical tension. The unweighted log-OLS specification delivers a small but statistically significant \emph{positive} coefficient of $0.0063$, which implies that the typical song's Spotify streams increased by about 0.63\% after the withdrawal, a finding that, taken at face value, would suggest TikTok \emph{substitutes} for paid streaming. This is qualitatively aligned with the headline result in \citet{cheng_value_2024}. The remaining three estimators all deliver negative coefficients. OLS in levels estimates a decline of about 4{,}661 streams per song per week, which is qualitatively consistent with the levels-based findings in \citet{bairathi_lambrecht_rao_2024}. Weighted log-OLS, PPML, and levels OLS yield proportional declines of $-2.9\%$, $-3.1\%$, and $-3.3\%$, respectively (where the levels estimate is scaled by treated songs' average pre-treatment streams). The sign reversal between log-OLS and the other three specifications, using the same data and panel structure, motivates a closer examination of what each specification estimates.

The source of divergence lies in the distributional structure documented in Section~\ref{subsec:concentration_viral_head}. An unweighted log-OLS specification gives equal weight to every song’s log change. Because 90\% of songs are in the long tail and contribute little TikTok-side exposure, this specification is dominated by songs for which the withdrawal changes little in economic terms. In contrast, levels OLS and PPML place more weight on the viral head, where the economically meaningful declines are concentrated: levels OLS through its sensitivity to large absolute changes, and PPML through the conditional mean in levels. We return to weighted log-OLS in Section~\ref{subsec:preferred_spec}, where we use it to separate the role of weighting from the role of functional form.

Panel B of \autoref{tab:did_pooled_vs_split_specs} confirms this interpretation directly by re-estimating the same four specifications while allowing the treated-post coefficient to differ between songs in the bottom nine TikTok-virality deciles, the long tail, and songs in the top decile, the viral head. Two patterns are immediately apparent. First, the treatment effect for songs in the top decile is uniformly negative across all four specifications. Second, in the bottom nine deciles, the log-OLS specification continues to deliver a positive and statistically significant coefficient, whereas the corresponding estimates from OLS in levels, weighted log-OLS, and PPML are small and statistically indistinguishable from zero. The positive pooled log-OLS estimate in Panel A is therefore not being driven by the most economically consequential songs. Instead, it reflects how unweighted log-OLS aggregates changes across the much larger mass of lower-virality songs, which can dominate a ``typical-song'' proportional-change summary even when those songs contribute little to platform-level outcomes.

\begin{figure}[!htbp]
  \centering
  \caption{Treatment Effects by Pre-Treatment TikTok Virality}
  \label{fig:decile_soc}
  \begin{subfigure}{0.75\textwidth}
    \centering
        \caption{Share of TikTok Creations}
        \vspace{-0.6em}
    \includegraphics[width=\textwidth]{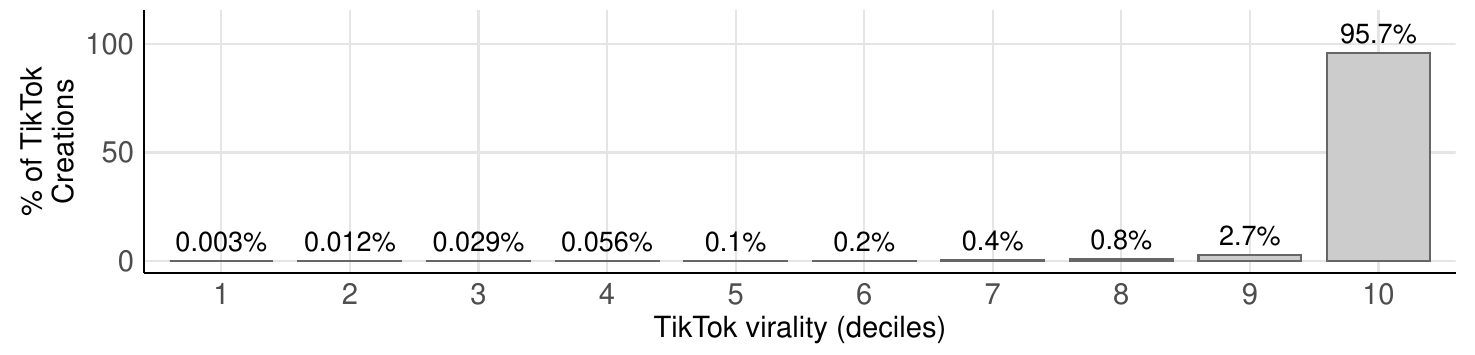}
  \end{subfigure}
  
  \begin{subfigure}{0.75\textwidth}
    \centering
     \label{fig:decile_sos}
        \caption{Share of Spotify Streams}
              \vspace{-0.6em}
    \includegraphics[width=\textwidth]{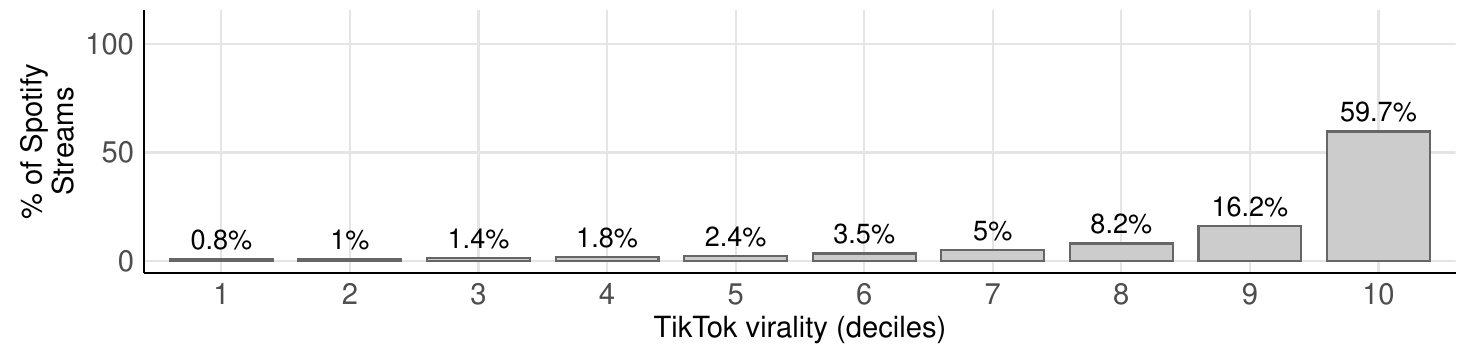}
  \end{subfigure}

  \begin{subfigure}{0.75\textwidth}
    \centering
     \label{fig:decile_treat_lev_ols}
        \caption{Levels OLS}
              \vspace{-0.6em}
    \includegraphics[width=\textwidth]{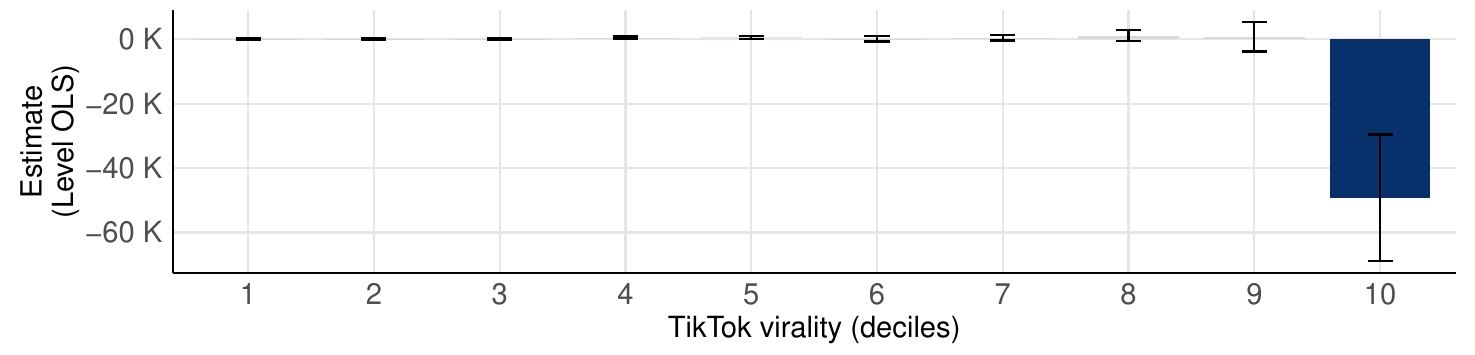}
  \end{subfigure}

   \begin{subfigure}{0.75\textwidth}
    \centering
     \label{fig:decile_treat_log_ols}
        \caption{Log OLS}
              \vspace{-0.6em}
    \includegraphics[width=\textwidth]{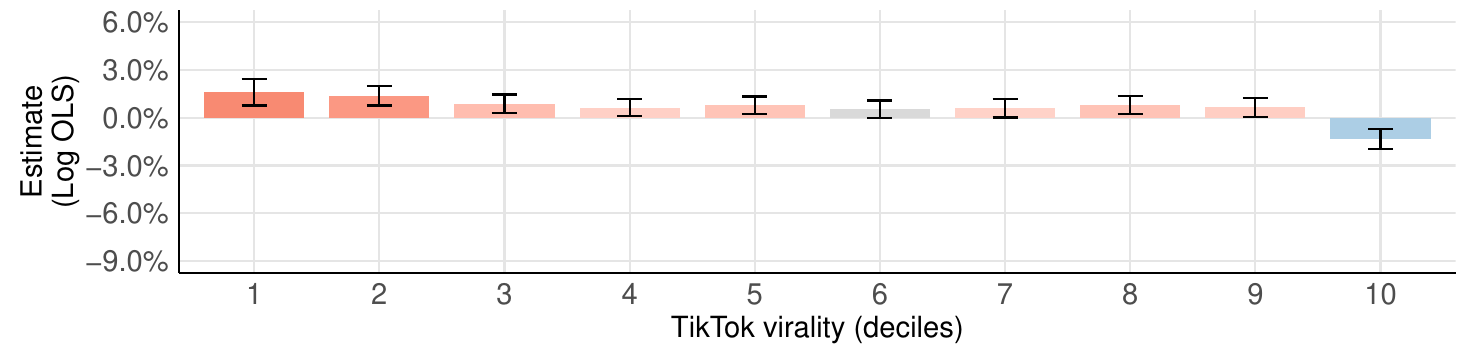}
  \end{subfigure}

   \begin{subfigure}{0.75\textwidth}
    \centering
     \label{fig:decile_treat_ppl}
        \caption{PPML}
              \vspace{-0.6em}
    \includegraphics[width=\textwidth]{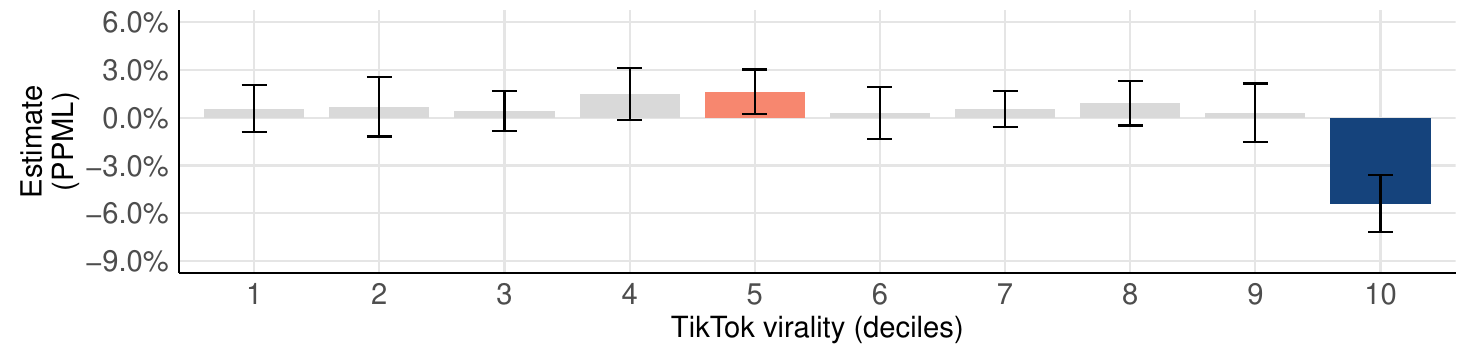}
  \end{subfigure}

  \vspace{0.01em}
  
  {\raggedright\scriptsize
  \parbox{\textwidth}{\textit{Notes}: Panels (a)--(b) show the pre-treatment distribution of TikTok creations and Spotify streams by virality decile. Panels (c)--(e) report decile-specific DiD estimates from fully interacted models with song-group and time fixed effects. Log and Poisson coefficients are shown as percentage effects. Error bars represent 95\% confidence intervals with standard errors two-way clustered by song and match-pair.}\par}
\end{figure}

\autoref{fig:decile_soc} sharpens this diagnosis. Panels (a)--(b) show again the economic stakes: the top TikTok-virality decile accounts for essentially all TikTok Creations and a majority of Spotify streams, whereas each of the bottom deciles contributes only a negligible share. Panels (c)--(e) report decile-specific DiD coefficients under levels OLS, log OLS, and PPML. The negative effect in the viral head is visible across all three specifications. The key divergence appears in the long tail. Unweighted log-OLS, panel (d), delivers positive and statistically significant coefficients across most bottom deciles, whereas levels OLS, panel (c), and PPML, panel (e), remain close to zero and mostly insignificant. Together, these decile-specific estimates show that the pooled disagreement across specifications is driven by how each estimator aggregates the long tail and the viral head.

\subsection{Preferred Estimator}
\label{subsec:preferred_spec}

The results above show that the choice of estimator is consequential. To choose among the candidate estimators, we focus on two questions: which counterfactual scale is most plausible, and which estimand matches the economic object of interest?

On the first question, streaming dynamics are more naturally described in proportional terms than in absolute changes. A shock that removes TikTok exposure is unlikely to add or subtract the same number of Spotify streams for a song with 2{,}000 weekly streams and a song with 2{,}000{,}000 weekly streams. More plausibly, the absolute change scales with baseline popularity. The same logic applies to untreated counterfactual dynamics: in a growing market, common proportional growth implies larger absolute changes for songs with higher baseline demand. This motivates a proportional specification. A related concern is that, when outcomes scale with baseline size, initial gaps between treated and control units can generate differential growth that levels OLS may attribute to treatment. In the Web Appendix~\ref{app:additive_vs_mult}, we use a simulation to show that this bias grows with both the baseline gap and the common growth rate, and can even reverse the sign of the estimated effect. Our matching procedure mitigates this concern by closely aligning treated and control songs on pre-treatment levels and trajectories, which shrinks the relevant baseline gap in the matched sample and largely closes this bias channel.

On the second question, the economically relevant margin in our setting is the aggregate, i.e., label-level, proportional change in streams, not the typical-song percent change. The typical song in our sample contributes negligibly to total listening and revenue. Unweighted log-OLS estimates the DiD in mean log streams, which gives equal weight to every song's log change and is therefore dominated by these economically marginal titles. A further complication is that log-OLS can deliver a biased estimate of the mean-stream treatment effect when the variance of log streams changes differentially around treatment. In that case, Jensen's inequality implies that the DiD in mean log streams, $\mathbb{E}[\log Y]$, no longer tracks the corresponding change in mean streams, $\log\mathbb{E}[Y]$. Web Appendix~\ref{app:hetero} documents this pattern in our data and shows through a calibrated simulation that it is quantitatively relevant.

We therefore focus on two proportional implementations that target aggregate-relevant estimands: PPML with a log link and weighted log-OLS using predetermined pre-treatment stream shares as importance weights.\footnote{The levels specification delivers a similar aggregate implication in our application because the matching procedure largely eliminates the bias channel that arises when treated and control units differ in baseline levels (see Web Appendix~\ref{app:additive_vs_mult} for details). By matching closely on pre-treatment streaming levels and trajectories, our design makes the additive and proportional counterfactuals locally similar over the relevant window.} PPML models the conditional mean in levels under a multiplicative structure, so it only requires the conditional mean to be correctly specified and is robust to heteroskedasticity in the variance \citep{log_gravity,SantosSilvaTenreyro2011}. Weighted log-OLS preserves the proportional interpretation but shifts the log-DiD toward songs that account for more baseline listening \citep{SolonHaiderWooldridge2015}. In our data, the two approaches deliver very similar results. This similarity is informative: holding the log functional form fixed, moving from unweighted log-OLS (column 2 of Table~\ref{tab:did_pooled_vs_split_specs}, $+0.0063$) to weighted log-OLS (column 3, $-0.0286$) flips the sign of the pooled estimate. Weighting alone, without changing the outcome transformation, aligns the log specification with the levels and PPML conclusions because it shifts effective weight away from the long tail and toward the viral head. In what follows, we present PPML as the main specification in the text and report weighted log-OLS in the Web Appendix~\ref{app:wgt_logols_main} as a transparency check.\footnote{Web Appendix~\ref{app:ppml_vs_wlog} compares the two estimands in more detail, clarifying that PPML identifies a semi-elasticity of the conditional mean while weighted log-OLS identifies a weighted DiD in the geometric mean, and explains why baseline-share weighting and the log-link mean structure deliver the same qualitative conclusions in our application. Weighted log-OLS targets a weighted average of song-level log changes, which in principle can still be affected by treatment-induced changes in log-scale variance. In our application this concern is attenuated because the largest instability arises in the lowest deciles, which receive little weight. PPML avoids this issue by construction, since it only requires the conditional mean to be correctly specified.}

\subsection{Results: Phase-by-Phase Estimates}
\label{sec:phase_dynamics}

\begin{figure}[ht!]
  \centering
  \caption{PPML Event-Study Estimates of the UMG Withdrawal Effects on Streaming}
  \label{fig:event_study}

  \begin{subfigure}{0.49\textwidth}
    \centering
    \caption{Pooled estimates}
    \includegraphics[width=\textwidth]{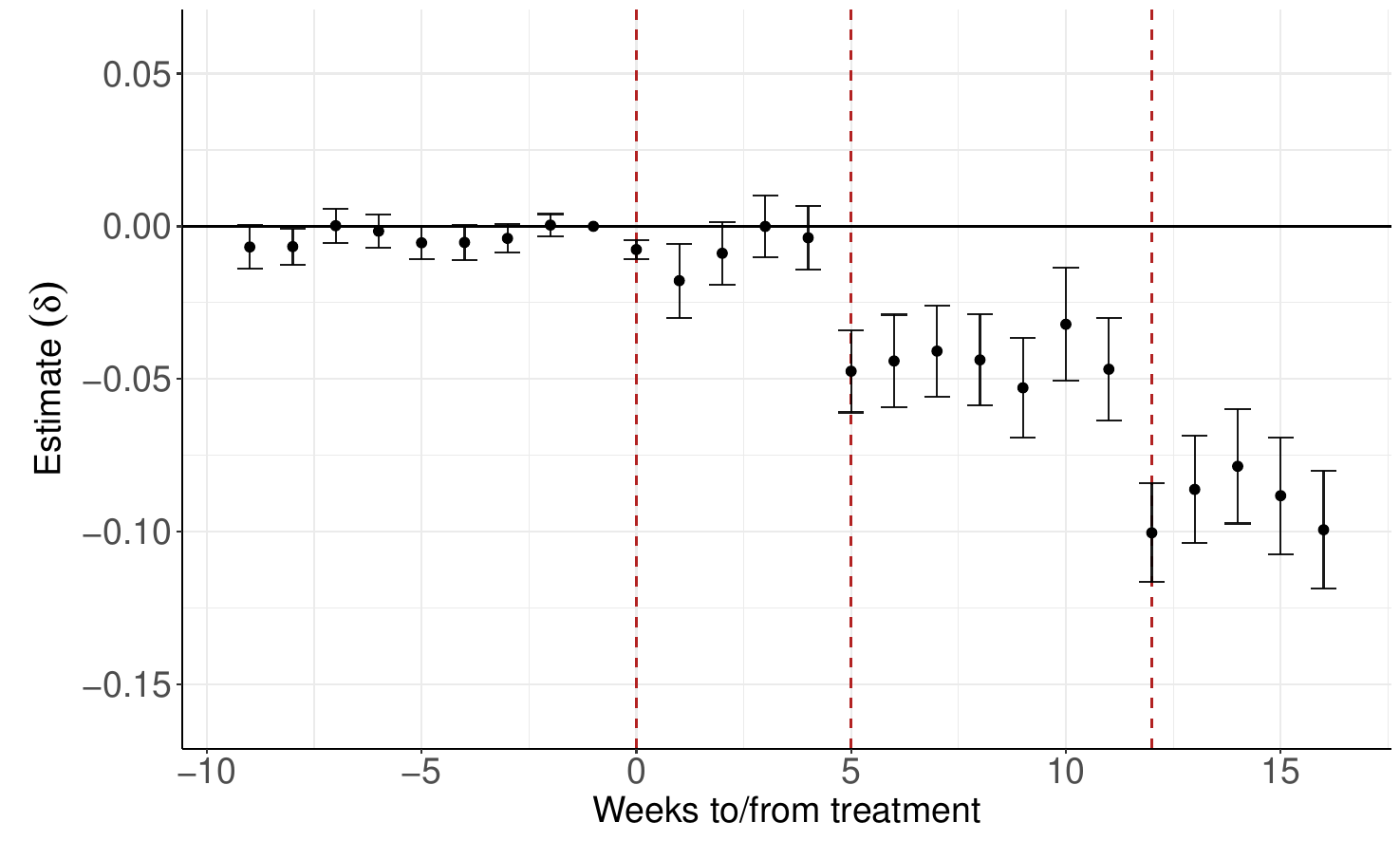}
    \label{fig:event_study_agg}
  \end{subfigure}%
  \hfill
  \begin{subfigure}{0.49\textwidth}
    \centering
    \caption{Estimates by baseline virality group}
    \includegraphics[width=\textwidth]{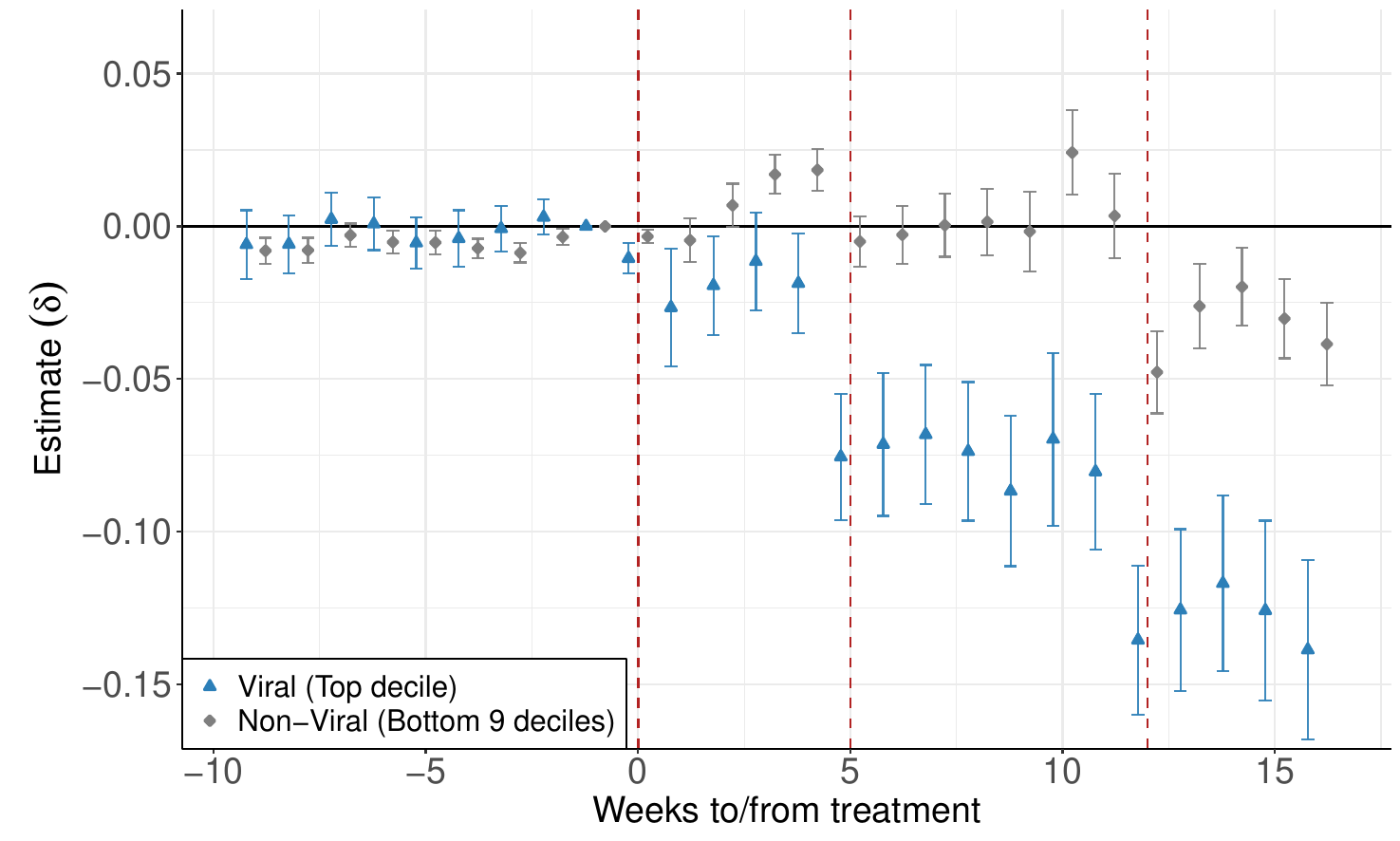}
    \label{fig:event_study_group}
  \end{subfigure}
\vspace{-0.5cm}
  \begin{justify}
\scriptsize{\textit{Notes:} The figure reports PPML event-study estimates. Coefficients are normalized to a pre-treatment reference week and reported in log points, which approximate percentage changes. Bars indicate 95\% confidence intervals. Vertical dashed lines mark the phase boundaries: Phase 2 begins with the UMG master-rights exit from TikTok, Phase 3 begins with the UMG publishing-rights exit, and Phase 4 begins with UMG re-entry (see \autoref{fig:tiktok_chart_global}). Panel (a) reports pooled estimates with song and week fixed effects. Panel (b) reports estimates from a split specification by baseline TikTok virality group, defined as the top decile versus the bottom nine deciles of pre-treatment creations. Standard errors are Huber--White sandwich standard errors, two-way clustered by song and match-pair.}
  \end{justify}
  \vspace{-0.5cm}
\end{figure}

Having selected PPML as our focal estimator, we now examine how the treatment effect evolves across the dispute phases. We estimate an event-study version of the DiD that interacts treatment status with indicators for each phase, while absorbing song and week fixed effects:
\begin{align}
\label{eq:ppml_eventstudy_stages}
\mathbb{E}\!\left[ Y_{it}\mid \cdot \right]
&=
\exp\!\Big(
\mu_i+\gamma_t
\Big)\cdot
\exp\!\Big(
\delta_{2}\,\big(\text{Treat}_i \times \mathbbm{1}[t\in \text{Phase 2}]\big)
\Big) \nonumber\\
&\quad\cdot
\exp\!\Big(
\delta_{3}\,\big(\text{Treat}_i \times \mathbbm{1}[t\in \text{Phase 3}]\big)
+\delta_{4}\,\big(\text{Treat}_i \times \mathbbm{1}[t\in \text{Phase 4}]\big)
\Big)
\end{align}

\noindent where $Y_{it}$ denotes weekly streams for song $i$ in week $t$. $\text{Treat}_i$ indicates treated songs, $\mu_i$ are song fixed effects, and $\gamma_t$ are week fixed effects. $\mathbbm{1}[t\in \text{Phase }s]$ is an indicator for week $t$ falling in Phase $s$ (with Phase 1 as the omitted reference period). In brief, Phase 1 is the pre-treatment period; Phase 2 begins with the UMG master-rights exit from TikTok; Phase 3 begins with the UMG publishing-rights exit; and Phase 4 begins with UMG re-entry to TikTok (see \autoref{fig:tiktok_chart_global} for the details). To generate \autoref{fig:event_study} we replace the phase-specific indicators with week indicators.

\begin{table}[htb]
    \centering
    \caption{PPML Estimates by Dispute Phase: Pooled vs. Split by Baseline Virality}
    \label{tab:ppml_eventstudy_phases_split}
    
    \begin{threeparttable}
    \renewcommand\TPTminimum{\textwidth}
    
    \small
    \setlength{\tabcolsep}{5pt}
    \renewcommand{\arraystretch}{1.0}
    
    \begin{tabularx}{\textwidth}{@{}X *{6}{>{\centering\arraybackslash}p{1.55cm}}@{}}
    \toprule
    & \multicolumn{2}{c}{Overall}
    & \multicolumn{2}{c}{Bottom 9 deciles}
    & \multicolumn{2}{c}{Top decile} \\
    \cmidrule(lr){2-3}\cmidrule(lr){4-5}\cmidrule(l){6-7}
    & Est. & S.E. & Est. & S.E. & Est. & S.E. \\
    \midrule
    
    Phase 2: Week 0--4
    & $-0.0044$ & $(0.0046)$
    & $0.0123^{***}$ & $(0.0028)$
    & $-0.0156^{*}$ & $(0.0073)$ \\
    
    Phase 3: Week 5--11
    & $-0.0477^{***}$ & $(0.0083)$
    & $0.0020$ & $(0.0059)$
    & $-0.0807^{***}$ & $(0.0126)$ \\
    
    Phase 4: Week 12--16
    & $-0.0849^{***}$ & $(0.0099)$
    & $-0.0233^{***}$ & $(0.0070)$
    & $-0.1251^{***}$ & $(0.0151)$ \\
    
    \addlinespace
    
    Song FE
    & \multicolumn{2}{c}{Yes}
    & \multicolumn{2}{c}{Yes}
    & \multicolumn{2}{c}{Yes} \\
    
    Week FE
    & \multicolumn{2}{c}{Yes}
    & \multicolumn{2}{c}{Yes}
    & \multicolumn{2}{c}{Yes} \\
    
    Observations
    & \multicolumn{2}{c}{2{,}795{,}156}
    & \multicolumn{2}{c}{2{,}509{,}936}
    & \multicolumn{2}{c}{285{,}220} \\
    
    Pseudo $R^2$
    & \multicolumn{2}{c}{0.9957}
    & \multicolumn{2}{c}{0.9930}
    & \multicolumn{2}{c}{0.9932} \\
    
    \bottomrule
\end{tabularx}
\begin{tablenotes}[para,flushleft]
\scriptsize\setstretch{1}\emph{Notes:} The table reports PPML estimates by dispute phase. Coefficients are reported in log points, which approximate percentage changes. Treatment status is interacted with mutually exclusive post-phase indicators, with Phase 1 as the omitted pre-treatment period. Phase 2 begins with the UMG master-rights exit from TikTok, Phase 3 begins with the UMG publishing-rights exit, and Phase 4 begins with UMG re-entry (see \autoref{fig:tiktok_chart_global}). The Overall column reports pooled estimates with song and week fixed effects. The Bottom 9 deciles and Top decile columns report estimates from a split specification by baseline TikTok virality group, defined by pre-treatment creations, with song fixed effects and group-by-week fixed effects. The observation period is 26 weeks. Standard errors are Huber--White sandwich standard errors, two-way clustered by song and match-pair. Significance levels: $^{*}p<0.05$, $^{**}p<0.01$, $^{***}p<0.001$.
\end{tablenotes}

\end{threeparttable}
\vspace{-0.5cm}
\end{table}

\autoref{fig:event_study} and \autoref{tab:ppml_eventstudy_phases_split} summarize the main results. In \autoref{fig:event_study} Panel (a), the pre-treatment coefficients for the pooled estimates are close to zero, and the post-treatment path becomes progressively more negative as the dispute moves from Phase 2 (master-rights exit) to Phase 3 (publishing-rights exit) and then into Phase 4 (re-entry). \autoref{tab:ppml_eventstudy_phases_split} quantifies this pattern in phase windows: the pooled effect is small in Phase 2 (about $-0.4\%$), becomes meaningfully negative in Phase 3 (about $-4.7\%$), and grows further in Phase 4 (about $-8.1\%$).

Panel (b) of \autoref{fig:event_study} shows that this pooled decline is driven by the viral head, and adds a temporal dimension to the heterogeneity documented in Section~\ref{subsec:understanding_divergence}. For the top decile, the phase-window estimates are negative already in Phase 2 and deepen substantially in Phases 3 and 4, whereas the bottom nine deciles exhibit at most small changes throughout the dispute. The losses in the viral head therefore not only dominate the pooled effect but also accumulate over time, growing larger the longer TikTok exposure remains disrupted. Web Appendix~\ref{app:mfe_by_decile} corroborates this pattern in a model-free way by plotting weekly mean streams for treated and matched control songs separately within each pre-treatment TikTok-virality decile; the matched series track each other tightly in the pre period across all deciles, and the post-period divergence is visually concentrated in the top decile, while the lower deciles show much smaller movements.

Three features of the documented patterns are worth noting. First, the decline becomes larger in Phase 3, even though the treated and control group definitions remain constant across Phases 2 and 3. For the treated UMG master recordings, the Phase 2 withdrawal had already removed official audio from TikTok, so the Phase 3 publishing exit imposed no additional mechanical restriction on these specific tracks. Our control group construction also excludes non-UMG masters with UMG publishing rights, which insulates the control group from the Phase 3 shock. The widening stream gap therefore cannot be explained by a change in the mechanical availability difference between treated and control songs. Instead, it points to mechanisms that may accumulate over time, including the erosion of TikTok-driven promotional momentum and Spotify-side amplification, which we examine in the next section.

Second, the Phase~3 escalation may also have increased cross-catalog reallocation. In early March 2024, the dispute extended to publishing rights, which removed tens of thousands of additional UMG-affiliated songs from TikTok. Although this did not further restrict the treated master recordings in our sample, and although the control group is insulated from the publishing-rights shock by construction, the escalation increased the overall amount of unavailable music on TikTok. This could have shifted more of creator or listener activity toward non-UMG content, which would raise control-group streams and widen the treated-minus-control gap. We discuss this interference concern and its economic interpretation in Section~\ref{sec:sutva}.

Finally, the persistence of sizable effects into Phase 4 suggests that re-entry does not mechanically undo what was missed during the withdrawal window. If TikTok exposure helps tracks convert short-run attention into Spotify-side distribution that accumulates over time, a temporary disruption can leave a gap that does not close immediately once the catalog returns, especially after the original viral moment has passed. A related possibility is that attention at re-entry shifts toward newer UMG releases outside the pre-period cohort, limiting the scope for older treated titles to regain exposure. We return to these interpretations in the next section.

\section{Interpreting the Treatment Effect}

\label{sec:identification}

The results above show that removing TikTok exposure lowers Spotify streams for highly TikTok-exposed UMG titles. We now turn to the interpretation of this effect. We first present evidence consistent with a discovery channel: losses are larger among the most TikTok-viral titles, songs with no pre-treatment TikTok exposure do not exhibit corresponding declines, and playlist evidence suggests that lost TikTok momentum propagates through Spotify-side distribution in a way that does not immediately reverse when UMG content is reintroduced. We then address the main identification concern created by the label-specific nature of the UMG withdrawal: cross-catalog reallocation may increase exposure and streams for non-UMG control songs. We show that this reallocation is limited in the creator-side data, and we clarify why the combined treated–control gap is economically meaningful when compensation depends on relative stream shares. Finally, we use the 2025 U.S.\ TikTok outage, which affects all labels symmetrically, to assess whether the promotional-channel interpretation persists when the label-specific reallocation channel is shut down.

\subsection{Evidence Consistent with Discovery}

\label{subsec:discovery}

The results above show that the negative treatment effects of the TikTok withdrawal are concentrated in the viral head. We now provide three pieces of evidence that help interpret this pattern as consistent with TikTok operating as a discovery channel.

\begin{figure}[ht!]
  \centering
 \caption{TikTok Exposure and Spotify-Side Amplification}
  \label{fig:discovery_amplification}
  \begin{subfigure}{0.49\textwidth}
    \centering
   \caption{Heterogeneity within the viral head}
    \includegraphics[width=\textwidth]{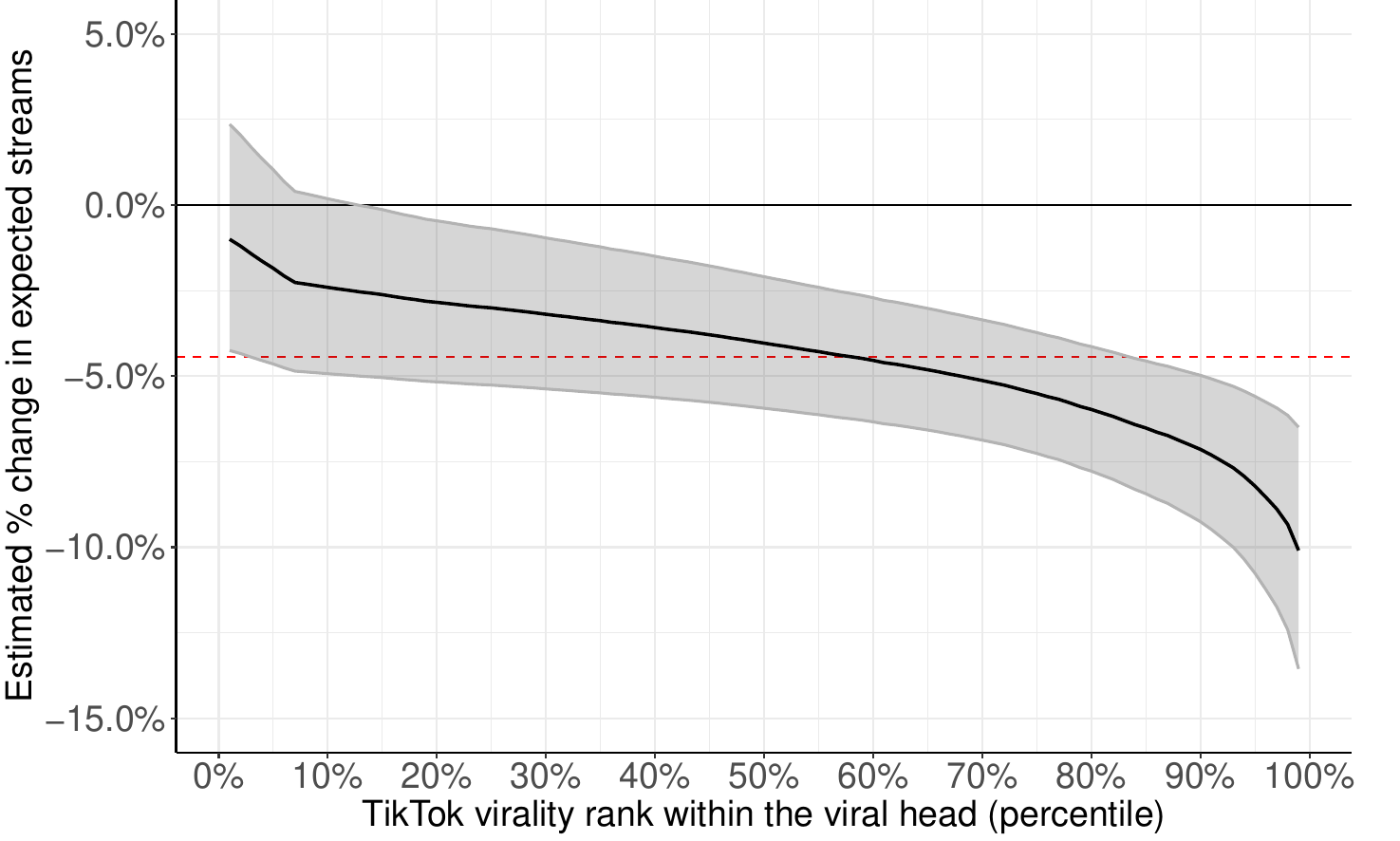}
    \label{fig:het_viral_head}
  \end{subfigure}%
  \hfill
  \begin{subfigure}{0.49\textwidth}
    \centering
   \caption{Curated-playlist followers}
    \includegraphics[width=\textwidth]{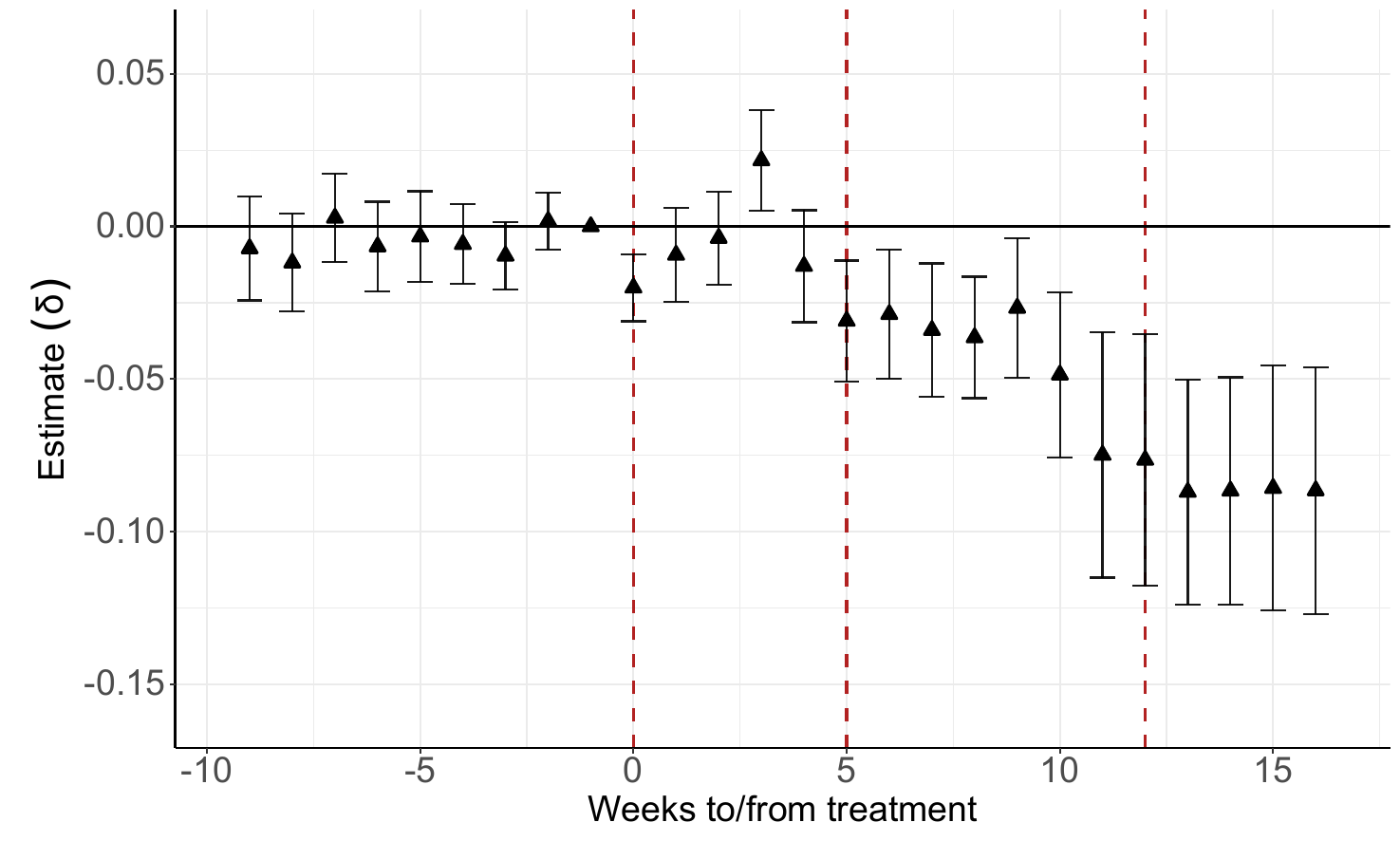}
    \label{fig:event_study_plfollowers}
  \end{subfigure}
  \vspace{-0.6cm}
  \begin{justify}
   \scriptsize{\textit{Notes:} Both panels focus on viral-head songs, defined as the top baseline TikTok-virality decile based on pre-treatment creations. Panel (a) visualizes heterogeneity in the streaming effect within this top decile. The x-axis ranks songs by pre-treatment TikTok virality within the viral head, in percentiles. The y-axis reports the implied total percent change in expected streams from the PPML DiD with an interaction between $\text{Treat}_i\times\text{Post}_t$ and mean-centered $\log(\text{TikTok\_creations})_i$. The black line plots the estimated total effect evaluated at each percentile of pre-treatment virality; the shaded area indicates 95\% confidence intervals. The red dashed horizontal line indicates the mean treatment effect across all viral-head songs. Panel (b) reports PPML event-study estimates for the effect of the UMG withdrawal on cumulative Spotify curated-playlist followers, again restricted to viral-head songs. Coefficients are normalized to the last pre-treatment week and reported in log points; whiskers indicate 95\% confidence intervals. Vertical dashed lines mark the phase boundaries (see \autoref{fig:tiktok_chart_global}). In both panels, standard errors are Huber--White sandwich standard errors, two-way clustered by song and match-pair.}
  \end{justify}
  \vspace{-0.5cm}
\end{figure}

First, we ask whether there is additional heterogeneity even within the viral head. \autoref{fig:discovery_amplification} Panel (a) summarizes the implied percent change in expected streams across the distribution of pre-treatment TikTok creation virality within the top decile. The effect is already negative at the lower end of the viral head and becomes progressively more negative as pre-treatment TikTok virality rises. Put differently, among songs that are all ``viral'' by our baseline definition, the largest streaming losses occur for the songs that generated the most TikTok UGC before the withdrawal.

Second, we examine whether the loss of TikTok exposure propagates through Spotify-side discovery. TikTok-driven momentum may affect Spotify demand not only through contemporaneous cross-platform discovery, but also through Spotify's own distribution system. Placements on large curated playlists and the associated follower accumulation are path-dependent and may update with delay, so missing the TikTok momentum window can reduce subsequent playlist exposure and generate a larger streaming gap in later weeks. \autoref{fig:discovery_amplification} Panel (b) shows that treated songs in the viral head experience a gradual and sustained decline in curated-playlist followers relative to controls, with the deterioration becoming more pronounced around Phase 3 and persisting even after UMG re-entry.

Third, we use a placebo-style test based on UMG songs with zero pre-treatment TikTok creations. For these titles, the withdrawal does not remove a meaningful pre-existing TikTok exposure channel because they were not used in UGC prior to the event. If the declines we document for TikTok-exposed songs reflect the loss of TikTok-based discovery and engagement, then songs that were never part of that channel should not exhibit a corresponding break in Spotify streaming around the dispute. We re-estimate the same phase-based specification on this subset, using the same matching approach as in the main analysis; the results are reported in the Web Appendix~\ref{app:placebo_no_tiktok_exposure}. The estimates are small and statistically indistinguishable from zero across phases.

Taken together, these patterns support the discovery interpretation. The streaming losses are largest for songs with the strongest pre-withdrawal TikTok exposure, the effects appear to propagate through Spotify-side playlist distribution, and songs with no prior TikTok exposure do not show systematic streaming changes around the withdrawal.

\subsection{Interference Across Catalogs}
\label{sec:sutva}

\begin{figure}[ht!]
  \centering
  \caption{Decomposition of the Treated--Control Gap Under Cross-Catalog Substitution}
  \label{fig:reallocation_schematic}
  \includegraphics[width=1\textwidth]{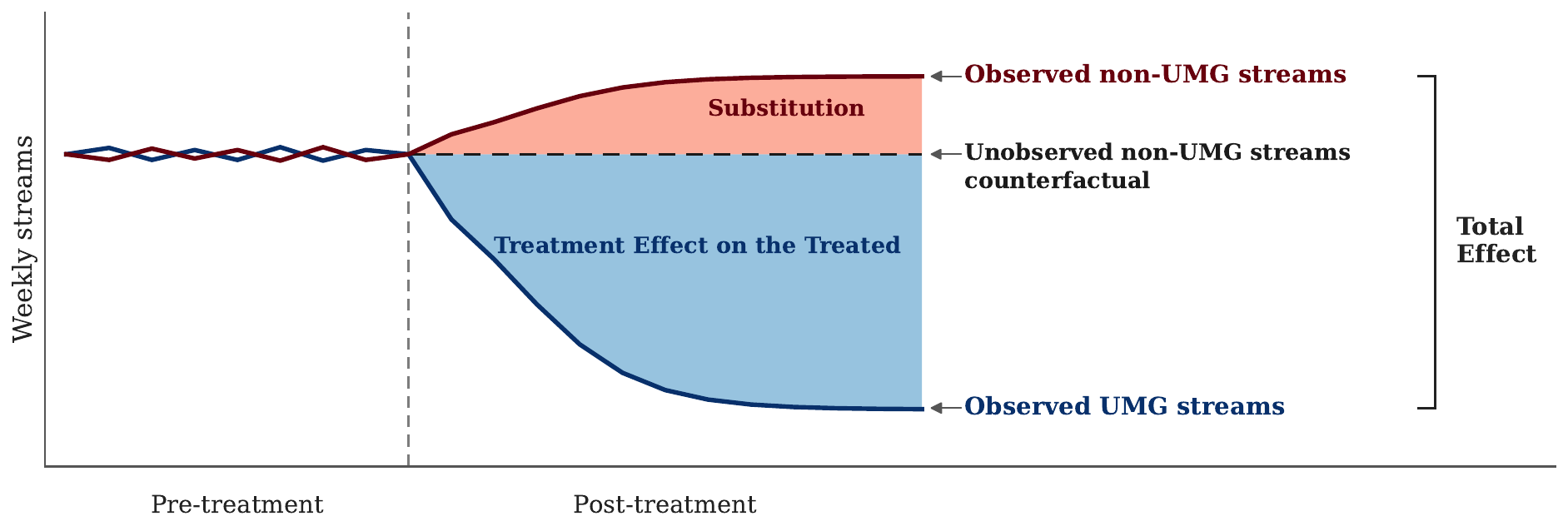}
  \vspace{-0.6cm}
  \begin{justify}
    \scriptsize{\textit{Notes:} The figure schematically decomposes the treated--control gap in the DiD design when the withdrawal reallocates attention or creation activity toward non-UMG music. Pre-treatment, UMG (treated) and non-UMG (control) streams follow parallel trends. Post-treatment, observed UMG streams fall relative to the no-withdrawal counterfactual, while observed non-UMG streams rise relative to their no-withdrawal counterfactual. The treated--control gap therefore combines two margins: the decline in UMG streams and the substitution-driven increase in control-group streams.}
  \end{justify}
\end{figure}

A central identification concern is that the UMG withdrawal is a label-specific shock. The no-interference component of the stable unit treatment value assumption (SUTVA) requires that a song's potential outcomes depend only on its own treatment status and not on the treatment status of other songs. In our setting, this assumption may fail if removing UMG audio from TikTok causally reallocates attention or creation activity toward non-UMG music. In that case, control songs may gain TikTok exposure and Spotify streams during the treatment period, and the treated-minus-control DiD gap combines two margins: the decline in UMG streams and any substitution-driven increase in control-group streams, as illustrated in \autoref{fig:reallocation_schematic}. This complicates a narrow own-catalog interpretation of the estimate, but the combined gap is also economically meaningful in a market where compensation depends on relative stream shares, as we discuss below.

\paragraph{Magnitude of reallocation.}
We cannot directly quantify all forms of cross-catalog reallocation. We therefore use creator-side posting data to evaluate the degree of plausible substitution toward non-UMG audio, which we use as a proxy for the broader interference concern.  Using public TikTok data from \url{tikapi.io}, we construct posting panels for 15{,}454 quasi-randomly drawn users and study how they adjust music posting when UMG audio becomes unavailable. We find two key patterns. First, overall music posting \emph{declines}, which is inconsistent with one-for-one substitution. Second, a back-of-the-envelope decomposition implies that only about 16\% of the missing UMG posting mass is reallocated into Sony and Warner catalogs. Under strong and conservative assumptions about how this additional TikTok exposure translates into incremental Spotify streams, attributing the full treated--control gap to forgone UMG streams would overstate the implied own-catalog effect by at most about 16\%. These results indicate that creator-side substitution is present, but limited in magnitude. The detailed TikTok creator-level analysis is reported in the Web Appendix~\ref{sec:sutva_substitution}.

\paragraph{Economic interpretation of the treated--control gap.}
The interpretation of this interference depends on the estimand of interest. If the target is a narrow own-catalog effect, defined as the decline in UMG streams holding competitor exposure fixed, then cross-catalog reallocation is a source of bias: it raises control-group streams and makes the treated-minus-control gap more negative than the own-UMG stream loss. However, from UMG's perspective, the withdrawal affects its competitive position in a market where streaming compensation depends on relative stream shares.

Spotify describes its royalty calculation as based on each rights holder's share of total streams: the platform first tallies total streams in a given month and then determines what proportion of those streams belong to music owned or controlled by a particular rights holder \citep{spotify_royalties}. Thus, a label's payout depends not only on the absolute number of streams its catalog receives, but also on how those streams compare with streams received by competing catalogs. When UMG content is removed from TikTok and some attention shifts toward Sony and Warner catalogs, two things happen simultaneously: UMG streams fall and competitors' streams may rise. Both margins reduce UMG's stream share, and therefore its revenue share.\footnote {A simple example makes the revenue-share logic transparent. Suppose there are two songs, one UMG and one non-UMG, and each initially receives 1{,}000 streams. If UMG loses 10\% of its streams and there is no spillover to the competing song, UMG's post-treatment stream share is $900/(900+1000)=47.37\%$. If instead half of the lost UMG streams shift to the competing song, UMG's post-treatment stream share is $900/(900+1050)=46.15\%$. The substitution component therefore directly lowers UMG's payout share, even though the number of UMG streams is the same in both scenarios.} For this reason, the treated-minus-control gap captures the combined effect of the UMG stream decline and any substitution-driven increase in non-UMG streams, which is the relevant estimand for UMG's financial position under the existing compensation scheme.

\subsection{The 2025 U.S.\ TikTok Outage}
\label{sec:tiktok_ban}

The UMG withdrawal is the main empirical setting because it removes TikTok exposure for a large and commercially important catalog over an extended period. Its label-specific nature, however, raises the possibility that part of the treated--control gap reflects reallocation toward songs that remained available on TikTok. We therefore use a second, shorter disruption to TikTok access, the 2025 U.S.\ TikTok outage, as corroborating evidence. Because the outage affected all labels symmetrically, it shuts down the label-specific reallocation channel and allows us to assess whether reductions in TikTok activity still lower off-platform streaming demand when relative catalog availability is unchanged.

Specifically, we use the 2025 disruption to TikTok access in the United States. A U.S.\ law required TikTok to be sold to a non-Chinese owner by January 19, 2025, or otherwise face a U.S.\ ban. Although the incoming Trump administration indicated that the law would not be enforced, TikTok temporarily shut down service in the United States for approximately 14 hours beginning on January 19, 2025. While the platform restored service quickly following federal intervention, the disruption was substantial: during the outage, TikTok’s U.S.\ traffic dropped sharply relative to the prior week’s levels \citep{donati2025cost}. Importantly, even after service resumed, the TikTok app remained unavailable for new installations and reinstallations due to removal from Apple’s App Store and Google’s Play Store for an additional 25 days, until February 13, 2025 \citep{maheshwari_tiktok_2025}. Although existing users retained access, this install restriction plausibly constrained platform activity. Consistent with that interpretation, TikTok-related domain traffic averaged about 30\% below pre-disruption levels in the first week and recovered only gradually over subsequent weeks \citep{donati2025cost}.

Because TikTok access was disrupted in the United States but not in other countries, we use cross-country variation in streaming demand to construct the counterfactual. We treat the United States as the exposed market and compare its streaming path to a set of nine countries that were not affected by the disruption: Australia, Brazil, Canada, Germany, Spain, France, Italy, Mexico, and the United Kingdom.

\paragraph{Data and Identification.}
We construct a daily song--country panel using data from Luminate, an industry measurement provider that aggregates music consumption data and whose metrics are used to compile Billboard's weekly charts.\footnote{Daily stream counts from Luminate cover all major services, while the Soundcharts data in our main analysis use Spotify-only streams.} For each of the 10 countries in our analysis, we start from the 50{,}000 most streamed tracks and obtain daily on-demand streams by country over the event window. We impose a balanced-panel restriction by retaining only song--country observations with strictly positive daily activity on each day of the window. This yields 42{,}850 U.S.\ tracks observed over the 28-day event window, and we apply the same restriction to the control countries.

By construction, this sample is concentrated in the upper tail of the popularity distribution. In the United States, the balanced-panel sample has a minimum of 8{,}538 daily streams, a mean of 52{,}131, and a maximum of 5{,}334{,}115. Relative to the distribution in our main analysis, these tracks fall primarily in the upper part of the streaming distribution, which is the part of the market where the UMG withdrawal produces the largest effects.\footnote{Web Appendix~\ref{app:tt_ban_results} provides additional details on the sample construction, the mapping to the main-study popularity distribution, and the matching procedure.}

We use a matching procedure analogous to the main analysis and estimate the same focal estimator: PPML with a log link. The outcome is daily on-demand streams per capita. The treated market is the United States, and the post period begins on January 19, 2025, when TikTok temporarily went offline in the United States and then remained unavailable for new installs for the next 25 days. We use a 28-day event window, with 14 days before and 14 days after January 19. The specification includes track-by-country fixed effects, which absorb time-invariant differences in a track’s baseline popularity across countries, and day-by-country fixed effects, which absorb day-specific shocks on the country level. The estimand is the proportional change in expected daily streams per capita in the United States during the disruption window relative to the matched cross-country counterfactual. To recover the event-time path, we replace the single post indicator with U.S.-by-event-day indicators, normalized to a pre-period reference day.

\begin{figure}[!htbp]
  \centering
  \caption{Event-study Estimates of the U.S.\ TikTok Outage on Streams}
  \label{fig:ban_eventstudy}
  \includegraphics[width=0.75\textwidth]{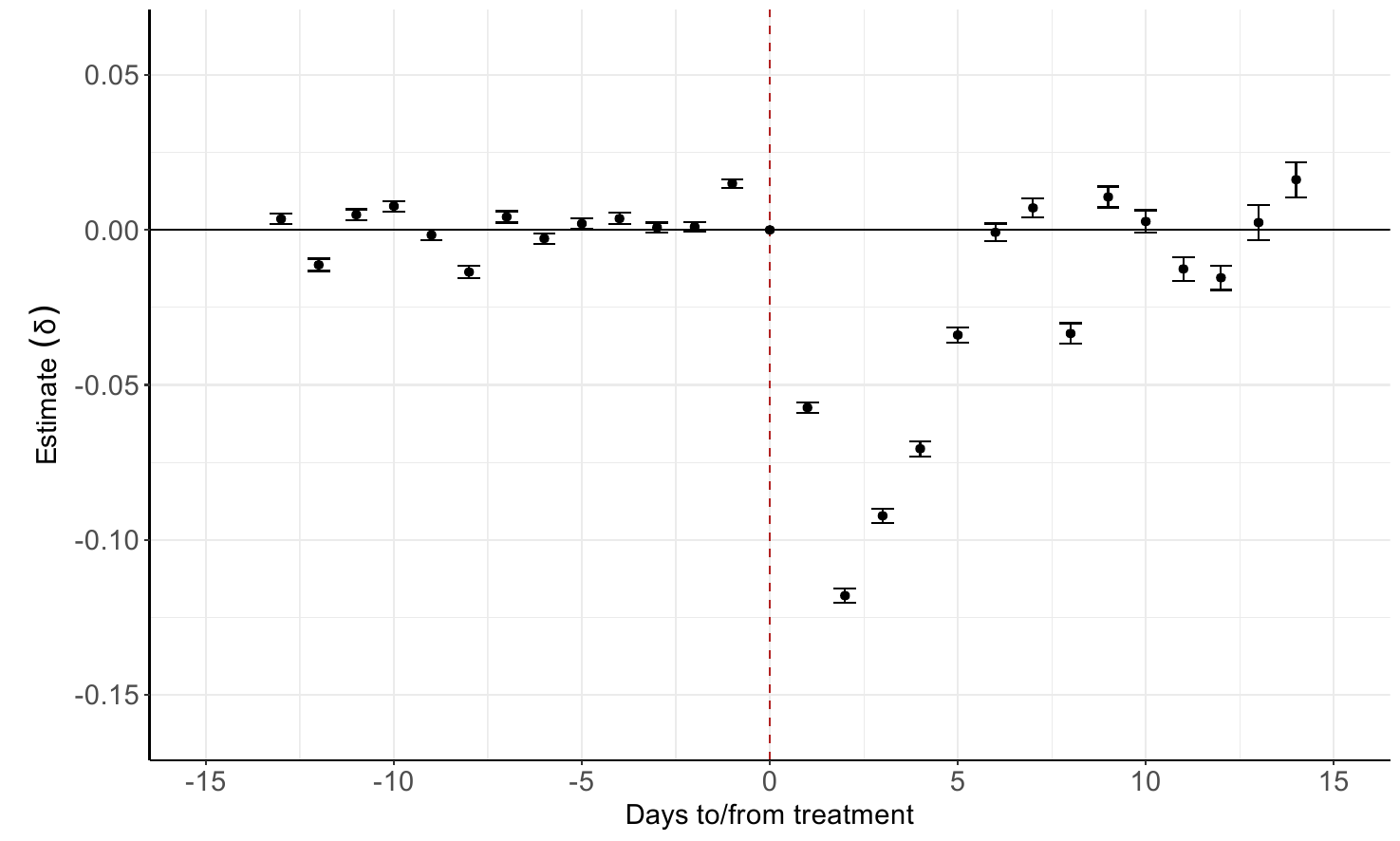}

  \vspace{2mm}
  \scriptsize
  \parbox{0.99\textwidth}{\textit{Notes:} The figure reports Poisson pseudo-maximum-likelihood (PPML) event-study estimates from a log-link model with daily U.S.\ song streams as the outcome. Points show coefficients in log points relative to the reference day $t = 0$ (January 18, 2025); whiskers are 95\% confidence intervals. The vertical dashed line marks January 18, 2025, the day before TikTok's temporary disruption. Standard errors are computed using a Huber--White (sandwich) variance estimator, two-way clustered by country-song and match-pair.}
  \vspace{-0.5cm}
\end{figure}

\paragraph{Results.}
\autoref{fig:ban_eventstudy} reports the event-study estimates. At the disruption date, the coefficients exhibit a clear negative break: beginning after the day of the outage, U.S.\ streams fall relative to the matched controls, and the decline deepens over the next day, peaking on day two at roughly $-0.12$ log points, or approximately an 11\% reduction. The effect then attenuates over subsequent days. A pooled PPML DiD specification over the 14-day (7-day) post window yields a statistically significant average effect of approximately $-2.8\%$ ($-5.1\%$) (see Web Appendix~\ref{app:tt_ban_results}). 

The magnitudes from the two experiments are not directly comparable. The TikTok outage sample is concentrated in the upper tail of the popularity distribution, the outage was short-lived, and the disruption reduced but did not eliminate TikTok activity because existing users retained access. By contrast, the UMG withdrawal lasted months and cut treated songs off from the TikTok promotional channel. In addition, the UMG estimates may include the cross-catalog substitution component discussed above. Taking these differences into account, the implied magnitudes are broadly consistent. Using the approximately 30\% reduction in TikTok traffic reported by \citet{donati2025cost}, the pooled outage estimate for the 14 day window implies a local effect of about $-0.0284/0.30=-0.095$, or roughly 9\%, for users whose TikTok access was disrupted.\footnote{Cloudflare \citep{tome2025tiktok} reported a decline between 85\% and 14\% in the initial period of the outage. Assuming an average reduction of $(85\% + 14\%) / 2 = 49.5\%$ yields a local effect of $-0.0284 / 0.495 = -0.057$, or roughly 5.6\%, which is broadly consistent with our overall effect for the top decile.} This magnitude is in line with the top-decile effects from the UMG withdrawal. 

Overall, the two experiments point to the same central implication: reductions in TikTok access lower off-platform streaming demand, consistent with TikTok playing a promotional role for paid streaming.

\section{Concluding Remarks}\label{sec:conclusion}
 
We study TikTok's role in the music economy using Universal Music Group's (UMG) global withdrawal from TikTok in February 2024 as a quasi-natural experiment. We find that removing TikTok exposure lowers Spotify demand for UMG titles, but the effect is highly heterogeneous. The typical long-tail track exhibits no economically meaningful change in paid streaming demand, while losses are concentrated among highly TikTok-exposed titles. Because listening is extremely concentrated in the viral head, these upper-tail declines dominate aggregate implications.

These heterogeneous effects help explain why prior work using the same natural experiment has reached conflicting conclusions. Through an estimand-based comparison of alternative DiD implementations, we show that the findings can be reconciled. \citet{cheng_value_2024} report a small positive effect using unweighted log-OLS, which corresponds to a ``typical song'' estimand that is dominated by the long tail, where effects are economically negligible. \citet{bairathi_lambrecht_rao_2024} document negative effects using levels and PPML specifications, which we also replicate in our data. The apparent contradiction dissolves once the estimands are made explicit: different functional forms place weight on different parts of the outcome distribution, and with heavy-tailed outcomes and heterogeneous treatment effects, these estimands can even differ in sign.
 
This reconciliation points to a broader methodological lesson for applied work. In settings with heavy-tailed outcome distributions or substantial treatment-effect heterogeneity, researchers should consider adding the PPML estimator to their toolkit \citep{log_gravity}. PPML targets the conditional mean in levels under a multiplicative structure, which is often the economically relevant estimand when the goal is to assess aggregate effects, and it only requires the conditional mean to be correctly specified. In our application, weighted log-OLS produces very similar results, although it is not immune to the variance-deflation problem that can bias unweighted log-OLS with respect to mean-treatment effects, i.e., when treatment compresses the log-scale variance, even a weighted log specification can in principle be affected, whereas PPML avoids this issue by construction.

On the substantive side, our findings have practical implications for understanding platform complementarity, discovery spillovers, and the bargaining dynamics between music rights holders and digital platforms. Overall, the evidence indicates that TikTok serves primarily as a complementary promotional channel for music streaming, especially for highly viral content. By facilitating music discovery and amplifying songs' reach, TikTok can drive downstream listening on revenue-generating platforms such as Spotify. The analysis reported in the Web Appendix~\ref{sec:be_revenue} reinforces the economic relevance of this mechanism: a back-of-the-envelope calculation implies roughly \$14--16 million in forgone Spotify royalties during the withdrawal window, with losses again concentrated in the viral head. We also document declines in Spotify curated-playlist followers, which suggests that lost TikTok exposure can propagate through Spotify-side discovery and distribution (Web Appendix~\ref{app:playlist} documents this analysis in detail). In addition, UMG’s Spotify Top-200 chart footprint falls after the withdrawal, which highlights the commercial stakes for labels when promotional momentum is disrupted (see Web Appendix~\ref{app:top200} for the analysis). From a strategic perspective, these results imply that removing music from TikTok can be costly for rights holders if the withdrawal dampens demand on platforms that generate royalties. This matters directly for licensing negotiations. UMG’s public position emphasized concerns that TikTok was not adequately compensating rights holders and might divert attention from monetized listening. Our results point to a different margin: during the window we study, TikTok appears to function as an upstream discovery channel that increases off-platform, on-demand streaming. This suggests that the combined TikTok-plus-streaming ecosystem can be larger with TikTok exposure than without it.

Consequently, when a rights holder considers withholding content to extract higher licensing payments, it must also account for the potential loss of promotional momentum and downstream streaming demand. This gives TikTok a distinct form of bargaining leverage. If the platform can credibly argue that it helps break songs and sustain demand beyond its own app, then some of its value to rights holders is realized off-platform and may not be immediately visible in TikTok-side payments alone.

For artists and content creators, the results also highlight the importance of maintaining visibility across platforms. Short-form video appears to be a powerful driver of music discovery and engagement, and its connection to paid streaming means that momentum on one platform can translate into demand on another. In line with \citet{bairathi_lambrecht_rao_2024}, our study provides evidence that TikTok is not merely a competing distraction for music listeners, but an important part of the music consumption ecosystem that can amplify demand on monetized platforms. This insight should inform ongoing negotiations between rights holders and digital platforms by encouraging agreements that recognize the promotional value created across platforms.

\newpage

\begin{singlespace}
    \putbib[main]
\end{singlespace}
 
\end{bibunit}

\clearpage
\pagenumbering{arabic}      
\phantomsection
\addcontentsline{toc}{section}{Methodological Online Companion}

\section*{Practitioner's Companion}
\label{sec:companion}

The practitioner's companion is a self-contained guide for selecting a difference-in-differences estimand and estimator when outcomes are heavy-tailed. It distills the paper's methodological argument into a decision workflow---from the target estimand implied by the research question, through diagnostics of the outcome distribution, to the choice of estimator and specification. It can be read independently of the main text.

\includepdf[pages=-]{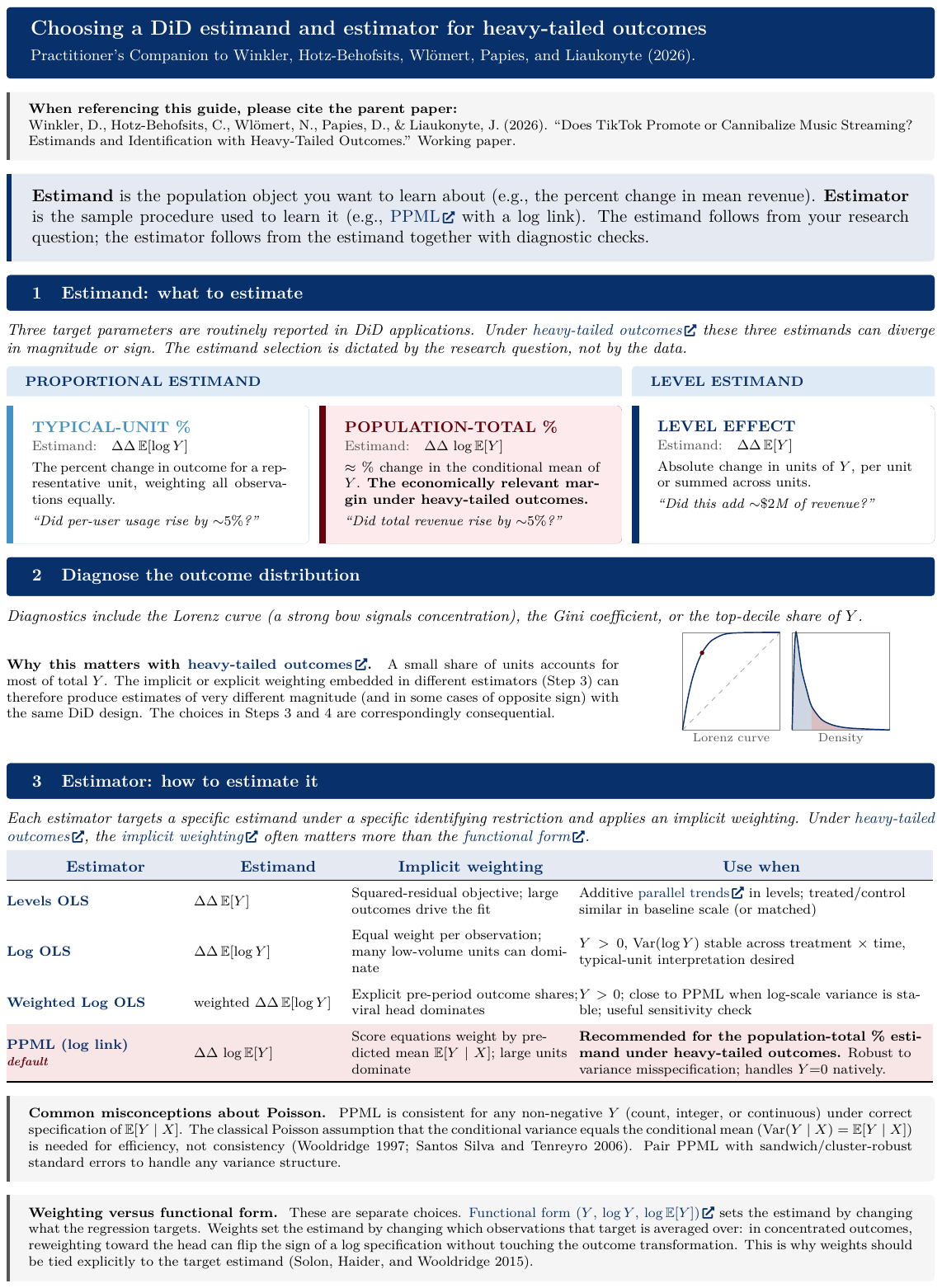}

 
\clearpage
\appendix

\setcounter{footnote}{0}
\setcounter{section}{0}
\setcounter{table}{0}
\setcounter{figure}{0}
\renewcommand{\thetable}{\thesection.\arabic{table}}
\renewcommand{\thefigure}{\thesection.\arabic{figure}}
\pagenumbering{arabic}

\singlespacing
\begin{bibunit}
\section*{Web Appendix}
\addcontentsline{toc}{section}{Web Appendix}

\startcontents[wa]
\printcontents[wa]{l}{1}{\setcounter{tocdepth}{2}}

\newpage

\section{Literature Overview}
\label{app:literature}
\autoref{tab:lit_review} provides an overview of empirical, quasi-experimental work on cross-platform spillover effects in content industries. The quasi-experimental literature offers a mixed but increasingly structured picture of cross-platform spillovers in content markets. Early evidence from music suggests that the same video platform can generate both substitution and promotion, depending on the level of analysis and the type of content. \citet{scott_hiller_sales_2016} finds that removing Warner content from YouTube increased album sales, especially for highly ranked albums, consistent with displacement at the top of the market. By contrast, \citet{kretschmer_video_2020} exploit both the 2009 GEMA blocking of YouTube videos in Germany and the 2013 entry of Vevo, finding that music video availability increased song sales and streams on other channels, consistent with a complementary discovery effect. More recent evidence sharpens this interpretation by emphasizing heterogeneity: \citet{wlomert_interplay_2024} show that YouTube UGC availability can stimulate demand for many songs but cannibalize demand for recent and hit releases, so that average song-level effects and aggregate revenue effects need not have the same sign.

Evidence from short-form video platforms points in a similar direction. \citet{yang_ugc_2024} show that removing user-generated condensed clips from a TikTok-style platform reduced demand for full-length TV originals, suggesting that derivative short-form content can increase visibility and stimulate downstream consumption. In music, however, recent work on TikTok and the UMG withdrawal reaches conflicting conclusions: \citet{cheng_value_2024} interpret the withdrawal as increasing Spotify streams, consistent with substitution, whereas \citet{bairathi_lambrecht_rao_2024} find declines in Spotify streams, consistent with complementarity. This tension motivates our study. We use the UMG withdrawal and the 2025 U.S.\ TikTok disruption to clarify when TikTok acts as a discovery channel rather than a substitute, and to show that the answer depends on treatment-effect heterogeneity and on whether the empirical estimand reflects the typical song or the economically dominant viral head.

\begin{table}[htbp]
\centering

\caption{Quasi-experimental Evidence on Cross-platform Spillover Effects in Content Industries}
\label{tab:lit_review}

\begin{threeparttable}

\setlength{\tabcolsep}{3pt}
\renewcommand{\arraystretch}{1}

{ \footnotesize
\begin{tabularx}{\textwidth}{@{} 
P{2.5cm}
P{2.5cm}
P{1.4cm}
P{1.5cm}
P{1.7cm}
P{1.9cm}
>{\raggedright\arraybackslash}X
@{}}
\toprule

Study & Platform / shock & Content & Treated units & Method & Outcome & Complement vs.\ substitute \\

\midrule

\citet{scott_hiller_sales_2016}
& YouTube (Warner Music removal, 2009)
& Music (albums)
& $\approx$266\tnote{a}
& DiD
& Album sales (iTunes)
& \textit{Heterogeneous}: displacement for best-sellers; promotional for lower-ranked \\

\addlinespace

\citet{kretschmer_video_2020}
& YouTube (GEMA blocking in Germany, 2009)
& Music (songs)
& 842
& DiD
& Song sales (physical \& digital)
& \textit{Complement}: blocking reduced recorded music sales \\

\addlinespace

\citet{kretschmer_video_2020}
& Vevo (entry in Germany, 2013)
& Music (songs)
& 759
& DiD
& Song sales and streams
& \textit{Complement}: entry increased sales and streams \\

\addlinespace

\citet{wlomert_interplay_2024}
& YouTube (GEMA resolution in Germany)
& Music (songs)
& 378{,}460
& DiD
& Streams on other platforms
& \textit{Heterogeneous}: promotional for niche; cannibalization for hits; overall negative \\

\addlinespace

\citet{yang_ugc_2024}
& TikTok-style platform (clip removal, China)
& TV shows
& 383
& DiD
& Streams of full-length originals
& \textit{Complement}: removal reduced demand by $\approx$3\% \\

\midrule

\addlinespace

\citet{cheng_value_2024}
& TikTok (UMG withdrawal, 2024)
& Music (songs)
& 35{,}837
& DiD (log OLS)
& Spotify streams
& \textit{Substitute}: small positive effect on Spotify streams \\

\addlinespace

\citet{bairathi_lambrecht_rao_2024}
& TikTok (UMG withdrawal, 2024)
& Music (songs)
& 21{,}526
& DiD (levels, PPML)
& Spotify streams
& \textit{Complement}: decline in streams after withdrawal \\

\addlinespace

This paper
& TikTok (UMG withdrawal, 2024)
& Music (songs)
& 53{,}753
& DiD (PPML, level, log-OLS, weighted log-OLS)
& Spotify streams
& \textit{Complement}: effect concentrated in viral head; long tail largely unaffected \\

\addlinespace

This paper
& TikTok (U.S.\ disruption, 2025)
& Music (songs)
& 42{,}850\tnote{b}
& DiD (PPML)
& on-demand streams
& \textit{Complement}: disruption reduced U.S.\ streams relative to control countries \\

\bottomrule
\end{tabularx}
}

\begin{tablenotes}[para,flushleft]
\scriptsize\setstretch{1}\selectfont\emph{Notes:} The table summarizes quasi-experimental studies on cross-platform spillover effects in content industries. ``Complement'' indicates that platform availability increases demand in other channels; ``Substitute'' indicates that it decreases demand. ``Heterogeneous'' indicates that the direction varies, e.g., by baseline popularity. All studies exploit content removal or restoration as a source of exogenous variation in platform availability. For \citet{kretschmer_video_2020}, the GEMA blocking and Vevo entry are reported separately because the paper studies two distinct natural experiments with different treated samples. (a) Hiller reports 1{,}796 albums in the sample and a Warner share of 0.148 in Table 1. The treated-unit count is therefore inferred as $1{,}796 \times 0.148 \approx 266$. (b) For the U.S.\ TikTok disruption, treated units are U.S.\ tracks in the balanced event-window sample.
\end{tablenotes}
\end{threeparttable}
\end{table}

 \FloatBarrier
\section{Sample Construction}
\label{app:sample}
 
We begin with 817,520 songs, which are essentially all relevant tracks that appear on Spotify curated and third-party playlists since 2018 (144,312 UMG; 673,208 non-UMG). We first require at least 1,000 weekly global streams, which removes songs with negligible activity and yields 517,932 songs (108,688 UMG; 409,244 non-UMG). We then exclude Holiday/Seasonal, Children, and Comedy, which drops categories with atypical seasonality and consumption patterns and leaves 505,532 songs (103,222 UMG; 402,310 non-UMG). To protect the validity of the control group, we drop non-UMG songs (songs for which UMG does not have master rights) but for which UMG holds publishing rights, which reduces the sample to 480,088 songs (103,222 UMG; 376,866 non-UMG). Finally, we exclude artists who release new albums during our observation window, since these releases generate large contemporaneous demand shocks, which leaves 381,370 songs (84,375 UMG; 296,995 non-UMG).\footnote{This exclusion also covers Taylor Swift, who is a UMG artist but released a new album during the observation period and independently returned her content to TikTok prior to the conclusion of UMG's licensing dispute. Including her songs would induce bias in the treated-versus-control comparison, since her TikTok presence was restored through a channel unrelated to the UMG–TikTok negotiation.} Within this cross section, 218,294 songs have strictly positive pre-treatment TikTok creations (57,109 UMG; 161,185 non-UMG), while 163,076 have zero pre-treatment creations (27,266 UMG; 135,810 non-UMG). We restrict the control group to non-UMG songs released by other major labels (Sony and Warner) and exclude independent artists to ensure greater comparability with UMG releases; this yields 60,454 control songs with pre-treatment TikTok creations. 
Finally, as part of the matching procedure described below, we further refine the sample by retaining only treated songs for which a closely comparable control song can be identified with sufficiently aligned pre-treatment streaming trends. Applying this additional restriction results in a final estimation sample of 53,753 treated songs and an equal number of matched control observations.\footnote{Related studies examining the same event analyze treated samples of 35{,}837 songs in \citet{cheng_value_2024} and 21{,}526 songs in \citet{bairathi_lambrecht_rao_2024}, which reflects differences in data access and sampling frames.}
 
We assemble daily data from 1 October 2023 to 1 July 2024. For estimation we focus on 29 November 2023 to 30 May 2024, since the weeks before and after show anomalies and coincide with other TikTok platform changes such as the launch of “Add to Music App,” which could confound identification outside the core event window. We also aggregate the daily data to weekly measures, which smooths the pronounced day-to-day volatility.
 
\section{Difference-in-Differences Diagnostics: Supplementary Material}
\label{app:diagnostics_supplement}

\subsection{Different Matching Procedures}
\label{app:matching_procedures}

\autoref{tab:model-comparison-reduced} shows the PPML and \citet{callawayDifferenceinDifferencesMultipleTime2021} estimates under different matching strategies. First, estimates without any matching are presented. The remaining rows present matching procedures with and without replacement as well as variants with and without exact matching on the TikTok popularity decile prior to treatment. The first three columns give an overview of the number of observations in the full model after matching. The estimate of the PPML model estimated with that sample is shown in column 4. Across all specifications the estimate is significantly negative. The final column presents alternative strategies proposed by \citet{callawayDifferenceinDifferencesMultipleTime2021}. The first row shows the estimate of the outcome regression using a conditional parallel trends assumption (conditioning on but not matching on the same variables used for matching). The other two estimates show the estimates using their doubly-robust estimator (both conditioning parallel trends on and inverse probability weighting based on the matching variables). Again all estimates are significantly negative.

\begin{table}[!ht]
    \centering
    \caption{Overall Treatment Effect: Treated $\times$ Post}
    \label{tab:model-comparison-reduced}
    \small
    \renewcommand{\arraystretch}{1.0}
    \begin{threeparttable}
        \begin{tabular}{lrrrcc}
        \toprule
        Specification & Obs. & Treated & Control & PPML & \makecell{Callaway\\Sant'Anna} \\
        \midrule
        Unmatched Benchmark & 2{,}586{,}386 & 57{,}109 & 60{,}454 & -0.0112* & -0.0263** \\
         &  &  &  & (0.0052) & (0.0083) \\
        \addlinespace[2pt]
        \makecell{With Replacement\\Bin Blocking} & 2{,}365{,}132 & 53{,}753 & 29{,}650 & -0.0310*** &  \\
         &  &  &  & (0.0063) &  \\
        \addlinespace[2pt]
        \makecell{Without Replacement\\Bin Blocking} & 2{,}038{,}828 & 46{,}337 & 46{,}337 & -0.0177*** & -0.0338*** \\
         &  &  &  & (0.0044) & (0.0079) \\
        \addlinespace[2pt]
        \makecell{With Replacement\\No Bin Blocking} & 2{,}445{,}652 & 55{,}583 & 32{,}013 & -0.0317*** &  \\
         &  &  &  & (0.0062) &  \\
        \addlinespace[2pt]
        \makecell{Without Replacement\\No Bin Blocking} & 2{,}331{,}252 & 52{,}983 & 52{,}983 & -0.0171*** & -0.0373*** \\
         &  &  &  & (0.0041) & (0.0074) \\
        \bottomrule
        \end{tabular}

        \begin{tablenotes}[para,flushleft]
            \scriptsize\emph{Notes:} Callaway and Sant'Anna doubly robust estimates are reported as without replacement. Controls can receive unequal weight as a result of the IPW. The unmatched benchmark for Callaway and Sant'Anna uses the outcome regression model and conditions parallel trends on the matching variables without reweighting controls. All models with matching are estimated on matched samples that retain only treated and control pairs with closely aligned pretreatment dynamics and exclude pairs for which the normalized match distance exceeds 0.1, consistent with the main analysis.
        \end{tablenotes}
    \end{threeparttable}
\end{table}

\subsection{Inference Under Different Clustering Dimensions}
\label{app:clustering_dimensions}

\autoref{tab:clustering-robustness} presents statistical inference for the main PPML model under different assumptions of correlation in the error term. Column 1 presents the most flexible structure allowing both correlations within song across match-pairs (for controls used multiple times) and time, and within match-pairs \citep[see][]{abadie_spiess_2022}. Column 2 allows correlated error terms only within song across match-pairs and time and column 3 allows error correlation only within match-pairs and time but not across match-pairs. For matching procedures without replacement columns 1 and 3 are equivalent as there is no additional correlation induced by the reuse of controls. In all cases the results remain statistically significant providing further evidence for the robustness of our results.

\begin{table}[!htb]
    \centering
    \footnotesize
    \renewcommand{\arraystretch}{1.0}
    \caption{PPML Clustering Robustness}
    \label{tab:clustering-robustness}
    \begin{threeparttable}
        \begin{tabular}{lccc}
        \toprule
        Specification & Song + Match-set & Song & Match-set \\
        \midrule
        With Replacement + Bin Blocking & -0.0310*** & -0.0310*** & -0.0310*** \\
         & (0.0063) & (0.0065) & (0.0047) \\
        Without Replacement + Bin Blocking & -0.0177*** & -0.0177*** & -0.0177*** \\
         & (0.0044) & (0.0049) & (0.0044) \\
        With Replacement, No Bin Blocking & -0.0317*** & -0.0317*** & -0.0317*** \\
         & (0.0062) & (0.0063) & (0.0045) \\
        Without Replacement, No Bin Blocking & -0.0171*** & -0.0171*** & -0.0171*** \\
         & (0.0041) & (0.0045) & (0.0041) \\
        \bottomrule
        \end{tabular}
        \begin{tablenotes}[para,flushleft]
            \scriptsize\textit{Notes:} Standard errors in parentheses. All models use PPML (Poisson pseudo-maximum likelihood) with fixed effects (song + period). Clustering varies by column. * $p < 0.05$, ** $p < 0.01$, *** $p < 0.001$ (two-tailed). 
        \end{tablenotes}
    \end{threeparttable}
\end{table}

\subsection{Assessing Counterfactual Trends: Additive vs.\ Multiplicative}
\label{app:additive_vs_mult}
 
Any DiD design relies on a maintained counterfactual trend assumption: absent treatment, the average outcome for treated units would have evolved in the same way as for control units. A first-order question in our setting is the scale on which this counterfactual restriction is most plausible. In particular, should untreated evolution be approximated in levels (common absolute changes) or in proportional terms (common growth rates)? This distinction is central because it governs both the maintained parallel-trends restriction and the population estimand that a DiD coefficient summarizes \citep{Roth_SantAnna_2023}.
 
There are intuitive reasons to expect proportional rather than additive dynamics in digital music consumption. First, baseline popularity varies by orders of magnitude. In such a heavy-tailed outcome setting, imposing parallel trends in absolute units would require implausible composition shifts. For example, with streaming growing by 9.5\% in 2024 \citep{IFPI2025gmr}, a model of common \emph{absolute} growth would imply that new listening is disproportionately allocated to long-tail titles and away from viral-head artists, purely so that both groups gain similar numbers of streams. Relatedly, streaming trajectories reflect amplification and feedback through recommendation, playlists, and episodic attention shocks. Intuitively, a boost from algorithmic recommendation typically multiplies a song's existing reach, generating a much larger absolute increase for a popular track than for a niche one. These mechanisms are more naturally interpreted as scaling an underlying demand intensity than as adding a constant increment.
 
To make the maintained restriction explicit, we contrast an additive conditional-mean structure,
\begin{equation}
\label{eq:did_additive}
\mathbb{E}\!\left[ Y_{it} \right]
=
\mu_i+\gamma_t+\tau\,\big(\text{Treat}_i\times \text{Post}_t\big),
\end{equation}
with a multiplicative (log-link) conditional-mean structure,
\begin{equation}
\label{eq:did_multiplicative}
\mathbb{E}\!\left[ Y_{it} \right]
=
\exp\!\left(\mu_i+\gamma_t\right)\cdot
\exp\!\left(\delta\,\big(\text{Treat}_i\times \text{Post}_t\big)\right).
\end{equation}
Under \eqref{eq:did_additive}, parallel trends is an additive restriction on the mean, and $\tau$ summarizes an average \emph{level} change in streams. Under \eqref{eq:did_multiplicative}, the restriction is imposed on ratios of conditional means, and $\exp(\delta)-1$ is interpreted as a proportional change in the conditional mean. This choice of maintained restriction and estimand is conceptually separate from the estimator used to implement it in the sample \citep{ciani_did_2019, mcconnell_cant_2024}. We therefore report diagnostics that assess whether an additive-in-levels description is a coherent approximation for streaming outcomes in our setting, and we clarify when and why a levels DiD can be mechanically misleading under proportional untreated dynamics.
 
\begin{figure}[ht!]
  \centering
  \caption{Diagnostics of Additive versus Multiplicative Conditional-Mean Structures}
  \label{fig:diagnostics_add_mult}
 
  \begin{subfigure}{0.49\textwidth}
    \centering
    \caption{Scale Dependence of Treatment Incidence Across the Baseline Outcome Distribution}
    \includegraphics[width=\textwidth]{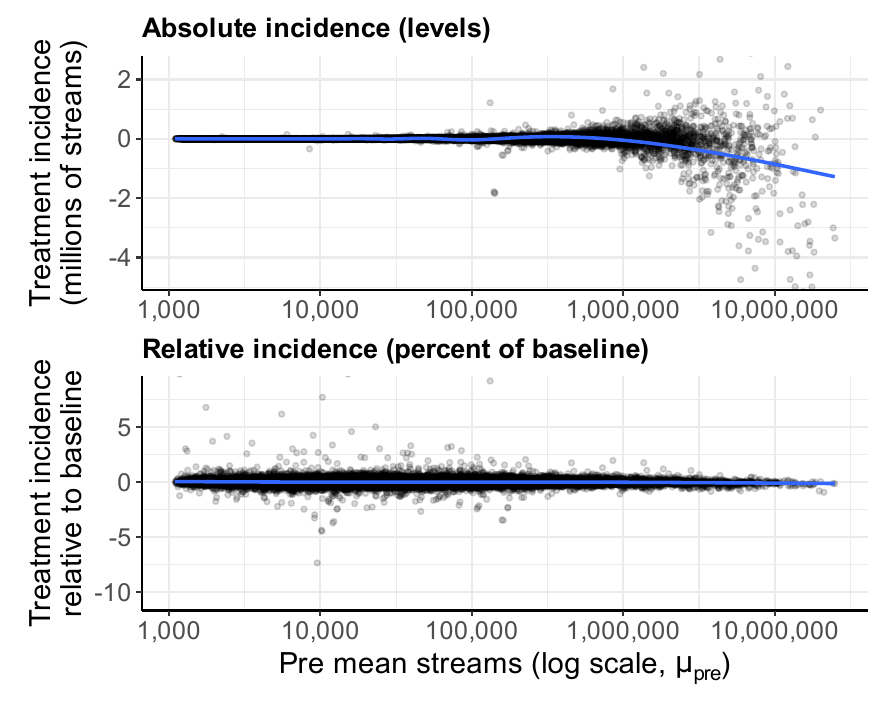}
    \label{fig:diagnosis_incidence}
  \end{subfigure}%
  \hfill
  \begin{subfigure}{0.49\textwidth}
    \centering
    \caption{Held-Out Proportional Prediction Error by Baseline Decile (Pre Period)}
    \includegraphics[width=\textwidth]{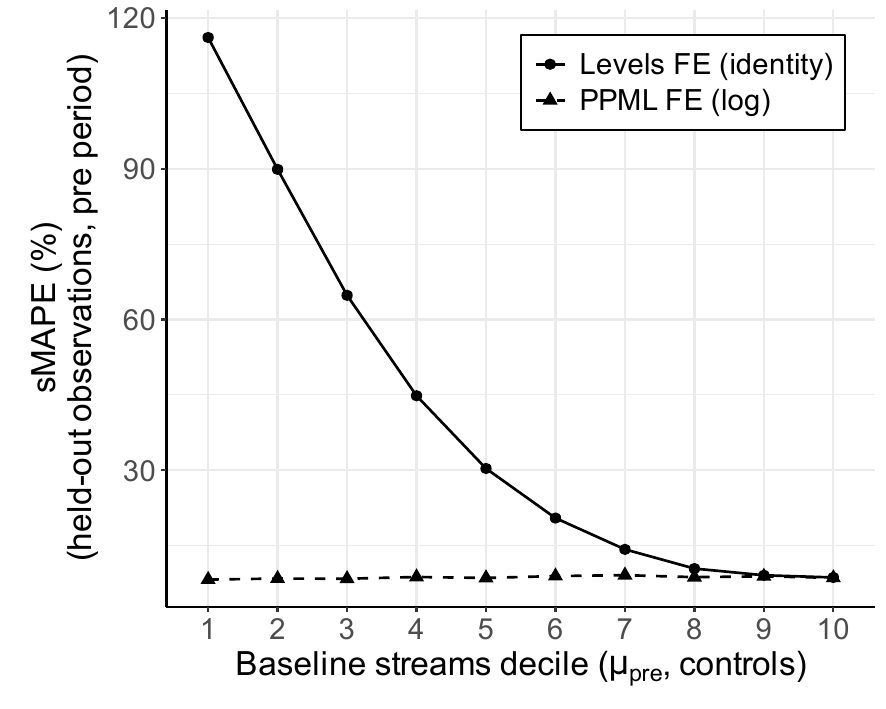}
    \label{fig:diagnosis_holdout}
  \end{subfigure}
\vspace{-0.8cm}
  \begin{justify}
    \scriptsize{\textit{Notes:} \textbf{(a)} The x-axis plots a song’s pre-period mean streams $\bar y_{i,pre}$ on a log scale; the y-axis reports the song-level DiD contrast expressed either in levels (top: absolute change in streams) or as a fraction of baseline (bottom: $\Delta_i/\bar y_{i,pre}$), with points showing individual songs and a smooth fit summarizing the average pattern. \textbf{(b)} The x-axis groups songs into deciles of $\bar y_{i,pre}$ (controls, pre period); the y-axis shows held-out proportional prediction error (sMAPE, in percent) for a levels TWFE model and a PPML TWFE model (log link).}
  \end{justify}
\end{figure}
 
We begin with a model-free diagnostic that uses our matched design and does not impose a parametric form on streaming dynamics. For each treated song $i$, let $m(i)$ denote song $i$'s matched control, and define the song-level DiD contrast as $\Delta_i \equiv (\bar y_{i,\text{post}}-\bar y_{i,\text{pre}})-(\bar y_{m(i),\text{post}}-\bar y_{m(i),\text{pre}})$, where $\bar y_{i,\text{pre}}$ and $\bar y_{i,\text{post}}$ are mean streams in the pre and post windows. This contrast has a natural levels interpretation (units of streams). To focus on scale, we also consider the corresponding relative contrast $\widetilde{\Delta}_i \equiv \Delta_i/\bar y_{i,\text{pre}}$. Panel (a) of \autoref{fig:diagnostics_add_mult} shows that $\Delta_i$ is not approximately constant over baseline size: absolute losses become more negative toward the upper part of the distribution. In contrast, the relative contrast $\widetilde{\Delta}_i$ is comparatively stable. In other words, the data look closer to a roughly constant \emph{percentage} effect than to a constant \emph{unit} effect, which supports multiplicative incidence.
 
We complement the figure with a simple baseline-scaling regression. Under constant-absolute incidence, an intercept-only approximation $\Delta_i=\alpha+\varepsilon_i$ should provide a reasonable summary. Under proportional incidence, the absolute DiD contrast should scale with baseline streams, which motivates $\Delta_i=a+b\,\bar y_{i,\text{pre}}+\varepsilon_i$. The estimated baseline-scaling coefficient is negative and precisely estimated, and nested-model tests and information criteria favor the baseline-scaling specification. \autoref{tab:horse_race_baseline_scaling} in the Web Appendix provides more details. Taken together, this diagnostic rejects constant-absolute treatment incidence and supports proportional-growth dynamics in untreated evolution in our data.
As a second diagnostic, \autoref{fig:diagnostics_add_mult} panel (b) evaluates pre-period coherence using a cross-fitting prediction exercise. We estimate two-way fixed effects models in the pre period on 80\% of observations and evaluate predictive accuracy on the remaining 20\% held-out observations within the same period. Because outcomes are heavy-tailed, we focus on a proportional error metric, which asks whether the model tracks untreated dynamics in relative terms across baseline popularity. The levels fixed-effects model implied by \eqref{eq:did_additive} delivers large proportional prediction errors in the long tail, whereas a log-link specification estimated by PPML, which corresponds to \eqref{eq:did_multiplicative}, is comparatively stable across baseline deciles. This pattern is consistent with proportional dynamics, where dispersion scales with baseline size.
 
\paragraph{When can a levels DiD be mechanically misleading under proportional dynamics?}
The diagnostics above indicate that untreated outcomes evolve approximately proportionally in our data. This is important because under proportional growth, a levels DiD can deliver a misleading additive estimate for a simple mechanical reason: the levels DiD implicitly subtracts off the control group’s absolute change, and under proportional growth absolute changes scale with baseline size. If treated and control units start from different baseline levels, then even with perfectly parallel \emph{growth rates} the groups will exhibit different \emph{absolute} increases absent treatment \citep{Roth_SantAnna_2023}. A levels DiD interprets this baseline-driven difference in absolute growth as part of the treatment effect, which can generate sizable distortion when baseline gaps are nontrivial and aggregate growth is meaningful. In extreme cases, the distortion can even reverse the sign of the implied levels effect relative to the true effect.

\begin{figure}[ht!]
  \centering
  \caption{Bias and Sign Flips of a Levels DiD under Proportional Growth}
  \label{fig:pareto_bias_heatmap}
  \includegraphics[width=0.75\linewidth]{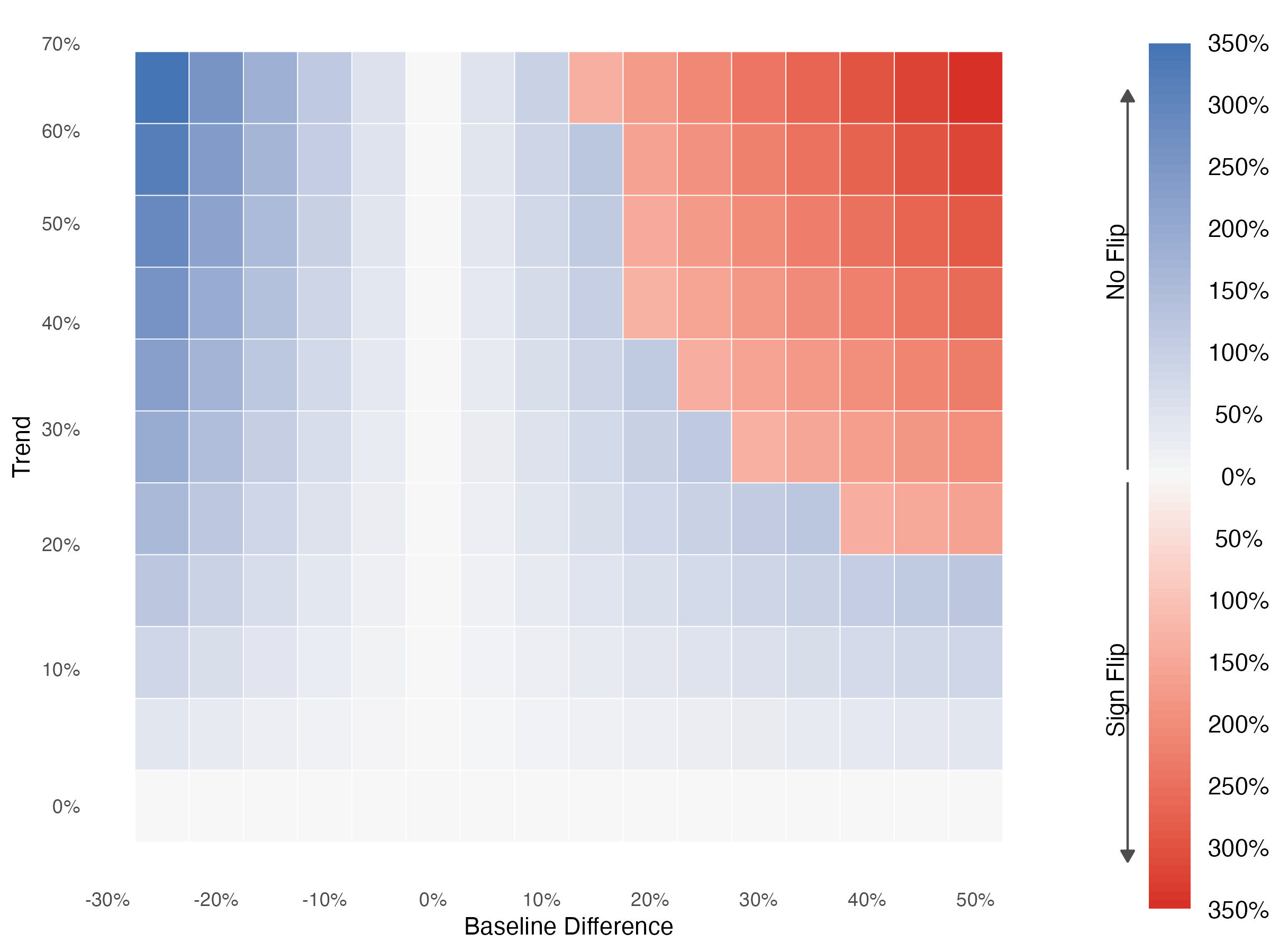}
\vspace{-0.4cm}
  \begin{justify}
  \scriptsize
  \textit{Notes:} Each cell reports Monte Carlo evidence on the bias of a conventional levels two-period DiD estimator when untreated outcomes evolve proportionally (common log-growth) but the analysis is conducted in levels. The x-axis varies the imposed treated--control baseline difference (in percent) in the pre period, implemented by rescaling treated outcomes so that the treated pre-period mean differs from the control mean by the indicated amount (focusing on the viral head only and with no heterogeneity across songs). The y-axis varies the common untreated trend, expressed as log growth per period (shown as percent growth). Across cells, we simulate a two-period DiD environment with proportional untreated growth and a constant proportional treatment effect, and then estimate a conventional DiD in levels. Color intensity reports the relative gap between the levels DiD estimate and the true treated effect in levels implied by the DGP, where blue indicates no sign flip and red indicates a sign flip. Note that a sign flip implies a bias of at least 100\%.
  \end{justify}
\end{figure}

\begin{figure}[ht!]
    \centering
    \caption{Bias Comparison Across Estimators under Proportional-Growth Dynamics}
    \label{fig:pareto_bias_heatmap_all}
    \includegraphics[width=0.92\linewidth]{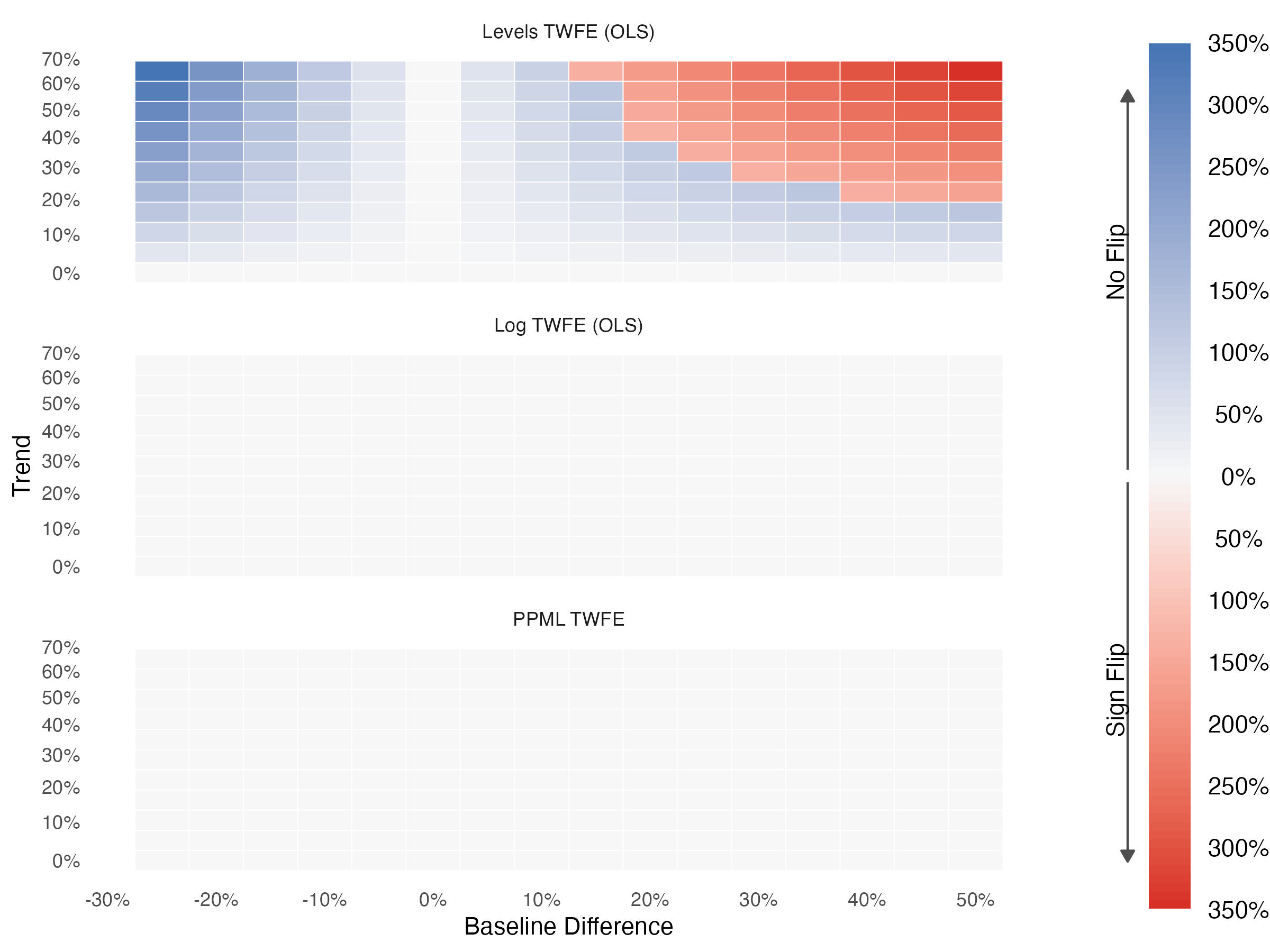}
    \begin{justify}
    \scriptsize{\textit{Notes:} The figure extends the Monte Carlo exercise to compare three common estimators under the same proportional-growth DGP, focusing on the viral head only and imposing a constant proportional treatment effect with no heterogeneity across songs. The x-axis varies the treated--control baseline difference (in percent) in the pre period, implemented by rescaling treated outcomes so that the treated pre-period mean differs from the control mean by the indicated amount. The y-axis varies the common untreated trend, expressed as log growth per period (shown as percent growth). Each panel reports the relative gap between the estimator’s implied effect in levels and the true treated effect in levels implied by the DGP; blue indicates bias in the direction of the true effect and red indicates a sign flip. }
    \end{justify}
\end{figure}

We quantify the magnitude of the mechanical distortion in a Monte Carlo simulation calibrated to the part of the distribution that drives our economic conclusions, the viral head. To keep the exercise simple, we impose no underlying treatment-effect heterogeneity: every simulated song in the viral head is assigned the same proportional treatment effect, calibrated to $\delta=-0.05$, which is close to the order of magnitude we estimate for highly exposed songs in our setting. The illustration below describes the data-generating process and implementation details. In brief, we simulate a two-period DiD panel of songs with a multiplicative (log-link) mean structure and a fixed proportional treatment effect, and we vary two primitives across cells: (i) the treated--control baseline gap in the pre period (implemented by rescaling treated outcomes to hit the desired baseline difference), and (ii) the common untreated growth rate between periods. For each cell, we estimate a conventional levels TWFE DiD and compare its coefficient to the true  effect implied by the multiplicative DGP.  We summarize the distortion using a scale-free relative-bias metric: for each cell, we report the average gap between the estimated levels DiD effect and the true levels ATT, expressed as a fraction of the true levels ATT. \autoref{fig:pareto_bias_heatmap} plots this relative bias, with color intensity capturing its magnitude of bias and sign. Two patterns stand out. First, the distortion grows quickly as baseline gaps and proportional growth become larger. Second, for sufficiently large baseline gaps, a levels DiD can imply the opposite sign, even though the underlying proportional treatment effect is held fixed. Importantly, the heatmap also makes the ``no-bias'' benchmark clear: along the zero-baseline-gap column (and at the zero-growth row), relative bias is essentially zero. Similarly, Appendix \autoref{fig:pareto_bias_heatmap_all} reports the analogous heatmaps for three estimators; consistent with the DGP, the corresponding bias disappears completely under PPML (log link) and log-OLS, which target proportional dynamics directly.
 
The simulation highlights a generic risk of levels DiD under proportional untreated dynamics when treated and control units differ in baseline levels. Our main research design sharply limits this channel by construction through 1:1 matching on pre-withdrawal streaming levels and trajectories (and related moments), which makes treated and matched control songs close in baseline scale within each pair. This largely shuts down the mechanical discrepancy that drives the large biases and sign flips in the simulation, which helps explain why levels and log-link specifications yield similar implications for aggregate, economically relevant outcomes in our data.
 
\paragraph{Illustration of levels DiD bias under proportional growth}

The following example illustrates a simple but important point for our setting: when untreated outcomes evolve proportionally over time, a level-difference DiD can be substantially biased if treated and control groups differ in baseline levels. The reason is mechanical: under proportional growth, the same common growth rate implies different absolute changes for units with different baselines. A level-parallel-trends restriction therefore fails as soon as treated and controls are not  aligned in levels, and the resulting bias can be large enough to overturn the sign of the estimated effect.
 
To illustrate this, we simulate a two-period DiD environment ($t\in\{0,1\}$) with $N=10{,}000$ units (``songs''), half treated and half controls. Outcomes are generated from a log-additive model and then exponentiated:
\[
\log Y_{it}
=
\alpha_i
+
g\cdot t
+
\delta \cdot \mathbbm{1}\{Treat_i=1\}\cdot \mathbbm{1}\{t=1\}
+
\log \varepsilon_{it},
\qquad
Y_{it}=\exp(\log Y_{it}),
\]
where $g$ is the common log-growth rate (the ``trend''), $\delta$ is the true log treatment effect in the post period, and $\varepsilon_{it}$ is a multiplicative shock.\footnote{In the simulation, we draw the multiplicative shock $\epsilon_{it}$ from a Pareto distribution with shape parameter $k=150.1$, scaled to have $\mathbb{E}[\epsilon_{it}]=1$. We choose a large shape parameter so that idiosyncratic dispersion is small, which keeps Monte Carlo noise low for a given number of replications. This choice reduces the computational burden, since increasing the number of simulations substantially would otherwise be required for stable heatmaps in our current implementation.} The key implication of this DGP is multiplicative treatment: for treated units in the post period, $Y_{i1}(1)=Y_{i1}(0)\exp(\delta)$.
 
To vary baseline differences between treated and controls, we rescale treated outcomes so that the treated-to-control mean ratio in the pre period equals a target value $(1+b)$, with $b\in[-0.30,0.50]$ (reported below in percentage points). Let
\[
R_{0}\equiv \frac{\bar Y_{1,0}}{\bar Y_{0,0}}
\]
denote the realized treated-to-control mean ratio in the pre period from the initial simulated panel. We then apply the multiplicative adjustment
\[
Y_{it}^{adj}
=
\begin{cases}
Y_{it}\cdot \dfrac{1+b}{R_{0}}, & \text{if } Treat_i=1,\\
Y_{it}, & \text{if } Treat_i=0,
\end{cases}
\]
so that the adjusted pre-period ratio equals $(1+b)$ by construction.
 
\paragraph{Estimator and true ATT in levels.}
For each simulated dataset we compute the conventional level DiD estimator,
\[
\widehat{\beta}^{lvl}
=
\Big(\bar Y_{1,1}-\bar Y_{1,0}\Big)-\Big(\bar Y_{0,1}-\bar Y_{0,0}\Big),
\]
where $\bar Y_{d,t}$ is the sample mean among units with $Treat_i=d$ at time $t$.
 
To quantify bias, we compare $\widehat{\beta}^{lvl}$ to the \emph{true} average treatment effect on the treated (ATT) in levels implied by the DGP. The level ATT is
\[
ATT^{lvl,true}
=
\mathbb{E}\!\left[Y_{i1}(1)-Y_{i1}(0)\mid Treat_i=1\right].
\]
Because treatment enters multiplicatively, $Y_{i1}(0)=Y_{i1}(1)\exp(-\delta)$ for treated units in the post period, so
\[
ATT^{lvl,true}
=
\mathbb{E}\!\left[Y_{i1}(1)\big(1-\exp(-\delta)\big)\mid Treat_i=1\right].
\]
In the simulation, we therefore compute the sample analog
\[
\widehat{ATT}^{lvl,true}
=
\bar Y_{1,1}\cdot\big(1-\exp(-\delta)\big),
\]
which uses the observed treated post mean $\bar Y_{1,1}$ (i.e., $Y(1)$) and backs out the implied counterfactual component via the DGP.
 
We  report a treatment-scale normalization that expresses bias as a fraction of the true levels ATT:
\[
RelBias(b,g)
\equiv
\mathbb{E}\!\left[\frac{\widehat{\beta}^{lvl}-\widehat{ATT}^{lvl,true}}{\widehat{ATT}^{lvl,true}}\right].
\]
In the simulation, \(\widehat{ATT}^{lvl,true}\) is the sample analog of
\(ATT^{lvl,true}= \mathbb{E}\!\left[Y_{1,1}\cdot(1-e^{-\delta})\mid Treat=1\right]\),
so each replication computes
\(\widehat{ATT}^{lvl,true}=\overline{Y}_{1,1}\,(1-e^{-\delta})\)
using treated post outcomes. We then average \(RelBias(b,g)\) across 10 Monte Carlo replications for each \((b,g)\) cell and report it in percent terms.
 
\autoref{fig:pareto_bias_heatmap} plots $RelBias(b,g)$ over a grid of baseline differences ($b$) and proportional trends ($g$). Two patterns stand out. As discussed in the paper, the magnitude of the bias rises with both baseline imbalance and the strength of proportional growth \citep{Roth_SantAnna_2023}. Intuitively, when outcomes grow proportionally, absolute changes scale with baseline levels; a level-parallel-trends restriction therefore becomes increasingly inaccurate as either (i) treated and control baselines diverge or (ii) the common proportional growth rate rises, which amplifies absolute differences over time.
 
Second, bias can be large enough to generate sign flips. We highlight the region where $\widehat{\beta}^{lvl}$ has the opposite sign from $\widehat{ATT}^{lvl,true}$. In these cells, a negative true effect (here $\delta=-0.05$) is estimated as positive by level DiD. The sign-flip region expands as $g$ increases and as treated units start sufficiently above controls in baseline levels ($b>0$), which is the mechanical case in which proportional growth generates larger absolute increases for the higher-baseline group and a level-based counterfactual misattributes part of this differential growth to treatment.

In \autoref{fig:pareto_bias_heatmap}, sign flips and large relative bias for a levels DiD arise when two things coincide: treated and control groups start from meaningfully different baseline levels, and the untreated process is proportional growth. In our setting, this corner is unlikely to be empirically relevant because our research design forces treated and matched controls to be extremely close in pre-withdrawal levels and trajectories. That near-balance shrinks the baseline-difference axis toward zero, which is where the heatmap shows the levels estimator behaving well and where sign flips disappear. In other words, matching in levels is not “making levels OLS correct” in general; it is creating a special case in which the additive and multiplicative counterfactuals are locally similar over the relevant window, so levels, log specifications, and log-link PPML line up in practice. 
 
This distinction also clarifies why our ``matching on levels'' design is a special case and why it need not carry over to approaches like synthetic DiD. Synthetic DiD chooses unit weights to balance pre-period outcomes and allows for an intercept shift, which means exact level matching is not required and residual baseline gaps can remain by design \citep{arkhangelsky2021synthetic}. When untreated dynamics are closer to proportional growth, leaving residual treated-minus-synthetic differences in pre-period levels, or in how changes in levels scale with baseline size, can mechanically matter for level-based contrasts. As a result, a levels-based SDID estimate can inherit the same kind of sensitivity to baseline gaps.
 
Finally, \autoref{fig:pareto_bias_heatmap_all} shows that, under proportional-growth dynamics, the sensitivity to baseline gaps is specific to estimating the DiD in levels. The top panel reproduces the earlier pattern in \autoref{fig:pareto_bias_heatmap}. In contrast, the log TWFE and PPML panels are essentially flat at zero across the grid, which is expected because these specifications impose a proportional-trend structure that is aligned with the DGP in the simulation.
 
\paragraph{Formal baseline-scaling test.}
To formally test whether treatment incidence is constant in absolute terms or scales with baseline popularity, we regress the song-level DiD contrast $\Delta_i$ on an intercept only (Column 1) and on the pre-period baseline mean (Column 2).
 
\autoref{tab:horse_race_baseline_scaling} reports the results. The statistically significant negative coefficient on the baseline mean in Column 2 indicates that absolute treatment effects are not constant but scale with song popularity (larger songs experience larger absolute declines). Information criteria (AIC) favor the baseline-dependent specification, indicating that the constant-absolute-incidence hypothesis is less likely. The linear specification in column (2) of \autoref{tab:horse_race_baseline_scaling} yields a positive intercept. This arises from fitting a linear approximation to a relationship where effects are concentrated in the extreme tail (the ``viral head''). While this linear approximation is imperfect for prediction in the lower deciles, the statistically significant negative slope is sufficient to reject the null hypothesis of constant-absolute treatment incidence.
 
\begin{table}[!htbp]
\centering
\footnotesize
\renewcommand{\arraystretch}{1.0}
\caption{Baseline Scaling Test}
\label{tab:horse_race_baseline_scaling}
\begin{threeparttable}
\begin{tabular}{p{5cm}cc}
\toprule
& (1) Intercept only & (2) Linear in baseline \\
\midrule
Intercept 
& $-4{,}661.2^{***}$ & $12{,}043.5^{***}$ \\
& $(736.0)$ 
& $(671.3)$ \\
Baseline mean $\bar y_{i,\mathrm{pre}}$
& 
& $-0.1172^{***}$ \\
& 
& $(0.0010)$ \\
\midrule
Observations 
& $53{,}753$ 
& $53{,}753$ \\
AIC
& $1{,}447{,}698$ & $1{,}435{,}362$ \\
\bottomrule
\end{tabular}
\begin{tablenotes}[para,flushleft]
\scriptsize\textit{Notes:} The dependent variable is the song-level DiD contrast $\Delta_i = (\bar y_{i,\text{post}}-\bar y_{i,\text{pre}}) - (\bar y_{m(i),\text{post}}-\bar y_{m(i),\text{pre}})$, where $m(i)$ denotes song $i$'s matched control. Column (1) fits $\Delta_i=\alpha+\varepsilon_i$. Column (2) fits $\Delta_i=a+b\,\bar y_{i,\text{pre}}+\varepsilon_i$. Standard errors in parentheses. Significance levels: $^{*}p<0.05$, $^{**}p<0.01$, $^{***}p<0.001$.
\end{tablenotes}
\end{threeparttable}
\end{table}

\subsection{Assessing Heteroskedasticity and Implications for Log-OLS Bias}
\label{app:hetero}
 
In the discussion above, we argued that the counterfactual trends of streams are more plausibly characterized in proportional terms than as a common additive level shift: absolute movements depend strongly on baseline popularity, whereas proportional movements are comparatively stable. This motivates focusing on DiD specifications that naturally impose proportional dynamics, in particular log specifications and PPML with a log link.

However, even when the maintained counterfactual is broadly ``multiplicative,'' these implementations need not agree, because they target different estimands. Log-OLS identifies a DiD in mean logs ($\mathbb{E}[\log Y]$), whereas PPML identifies a DiD in log mean levels ($\log \mathbb{E}[Y]$). These two estimands coincide only under restrictive conditions. In particular, when the variance of $\log Y$ changes differentially for treated songs in the post period, Jensen's inequality implies that $\mathbb{E}[\log Y]$ and $\log \mathbb{E}[Y]$ can move in different directions. This is exactly the heteroskedasticity concern that is first-order in heavy-tailed outcomes: treatment can compress or expand the log-scale variance, which can mechanically shift the log-OLS estimator relative to the PPML estimator.

To assess whether heteroskedasticity on the log scale is empirically relevant in our setting, we directly test whether the variance of $\log Y_{it}$ changes differentially for treated songs in the post period. To this end, we follow \citet{ciani_did_2019}, and implement this as a two-step residual-variance diagnostic. First, we estimate the standard log-OLS DiD with song and week fixed effects and calculate fitted residuals, $\widehat{\varepsilon}_{it}$. Second, we use the squared residuals, $\widehat{\varepsilon}_{it}^{\,2}$, as an observation-level proxy for the conditional variance of $\log Y_{it}$ and estimate the following DiD specification:
\begin{equation}
\label{eq:did_sqresid}
\widehat{\varepsilon}_{it}^{\,2}
=
\theta\,(\text{Treat}_i \times \text{Post}_t)
+\mu_i+\gamma_t+\nu_{it},
\end{equation}
where $\theta$ captures whether the log-scale residual variance changes disproportionately for treated songs in the post period.

As reported in \autoref{tab:did_sqresid} below, $\widehat{\theta}$ is negative and statistically significant, which indicates that the variance of $\log Y_{it}$ falls disproportionately for treated songs in the post period. Intuitively, this estimate is consistent with a setting in which the UMG song withdrawal reduces the incidence of short-lived viral spikes among treated songs, which compresses log-scale volatility and can shift the log-OLS estimand upwards. 
 
This is why the DiD in mean logs, $\mathbb{E}[\log Y]$, can be biased upward relative to the DiD in log mean levels, $\log \mathbb{E}[Y]$. The intuition is Jensen’s inequality: when the log-scale variance changes around treatment, the mapping between ``average logs'' and the ``log of averages'' is no longer stable.\footnote{Let $Z\equiv \log Y$. A standard cumulant expansion gives $\log \mathbb{E}[\exp(Z)] = \mathbb{E}[Z] + \tfrac{1}{2}\mathrm{Var}(Z) + \text{(higher cumulants)}$, so $\mathbb{E}[\log Y] \approx \log \mathbb{E}[Y] - \tfrac{1}{2}\mathrm{Var}(\log Y) - \text{(higher cumulants)}$. Taking Differences-in-Differences yields $\Delta\Delta\,\mathbb{E}[\log Y] \approx \Delta\Delta\,\log \mathbb{E}[Y] - \tfrac{1}{2}\Delta\Delta\,\mathrm{Var}(\log Y) - \Delta\Delta\,\text{(higher cumulants)}$. Thus, if $\mathrm{Var}(\log Y)$ falls disproportionately for treated songs in the post period (as in \autoref{tab:did_sqresid}), the term $-\tfrac{1}{2}\Delta\Delta\,\mathrm{Var}(\log Y)$ is positive, which pushes the mean-log DiD upward relative to the log-mean DiD even when mean demand falls.} As a result, even under a true null treatment effect on mean streams, log-OLS can deliver a positive DiD estimate purely due to heteroskedasticity. We illustrate this mechanism below using a simulation exercise calibrated to our setting.

\begin{table}[htbp]
    \centering
    \caption{DiD Test for Treated-Post Changes in Log-Scale Residual Variance}
    \label{tab:did_sqresid}
        
    \footnotesize
    \renewcommand{\arraystretch}{1.0}
    \begin{threeparttable}
        \begin{tabular}{p{7cm} cc}
            \toprule
             & Est. & S.E. \\
            \midrule
            $\text{Treat}_i \times \text{Post}_t$ ($\theta$)
            & $-0.0016^{***}$ & $(0.0002)$ \\
            \hspace{0.5em}Song FE & \multicolumn{2}{c}{Yes} \\
            \hspace{0.5em}Week FE & \multicolumn{2}{c}{Yes} \\
            \addlinespace
            Observations & \multicolumn{2}{c}{2{,}364{,}736} \\
            $R^{2}$ & \multicolumn{2}{c}{0.4464} \\
            \bottomrule
        \end{tabular}%
        \begin{tablenotes}[para,flushleft]
        \scriptsize\emph{Notes:} The table reports $\widehat{\theta}$ from the DiD regression in \autoref{eq:did_sqresid}. A negative $\widehat{\theta}$ implies that the variance of $\log Y_{it}$ falls disproportionately for treated songs in the post period. Standard errors are computed using a Huber–White (sandwich) variance estimator, two-way clustered by song and match-pair. 
        \end{tablenotes}
    \end{threeparttable}
\end{table}

\paragraph{Simulation }To gauge whether the treated-post variance compression documented in \autoref{tab:did_sqresid} is a plausible mechanism that can bias log-OLS estimates relative to the mean-stream treatment effect in parts of the distribution, we conduct a simulation calibrated to the key features of our setting: (i) a heavy-tailed distribution in baseline streams and virality across songs, (ii) multiplicative time dynamics, and (iii) treatment heterogeneity concentrated in the right tail.
 
We simulate a panel of 10{,}000 songs over 20 periods (10 pre, 10 post), randomly assign half of the songs to treatment, and generate streams as $Y_{it}=m_{it}U_{it}$, where $m_{it}$ evolves proportionally over time. Songs are also assigned a highly skewed ``virality'' index, which defines baseline-virality deciles. Treatment effects are heterogeneous by construction: only the top virality decile experiences a negative post-period effect on mean streams, while deciles 1--9 have a true null effect. Finally, we allow the treated-post variance of $\log U_{it}$ to compress by an amount chosen to match the magnitude implied by $\widehat{\theta}$ in \autoref{tab:did_sqresid}.

Formally, for song $i=1,\dots,N$ and period $t=1,\dots,T$, let $Treat_i\in\{0,1\}$ indicate treatment and $\text{Post}_t=\mathbbm{1}\{t\ge t_0\}$ indicate the post period. We generate streams according to a multiplicative model,
\begin{equation}
Y_{it}=m_{it}\,U_{it}, \qquad Y_{it}>0,
\label{eq:sim_y}
\end{equation}
where $m_{it}$ governs the conditional mean and $U_{it}$ is a multiplicative shock with mean one. The mean component evolves proportionally over time:
\begin{equation}
\log m_{it}=q_i+\gamma_t+\tau_{it}, 
\qquad 
\gamma_t=g\,(t-1),
\label{eq:sim_m}
\end{equation}
where $q_i$ is song-specific baseline popularity and $\gamma_t$ is a common growth component. Treatment effects are heterogeneous by a baseline-virality decile indicator $G_i\in\{1,\dots,10\}$ and are concentrated in the top decile:
\begin{equation}
\tau_{it}=\tau_{10}\cdot \mathbbm{1}\{Treat_i=1\}\cdot \mathbbm{1}\{\text{Post}_t=1\}\cdot \mathbbm{1}\{G_i=10\}.
\label{eq:sim_tau}
\end{equation}
Thus, by construction, the true mean effect is zero for deciles $G_i\in\{1,\dots,9\}$ and negative only for $G_i=10$.
 
Shocks are lognormal with mean one,
\begin{equation}
\log U_{it}\sim \mathcal N\!\left(-\tfrac{1}{2}\sigma_{it}^2,\;\sigma_{it}^2\right)
\quad\Rightarrow\quad \mathbb E[U_{it}\mid \cdot]=1,
\label{eq:sim_u}
\end{equation}
and we allow the log-scale standard deviation to change in treated-post:
\begin{equation}
\sigma_{it}=
\begin{cases}
\sigma_{\text{base}}\cdot \sigma_{\text{mult}}, & \text{if } Treat_i=1 \text{ and } \text{Post}_t=1,\\
\sigma_{\text{base}}, & \text{otherwise.}
\end{cases}
\label{eq:sim_sigma}
\end{equation}

We calibrate the simulation as follows: we set the panel size to $N=10{,}000$ songs observed over $T=20$ periods with the post period beginning at $t_0=11$, assign treatment to one half of songs, draw baseline popularity $q_i$ and baseline virality from Pareto-like distributions with tail parameters $k_q=2.5$ and $k_v=1.3$ and correlation $\rho=0.25$ in logs, set the common proportional growth rate to $\gamma_t=g(t-1)$ with $g=0.02$, impose heterogeneous treatment effects with $\tau_{10}=-0.0556$ for treated songs in the top virality decile ($G_i=10$) and $\tau_{it}=0$ for deciles $1$–$9$, set the baseline shock dispersion to $\sigma_{\text{base}}=0.11$, compress treated-post log-shock dispersion by $\sigma_{\text{mult}}=0.94$, and run $N_{\text{sims}}=200$ Monte Carlo replications. \autoref{tab:sim_results_placeholder} reports the simulation estimates aggregated across replications. Two central patterns emerge. First, despite a true null effect for deciles $1$--$9$, log-OLS delivers positive and statistically significant DiD estimates in the long tail (bottom 9 deciles). Second, PPML and weighted log-OLS remain statistically indistinguishable from zero in those same deciles, which is the true effect by construction. At the same time, all estimators recover a negative effect in the top decile, where a negative mean effect is present by construction. These results illustrate that, under multiplicative counterfactual dynamics, a modest treated-post compression of log-scale variance can mechanically push the log-OLS DiD in mean logs upward. This pattern likely contributes to the divergence between log OLS and PPML documented in \autoref{fig:decile_soc}, panels (d) and (e).
 
\begin{table}[!htbp]
    \centering
    \renewcommand{\arraystretch}{1.0}
    \footnotesize
    \caption{Simulation Results: Mean DiD Estimates by Baseline-Virality Group}
    \label{tab:sim_results_placeholder}
    
    \begin{threeparttable}
        \begin{tabular}{lccc}
            \toprule
             & Log-OLS & Weighted log-OLS & PPML \\
            \midrule
            True effect (Bottom 9 deciles) & $0$ & $0$ & $0$ \\
            Estimated effect (Bottom 9 deciles) & $0.0008^{***}$ & $0.0006$ & $-0.0001$ \\
            &                                              $(0.0001)$ & $(0.0005)$ & $(0.0005)$ \\
            \addlinespace
            True effect (Top decile)  & $-0.0556$ & $-0.0556$ & $-0.0556$ \\
            Estimated effect (Top decile)
            & $-0.0552^{***}$ & $-0.0553^{***}$ & $-0.0560^{***}$ \\
            & $(0.0002)$ & $(0.0015)$ & $(0.0015)$ \\
            \midrule
            Simulation replications & \multicolumn{3}{c}{200} \\
            Songs ($N$) & \multicolumn{3}{c}{10{,}000} \\
            Periods ($T$) & \multicolumn{3}{c}{20 (10 pre, 10 post)} \\
            \bottomrule
        \end{tabular}
        
        \begin{tablenotes}[para,flushleft]
        \scriptsize\emph{Notes:} The table reports Monte Carlo evidence on the finite-sample behavior of alternative DiD estimators under a calibrated data-generating process designed to mirror key features of our setting, including heavy-tailed baseline heterogeneity, multiplicative time dynamics, and effect heterogeneity concentrated in the right tail. Each cell reports the average estimate across simulation replications, with standard errors in parentheses computed as the standard deviation of replication-specific averages divided by $\sqrt{N_{\text{sims}}}$. By construction, the true effect is zero outside the top virality decile, so any nonzero estimates in lower deciles reflect mechanical bias rather than causal effects. See Web Appendix \ref{app:hetero} for details on the data-generating process, calibration, and parameter choices.
        \end{tablenotes}
    \end{threeparttable}
\end{table}

\newpage

\subsection{Proportional Specifications: PPML vs.\ Weighted Log-OLS}\label{app:ppml_vs_wlog}
 
The diagnostics in Sections~\ref{app:additive_vs_mult} and~\ref{app:hetero} narrow the set of specifications that are defensible as primary in a heavy-tailed outcome settings. A levels TWFE implementation is hard to defend as a baseline in our setting because it imposes additive untreated dynamics and delivers highly uneven fit in proportional terms across baseline size, with especially large proportional errors in the long tail. Unweighted log-OLS, while interpretable as a DiD in mean logs, targets an estimand that is not central in our application, because it weights all songs equally and therefore need not map cleanly into economic outcomes such as changes in total streams or revenue. In addition, log-OLS can move for purely distributional reasons when treated-post dispersion in $\log Y$ changes, since $\mathbb{E}[\log Y]$ need not track $\log \mathbb{E}[Y]$ under heteroskedasticity and changing higher moments \citep{ManningMullahy2001,log_gravity}. 
 
We therefore focus on two proportional implementations that behave similarly in our data: PPML with a log link and weighted log-OLS using predetermined pre-treatment stream shares as importance weights. These approaches share a proportional ``percent-change’’ interpretation, but they correspond to closely related yet different estimands. PPML identifies the treated-post semi-elasticity in a log-link conditional-mean model, so $\exp(\delta)-1$ is directly interpretable as a percent change in the conditional mean of streams, whereas weighted log-OLS estimates a (weighted) DiD in $\mathbb{E}[\log Y\mid\cdot]$, which corresponds to a weighted proportional change in the geometric mean. 
 
Baseline-share weights change the implicit averaging: they downweight the many low-volume songs and upweight the relatively small set of songs that account for a large share of pre-treatment streams. This is a standard use of weighting when the goal is to summarize treatment effects in economically relevant terms \citep{SolonHaiderWooldridge2015,autor_fall_2020,winkler_separating_2026}. At the same time, weighting does not change the fact that a log-transformed specification targets mean logs, so changes in log-scale dispersion can in principle shift a mean-log estimand relative to an estimand defined on mean levels \citep{ManningMullahy2001,log_gravity}. In our application, the practical relevance of this concern is attenuated because the largest instability in proportional fit arises in the lowest deciles that receive little weight under pre-treatment stream-share weighting. 
 
PPML, in contrast, is designed to estimate log-link conditional-mean models and only requires that the conditional mean be correctly specified. To safeguard against variance misspecification and other forms of heteroskedasticity in PPML, we report cluster-robust standard errors which allow for heteroskedasticity \citep{White1982,log_gravity,SantosSilvaTenreyro2011,mackinnon_cluster_2023}. 
 
Because PPML and weighted log-OLS lead to the same qualitative conclusions in our setting, we do not frame the choice as one estimator uniformly dominating the other. We present PPML as the main specification below, and, in parallel, we report weighted log-OLS in the Web Appendix.

\section{Descriptive Statistics and Extensions to the Main Analysis}

\subsection{Descriptive statistics}
\label{app:descriptives}

Table \ref{tab:descriptive_tab} presents descriptive statistics for the full sample. 

\begin{table}[ht!]
\caption{Descriptive Statistics: Matched Sample (Pre- and Post Periods)}
\label{tab:descriptive_tab}
\centering
\renewcommand{\arraystretch}{1.0}
\resizebox{\linewidth}{!}{%
\begin{threeparttable}
\begin{tabular}{lrrrrrrrc}
\toprule
 & Mean & SD & Min & P25 & P50 & P75 & Max & N \\
\midrule
\multicolumn{9}{l}{\textit{Time-varying (song--week observations)}} \\
Spotify streams per week & 147{,}651.13 & 704{,}527.43 & 1{,}000 & 9{,}026 & 23{,}026 & 70{,}646 & 44{,}116{,}676 & 2{,}795{,}156 \\
Spotify playlist followers (weekly total) & 635{,}638.50 & 3{,}220{,}009.72 & 0 & 1{,}978 & 47{,}106 & 251{,}994 & 176{,}338{,}862 & 2{,}795{,}156 \\
\addlinespace
\multicolumn{9}{l}{\textit{Time-constant (measured at treatment, song level)}} \\
TikTok creations at treatment & 16{,}018.39 & 277{,}681.75 & 1 & 38 & 180 & 1{,}050 & 45{,}980{,}273 & 107{,}506 \\
Song age at treatment (weeks) & 920.84 & 846.35 & 10 & 247 & 683 & 1{,}361 & 6{,}473 & 107{,}506 \\
\bottomrule
\end{tabular}
\begin{tablenotes}[para,flushleft]
\footnotesize\emph{Notes:} Descriptive statistics for the matched sample used in the main analysis. The sample covers the period from 2023-11-30 to 2024-05-23 (26 weeks), comprising 53{,}753 treated UMG songs and 53{,}753 matched control songs (107{,}506 songs in total). Time-varying variables are summarized over all 2{,}795{,}156 song--week observations. Time-constant variables are summarized at the song level (one observation per song, measured at the time of treatment). Song age is measured in weeks since release at the time of treatment.
\end{tablenotes}
\end{threeparttable}
}%
\end{table}

\subsection{Decile cutoffs}
\label{app:decile_cutoffs}

Please refer to Table \ref{tab:decile_thresholds_joint} for an overview of decile thresholds regarding the TikTok creations and Spotify streams variables.

\begin{table}[htb]
    \centering
    \renewcommand{\arraystretch}{1.0}
    \caption{Decile Thresholds and Within-Decile Means (Pre-Treatment, Treated Songs)}
    \label{tab:decile_thresholds_joint}
    
    \resizebox{0.9\linewidth}{!}{%
    \begin{threeparttable}
        \begin{tabular}{l
                        S[table-format=8.0] S[table-format=8.0] S[table-format=9.2]
                        S[table-format=8.0] S[table-format=9.0] S[table-format=9.0]}
        \toprule
        & \multicolumn{3}{c}{TikTok creations at treatment} & \multicolumn{3}{c}{Spotify streams per week} \\
        \cmidrule(lr){2-4}\cmidrule(lr){5-7}
        Decile & {Lower} & {Upper} & {Mean} & {Lower} & {Upper} & {Mean} \\
        \midrule
        D1 & 1 & 9 & 4.11 & 1099 & 4236 & 2790 \\
        D2 & 9 & 24 & 15.7 & 4237 & 7168 & 5667 \\
        D3 & 24 & 50 & 36.4 & 7169 & 10892 & 8954 \\
        D4 & 50 & 95 & 70.8 & 10893 & 15689 & 13172 \\
        D5 & 95 & 172 & 130 & 15689 & 22704 & 18933 \\
        D6 & 172 & 325 & 239 & 22705 & 33664 & 27711 \\
        D7 & 325 & 662 & 469 & 33665 & 53312 & 42341 \\
        D8 & 662 & 1621 & 1044 & 53316 & 94816 & 70937 \\
        D9 & 1621 & 6508 & 3330 & 94855 & 223120 & 144606 \\
        D10 & 6512 & 14566721 & 118723 & 223140 & 24940777 & 1090419 \\

        \bottomrule
    \end{tabular}
    \begin{tablenotes}[para,flushleft]
    \scriptsize\emph{Notes:} The table reports empirical decile cutoffs (lower and upper bounds) and within-decile means for treated songs, based on pre-treatment TikTok creations and pre-treatment Spotify streams. Bounds correspond to the minimum and maximum values observed within each decile. Due to ties, adjacent deciles may share boundary values. The top-decile upper bound equals the sample maximum.
    \end{tablenotes}
    \end{threeparttable}
    }%
\end{table}

\subsection{Model-Free Evidence by TikTok-Virality Decile}
\label{app:mfe_by_decile}

\begin{figure}[!htbp]
	\centering
	\caption{Weekly Mean Spotify Streams by Pre-Treatment TikTok-Virality Decile}
	\label{fig:mfe_by_decile}
	\includegraphics[width=.95\textwidth]{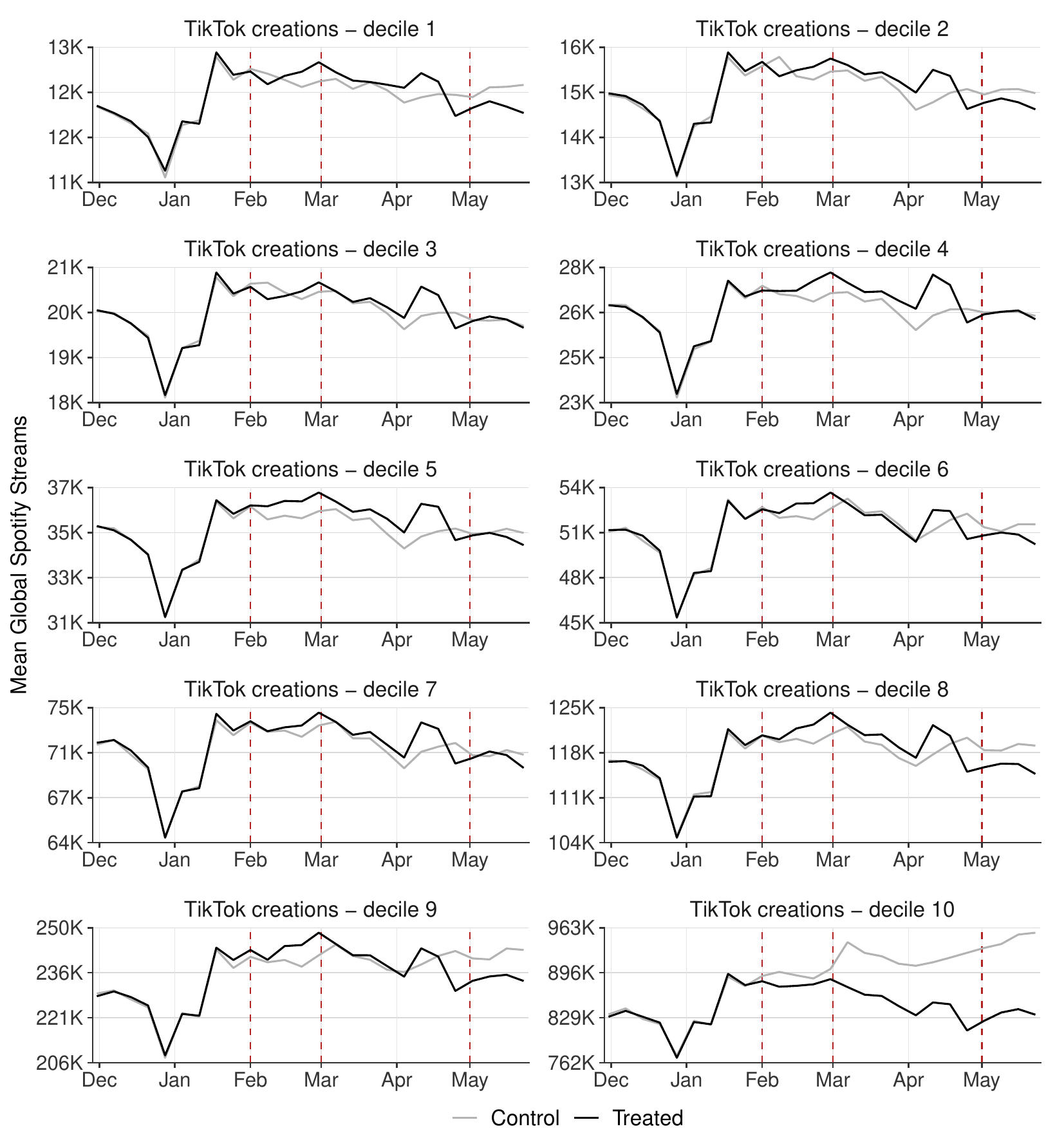}
	\vspace{0.5em}
	\centering
    \parbox{\textwidth}{\scriptsize\textit{Notes}: The plots show mean streams for treated songs (black) and matched control songs (grey) after 1:1 matching. Each subplot corresponds to a decile of pre-treatment TikTok virality. Vertical dashed lines mark key events and phase boundaries: Phase 1 is the pre-treatment period; Phase 2 begins with UMG's master-rights withdrawal; Phase 3 begins with the publishing-rights withdrawal; and Phase 4 begins with UMG's re-entry on TikTok. Means are computed over matched pairs. Note that y-axes differ across panels.}
\end{figure}

\autoref{fig:mfe_by_decile} plots weekly mean Spotify streams for treated songs and their matched controls separately by pre-treatment TikTok-virality decile. The figure serves two purposes. First, it provides a transparent check that the matched-control design delivers close pre-treatment alignment within each decile. Second, it highlights why decile-level heterogeneity is economically consequential in this setting. On the level scale, the post-period divergence is visually concentrated in the viral head: in the top decile, treated songs exhibit a clear relative decline after the withdrawal, whereas the lower deciles show much smaller movements. This descriptive pattern is informative for interpretation even before imposing any functional-form restrictions.

\subsection{Main results using weighted log OLS}
\label{app:wgt_logols_main}
 
\begin{figure}[ht!]
  \centering
  \caption{Weighted log-OLS Event-Study Estimates of the UMG Withdrawal Effects on Streaming}
  \label{fig:event_study_wgt_logols}
 
  \begin{subfigure}{0.49\textwidth}
    \centering
    \caption{Pooled estimates}
    \includegraphics[width=\textwidth]{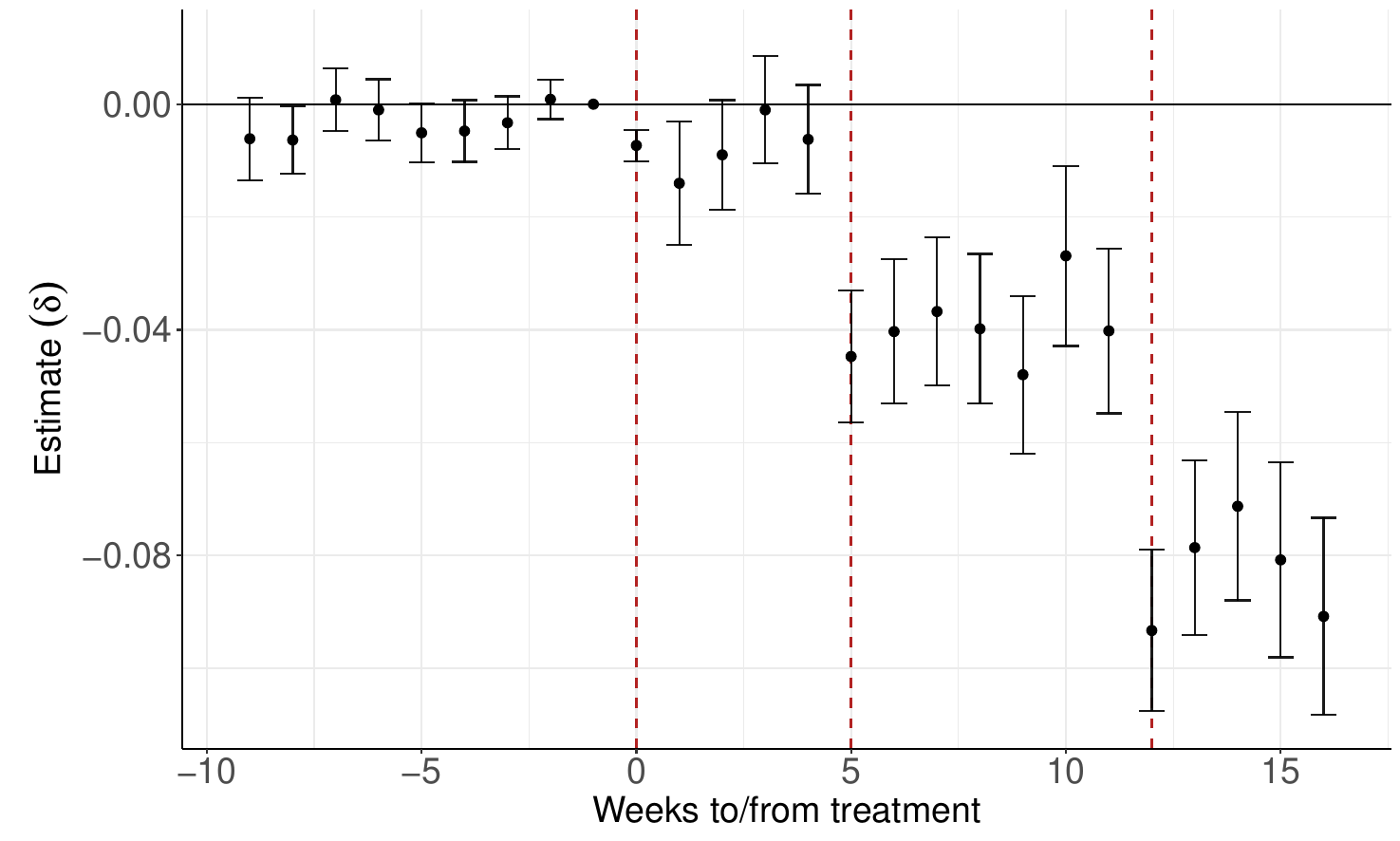}
    \label{fig:event_study_logols_agg}
  \end{subfigure}%
  \hfill
  \begin{subfigure}{0.49\textwidth}
    \centering
    \caption{Estimates by baseline virality group}
    \includegraphics[width=\textwidth]{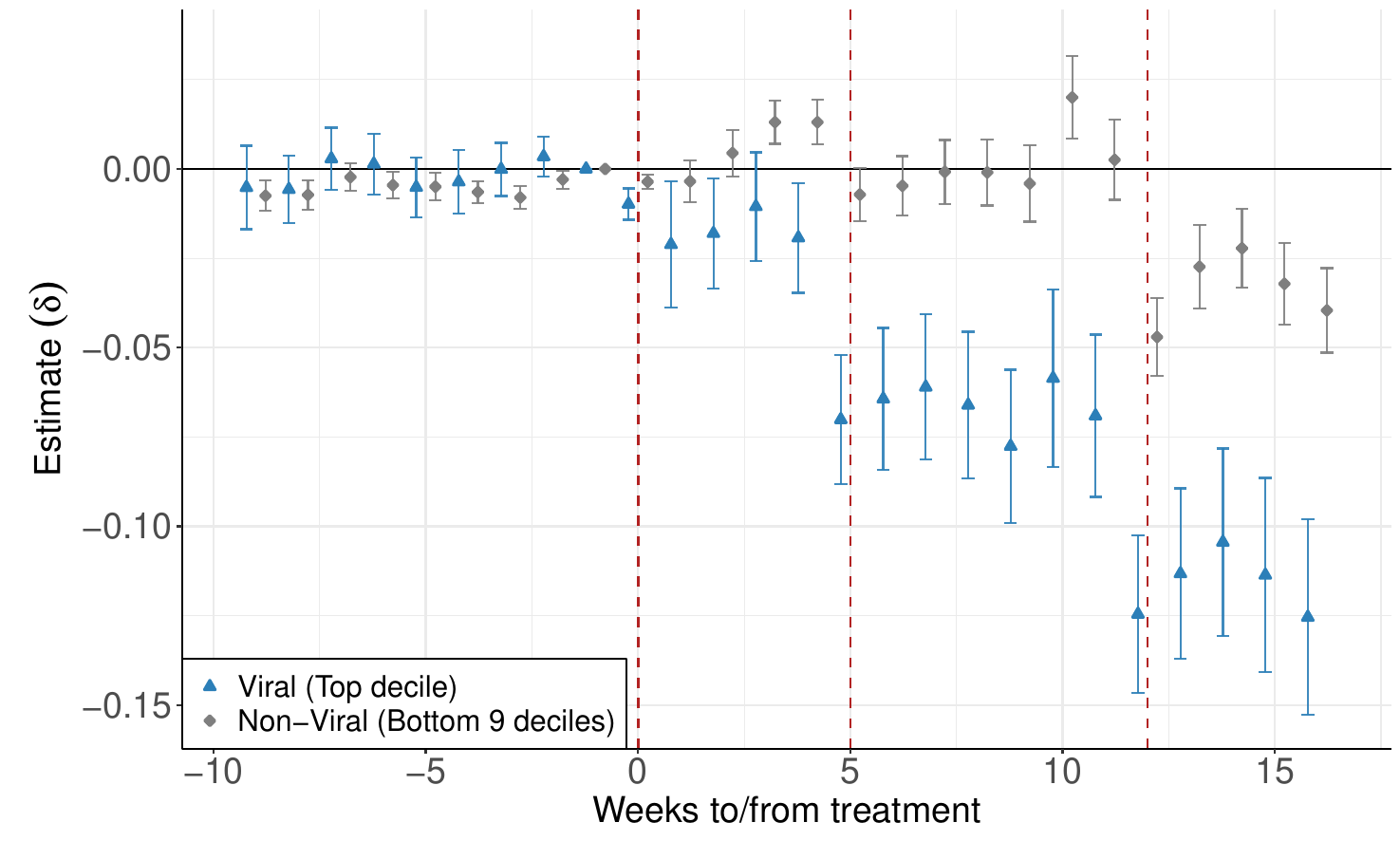}
    \label{fig:event_study_logols_group}
  \end{subfigure}
 
\vspace{-0.5cm}
  \begin{justify}
    \scriptsize{\textit{Notes:} The figure reports weighted log-OLS event-study estimates where the dependent variable is $\log(\text{streams})$. Coefficients are normalized to a pre-treatment reference week and are reported in log points, which can be interpreted as approximate percent effects; exact percent effects are $\exp(\widehat{\delta})-1$. Vertical dashed lines mark the phase boundaries. Phase 2 begins with the UMG master-rights exit from TikTok, Phase 3 begins with the UMG publishing-rights exit, and Phase 4 begins with UMG re-entry (see \autoref{fig:tiktok_chart_global} for the phase timeline). Panel (a) reports pooled estimates from a model with song fixed effects and week fixed effects. Panel (b) reports estimates from a split specification by baseline TikTok virality group (top decile versus bottom nine deciles, defined by pre-treatment creations) that includes song fixed effects and week fixed effects and is estimated separately by subgroup. Observations are weighted by the treated song's pre-treatment market share within each matched pair. Standard errors are computed using a Huber--White (sandwich) variance estimator, two-way clustered by song id and match-pair id.}
  \end{justify}
\end{figure}

In this appendix subsection, we re-estimate the main event-study using weighted log-OLS, which applies predetermined pre-treatment stream-share weights to summarize proportional responses in a way that is more aligned with aggregate listening. \autoref{fig:event_study_wgt_logols} reports the event-time path, and \autoref{tab:wgt_logols_eventstudy_phases_split} summarizes the same estimates in dispute-phase windows, mirroring the main PPML presentation in the paper.
 
\begin{table}[htb]
    \centering
    \caption{Weighted Log-OLS Estimates by Dispute Phase: Pooled vs. Split by Baseline Virality}
    \label{tab:wgt_logols_eventstudy_phases_split}
    \renewcommand{\arraystretch}{1.0}
    \resizebox{0.95\linewidth}{!}{%
    \begin{threeparttable}
    \begin{tabular}{lcccccc}
    \toprule
    & \multicolumn{2}{c}{Overall} & \multicolumn{2}{c}{Bottom 9 deciles} & \multicolumn{2}{c}{Top decile} \\
    \cmidrule(lr){2-3}\cmidrule(lr){4-5}\cmidrule(lr){6-7}
     & Est. (log point) & S.E. & Est. (log point) & S.E. & Est. (log point) & S.E. \\
    \midrule
    Phase 2: Week 0--4 
    & $-0.0047$        & $(0.0044)$
    & $0.0096^{***}$   & $(0.0026)$
    & $-0.0144^{*}$    & $(0.0070)$ \\
     
    Phase 3: Week 5--11 
    & $-0.0435^{***}$  & $(0.0073)$
    & $-0.0004$        & $(0.0051)$
    & $-0.0726^{***}$  & $(0.0112)$ \\
     
    Phase 4: Week 12--16  
    & $-0.0776^{***}$  & $(0.0089)$
    & $-0.0254^{***}$  & $(0.0061)$
    & $-0.1128^{***}$  & $(0.0139)$ \\
     
    \addlinespace
    Song FE 
    & \multicolumn{2}{c}{Yes} 
    & \multicolumn{2}{c}{Yes} 
    & \multicolumn{2}{c}{Yes} \\
     
    Week FE 
    & \multicolumn{2}{c}{Yes} 
    & \multicolumn{2}{c}{Yes} 
    & \multicolumn{2}{c}{Yes} \\
     
    Observations 
    & \multicolumn{2}{c}{2{,}795{,}156} 
    & \multicolumn{2}{c}{2{,}509{,}936} 
    & \multicolumn{2}{c}{285{,}220} \\
     
    $R^2$ 
    & \multicolumn{2}{c}{0.9956} 
    & \multicolumn{2}{c}{0.9936} 
    & \multicolumn{2}{c}{0.9915} \\
     
    \bottomrule
    \end{tabular}
 
    \begin{tablenotes}[para,flushleft]
        \footnotesize\emph{Notes:} The table reports weighted log-OLS estimates where the dependent variable is $\log(\text{streams})$. Specifications include song fixed effects and week fixed effects. Treatment status is interacted with mutually exclusive post phases, with the pre-treatment phase as the omitted reference category. Phase 2 begins with the UMG master-rights exit from TikTok, Phase 3 begins with the UMG publishing-rights exit, and Phase 4 begins with UMG re-entry (see \autoref{fig:tiktok_chart_global} for the phase timeline). The Overall column reports pooled estimates; the Bottom 9 deciles and Top decile columns report estimates from separate subsamples defined by baseline TikTok virality (pre-treatment creations). Observations are weighted by the treated song's pre-treatment market share within each matched pair. Standard errors (in parentheses) use a Huber--White (sandwich) variance estimator and are two-way clustered by song id and matched-pair id. Coefficients are reported in log points and can be interpreted as approximate percent effects; exact percent effects are $\exp(\widehat{\delta})-1$. Significance levels: $^{*}p<0.05$, $^{**}p<0.01$, $^{***}p<0.001$.
    \end{tablenotes}
 
\end{threeparttable}
}%
\end{table}

\subsection{Playlist Amplification and Reach}
\label{app:playlist}
 
Another mechanism that is consistent with our results is that TikTok affects Spotify demand  not only through contemporaneous cross-platform discovery, but also through a playlist amplification channel. Spotify playlists, especially large curated playlists such as “Viral Hits” with more than 4 million followers (as of December 2025) are a major source of listening, and placements on these playlists can substantially expand reach for tracks that are already gaining momentum \citep[e.g.,][]{aguiar_platforms_2021,wlomert_vanheerde_papies}. Consistent with this, Spotify reports that curation and recommendations drive close to 50\% of all users’ streams, which indicates that playlists play a central role in shaping where listening flows on the platform \citep{spotify_streamon_2023}.

\begin{figure}[!htbp] 
	\centering
	\caption{Event-Study Estimates of the UMG Withdrawal Effects on Spotify Curated-Playlist Followers} 
	\label{fig:effects_playlist_followers}
    \begin{subfigure}{0.5\textwidth}
		\centering
		\caption{Pooled estimates}
        \label{fig:effects_playlist_followers_pooled}
        \includegraphics[width=\textwidth]{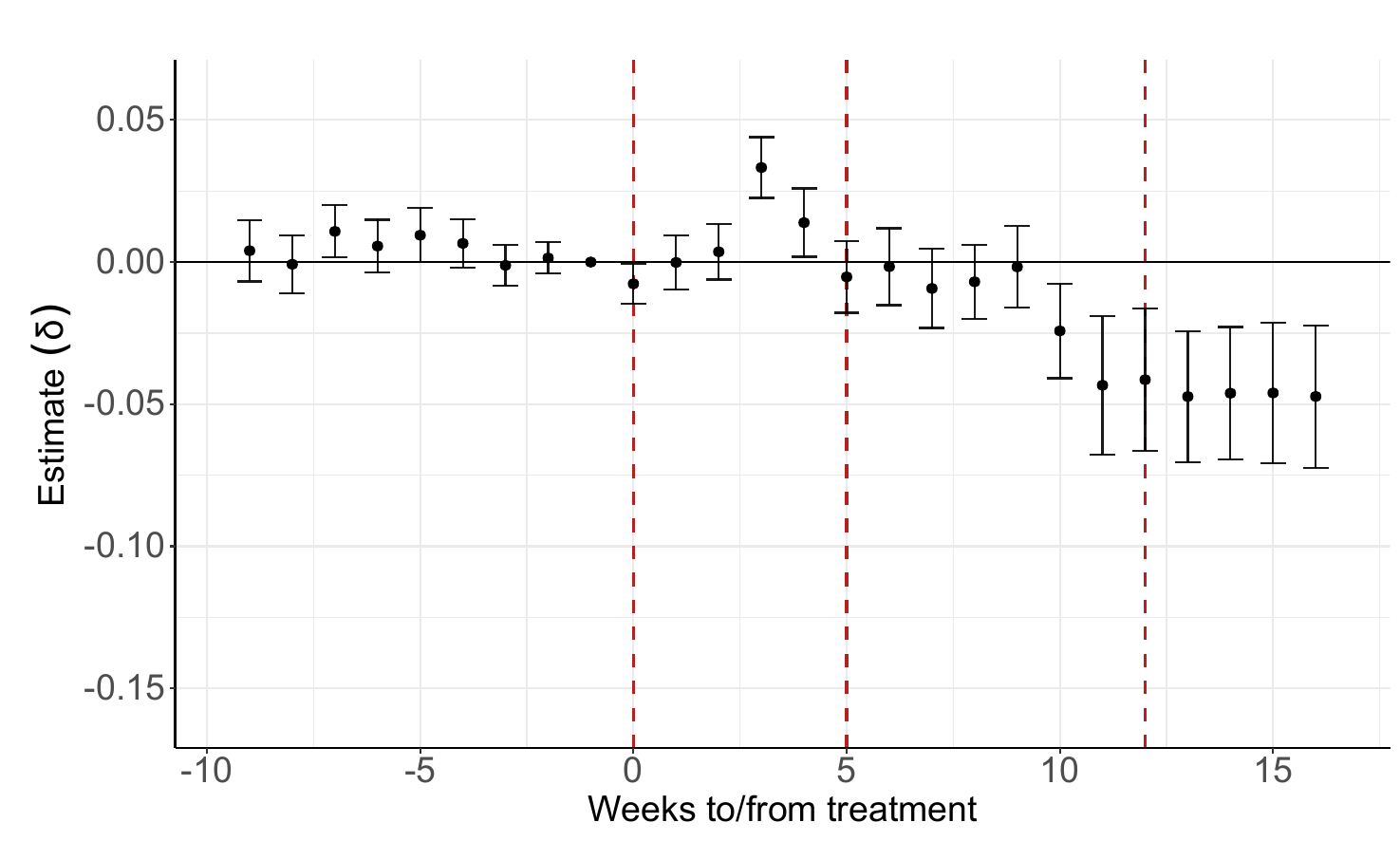}
	\end{subfigure}%
	\hfill
	\begin{subfigure}{0.5\textwidth}
		\centering
		\caption{By baseline TikTok virality group}
        \label{fig:effects_playlist_followers_split}
		\includegraphics[width=1\textwidth, trim=0 0 0 0cm, clip]{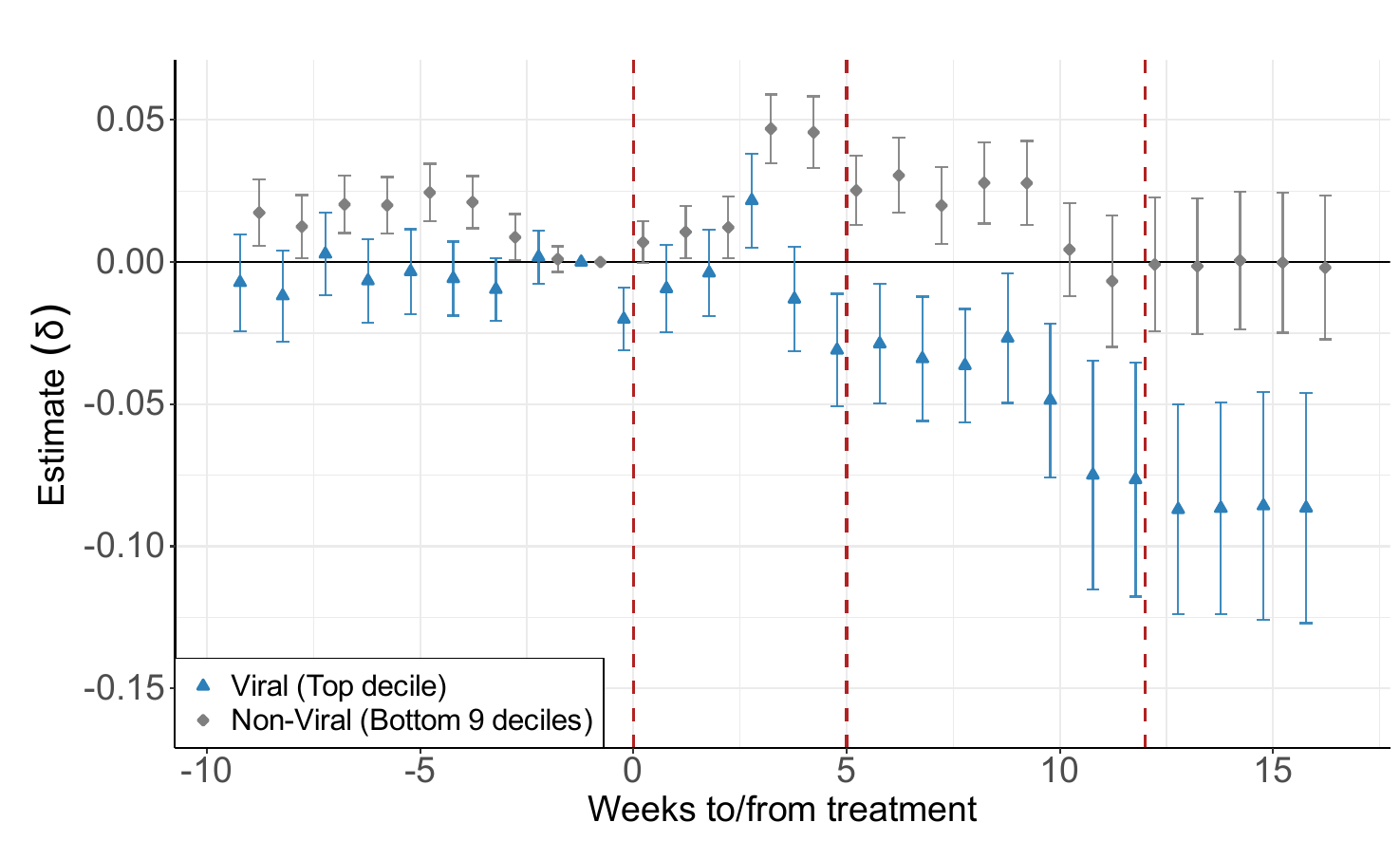}
	\end{subfigure}
	\raggedright\scriptsize
	\parbox{1\textwidth}{\textit{Notes:} The figure reports Poisson pseudo-maximum-likelihood (PPML) event-study estimates from a log-link conditional-mean model in which the dependent variable is a song's cumulative number of followers across Spotify curated playlists in week $t$. Coefficients are normalized to a pre-treatment reference week and are reported in log points. Vertical dashed lines mark the phase boundaries shown in \autoref{fig:tiktok_chart_global} (Phase 2: master-rights exit; Phase 3: publishing-rights exit; Phase 4: re-entry). Panel (a) reports pooled estimates and panel (b) reports estimates from a split specification by baseline TikTok virality group (top decile versus bottom nine deciles, defined by pre-treatment creations). Standard errors are computed using a Huber--White (sandwich) cluster-robust variance estimator, two-way clustered by song and match-pair.}
\end{figure}
 
To evaluate this mechanism, we re-estimate our focal proportional DiD specification using PPML, but with a song’s cumulative number of followers across Spotify curated playlists in week $t$ as the dependent variable.  \autoref{fig:effects_playlist_followers} reports the corresponding event-study estimates. The patterns are twofold. First, the decline in playlist followers emerges with a lag: the pooled series is close to zero early in the dispute and becomes meaningfully negative only later, which is consistent with a snowball process in which reduced TikTok exposure weakens the accumulation of Spotify-side distribution over time rather than shifting it immediately. Second, this effect is concentrated in the viral head. In panel (b), the follower losses are driven almost entirely by the top TikTok-virality decile, while the bottom nine deciles exhibit little systematic movement. Taken together, these results suggest that the withdrawal reduced the Spotify-side amplification of the most TikTok-exposed titles, which is consistent with TikTok-based momentum feeding into Spotify’s playlist ecosystem.
 
This mechanism also helps interpret why the main streaming effects persist into Phase 4, when the UMG catalog returns to TikTok. Playlist exposure and follower accumulation are path-dependent: missing the window when a track is accumulating momentum can translate into fewer playlist placements and a smaller follower base, and that gap does not mechanically close at re-entry. In that sense, the follower declines in \autoref{fig:effects_playlist_followers} are suggestive of an amplification channel that operates through Spotify’s own distribution, which helps explain both the delayed onset of the effects and why the streaming losses do not immediately reverse when UMG is reinstated.
 
To summarize, in our setting we view TikTok as an upstream source of discovery, and Spotify playlists as a downstream amplification layer. The withdrawal can therefore reduce streams both directly, by removing a discovery channel, and indirectly, by weakening the Spotify-side propagation of TikTok-generated momentum through playlist placement, follower accumulation, and their inertial listening. We do not attempt to separate these two margins; instead, we interpret the evidence as consistent with a combined mechanism in which a short-run exposure shock translates into a longer-run distribution gap.
  
\FloatBarrier

\subsection{Placebo Test: Songs with No Prior TikTok Exposure}
\label{app:placebo_no_tiktok_exposure}

This appendix reports a placebo-style test using UMG songs with zero pre-treatment TikTok creations. For these songs, the withdrawal does not remove an active TikTok exposure channel, so we should not expect a corresponding decline in Spotify streams if the main effects operate through TikTok-based discovery. \autoref{tab:ppml_placebo_phases_split} presents the results of a phase-by-phase analysis with coefficients that are statistically indistinguishable from zero across all phases.

\begin{table}[htbp]
\centering
\caption{Placebo: PPML Estimates by Dispute Phase for Songs With no Prior TikTok Exposure}
\label{tab:ppml_placebo_phases_split}
\footnotesize
\renewcommand{\arraystretch}{1.0}
\begin{threeparttable}
\begin{tabular}{p{7.5cm} c c}
\toprule
 & Est. (log point) & S.E. \\
\midrule
Phase 2: Week 0--4   & $0.0127$ & $(0.0073)$ \\
Phase 3: Week 5--11  & $0.0155$     & $(0.0106)$ \\
Phase 4: Week 12--16  & $-0.0081$    & $(0.0133)$ \\
\addlinespace
Song FE & \multicolumn{2}{c}{Yes} \\
Week FE        & \multicolumn{2}{c}{Yes} \\
Observations   & \multicolumn{2}{c}{1{,}304{,}836} \\
Pseudo $R^2$   & \multicolumn{2}{c}{0.9933} \\
\bottomrule
\end{tabular}
\begin{tablenotes}[para,flushleft]
    \scriptsize\emph{Notes:} PPML estimates from a log-link model, coefficients are in log points. The analysis is run on the matched sample and retains only treated--control pairs with closely aligned pre-treatment dynamics, excluding pairs for which the normalized match distance exceeds 0.1, consistent with the main analysis. This yields 25{,}093 treated songs (and 25{,}093 pair instances) and 15{,}539 unique control songs. Standard errors are computed using a Huber–White (sandwich) variance estimator, two-way clustered by song and match-pair.
\end{tablenotes}
\end{threeparttable}
\end{table}

\subsection{Back-of-the-Envelope Revenue Calculations}
\label{sec:be_revenue}
 
In this section we provide a back-of-the-envelope calculation to gauge the revenue impact of the withdrawal for the UMG catalog. \autoref{tab:impact_phase23_pooled_vs_topdecile} translates the PPML phase-specific estimates into an economic magnitude by (i) constructing baseline weekly treated streams $Y_0$ from the pre-period data (pooled: $Y_0=7.662$B streams/week; top decile: $Y_{0,10}=4.573$B streams/week) and then (ii) using the log-link implication that phase-$s$ losses equal $\widehat{\text{LostStreams}}_s = Y_0\big(1-\exp(\hat\delta_s)\big)$, aggregated over Phase 2 (5 weeks) and Phase 3 (7 weeks). In the pooled approach, Phase 2 implies 33.9M lost streams/week and Phase 3 implies 357.1M lost streams/week, summing to 2.67B lost streams over Phases 2--3; multiplying by a payout of $v=\$0.0035$ per stream yields an implied loss of $\$9.3$M. Focusing only on the top decile delivers 70.9M (Phase 2) and 354.7M (Phase 3) lost streams per week and a total of 2.84B lost streams over Phases 2--3, corresponding to $\$9.9$M at $v=\$0.0035$. This similarity in magnitude again highlights that the economically meaningful losses are concentrated in the head of the distribution. 
 
\begin{table}[!htb]
    \centering
    \footnotesize
    \renewcommand{\arraystretch}{1.0}
    \caption{Implied Lost Spotify Streams from PPML Phase Estimates (Phase 2--4): Pooled vs.\ Top-Decile-Only}
    \label{tab:impact_phase23_pooled_vs_topdecile}
    \begin{threeparttable}
        \begin{tabular}{lcc}
            \toprule
            & All treated songs (pooled) & Top decile only \\
            \midrule

            Baseline streams per week, $Y_0$
            & 7{,}662{,}240{,}800 & 4{,}572{,}983{,}052 \\
            \addlinespace
            \addlinespace
            \multicolumn{3}{l}{\emph{Phase 2}} \\
            PPML estimate, $\hat\delta_2$ & $-0.0044$ & $-0.0156$ \\
            Implied \% change, $\exp(\hat\delta_2)-1$ & $-$0.44\% & $-$1.55\% \\
            Lost streams per week, $Y_0\big(1-\exp(\hat\delta_2)\big)$ & 33{,}912{,}685 & 70{,}920{,}626 \\
            Weeks in phase, $w_2$ & 5 & 5 \\
            Total lost streams in phase, $w_2\,Y_0\big(1-\exp(\hat\delta_2)\big)$ & 169{,}563{,}426 & 354{,}603{,}131 \\
            \addlinespace
            \multicolumn{3}{l}{\emph{Phase 3}} \\
            PPML estimate, $\hat\delta_3$ & $-0.0477$ & $-0.0807$ \\
            Implied \% change, $\exp(\hat\delta_3)-1$ & $-$4.66\% & $-$7.76\% \\
            Lost streams per week, $Y_0\big(1-\exp(\hat\delta_3)\big)$ & 357{,}078{,}075 & 354{,}657{,}120 \\
            Weeks in phase, $w_3$ & 7 & 7 \\
            Total lost streams in phase, $w_3\,Y_0\big(1-\exp(\hat\delta_3)\big)$ & 2{,}499{,}546{,}525 & 2{,}482{,}599{,}842 \\
            \addlinespace
            \multicolumn{3}{l}{\emph{Phase 4}} \\
            PPML estimate, $\hat\delta_4$ & $-0.0849$ & $-0.1251$ \\
            Implied \% change, $\exp(\hat\delta_4)-1$ & $-$8.14\% & $-$11.76\% \\
            Lost streams per week, $Y_0\big(1-\exp(\hat\delta_4)\big)$ & 623{,}732{,}148 & 537{,}605{,}700 \\
            Weeks in phase, $w_4$ & 5 & 5 \\
            Total lost streams in phase, $w_4\,Y_0\big(1-\exp(\hat\delta_4)\big)$ & 3{,}118{,}660{,}741 & 2{,}688{,}028{,}500 \\
            \midrule

            \multicolumn{3}{l}{\emph{Phases 2+3}} \\
            Total lost streams, $\sum_{s\in\{2,3\}} w_s\,Y_0\big(1-\exp(\hat\delta_s)\big)$ & 2{,}669{,}109{,}950 & 2{,}837{,}202{,}973 \\
            Implied lost revenue (\$), $v=\$0.0035$/stream & 9{,}341{,}885 & 9{,}930{,}210 \\
            \midrule

            \multicolumn{3}{l}{\emph{Phases 2+3+4}} \\
            Total lost streams, $\sum_{s\in\{2,3,4\}} w_s\,Y_0\big(1-\exp(\hat\delta_s)\big)$ & 5{,}787{,}770{,}691 & 5{,}525{,}231{,}473 \\
            Implied lost revenue (\$), $v=\$0.0035$/stream & 20{,}257{,}197 & 19{,}338{,}310 \\
            \bottomrule
        \end{tabular}

        \begin{tablenotes}[para,flushleft]
        \scriptsize
\emph{Notes:} Lost streams are computed as $Y_0\big(1-\exp(\hat\delta_s)\big)$, where $Y_0$ is baseline weekly streams summed across all songs in the relevant sample and $\hat\delta_s$ is the PPML phase estimate (log points) from \autoref{tab:ppml_eventstudy_phases_split}. Total losses multiply weekly losses by phase length ($w_2=5$ weeks, $w_3=7$ weeks, $w_4=5$ weeks). Revenue losses equal stream losses multiplied by $v=\$0.0035$ per stream (USD/stream), rounded to dollars.
        \end{tablenotes}
    \end{threeparttable}
\end{table}

\subsection{UMG Spotify Top-200 Chart Footprint}
\label{app:top200}

Our main song-level analysis shows substantial treatment-effect heterogeneity. Most songs exhibit little change when TikTok exposure is disrupted, whereas economically meaningful losses are concentrated among tracks with high pre-treatment TikTok virality. Spotify Top-200 charts provide a natural complementary outcome because they are, by construction, concentrated in successful, high-exposure songs, which is precisely the part of the distribution where our main results predict negative effects for UMG following the withdrawal.
 
Chart presence is economically meaningful for rights holders because it is a winner metric that is used to diagnose which songs have crossed a commercially relevant visibility threshold, and that feeds directly into decisions about promotion, distribution, and bargaining. In practice, charts are not only a retrospective summary of listening, but also a coordination device for the ecosystem: charting status is used to justify incremental marketing support, to trigger additional playlist pitching and media attention, and to sustain momentum for releases that are already breaking through. Consistent with this, Spotify’s own chart framing treats chart position as a salient, shareable signal of success, which aligns with the broader logic that rankings can shift attention and choices by making “what others chose” visible and thereby creating amplification through observational learning \citep{salganik_experimental_2006, CaiChenFang2009ObservationalLearning}.
 
We collect weekly Spotify Top-200 chart data from Spotify’s chart portal for 69 countries over 20 weeks (2023-11-26 to 2024-04-07).\footnote{Spotify Top-200 chart data were obtained from publicly accessible platform pages at the time of data collection. We end our observation period on April 7 because Taylor Swift released a new album in the week after, which complicates causal attribution because she is a UMG artist who re-introduced her content to TikTok before the official new deal between UMG and TikTok.} We assign each charting song to one of four distributor groups (UMG, Sony, Warner, and independent/other) and aggregate to a label--country--week panel. The dependent variable is the number of unique songs per label per country in the weekly Top 200. This outcome captures the label’s extensive-margin chart footprint within each market and week, and it naturally includes new releases.
 
Following our baseline approach, we estimate a Poisson fixed-effects DiD model with a log link. We include label–country fixed effects, which absorb time-invariant differences in baseline chart presence across labels within a country, and country–week fixed effects, which absorb all shocks common to labels within a given country-week. Unlike in our song-level design, we do not implement matching here because the aggregated label–country series exhibit closely comparable pre-treatment trends, and with label–country and country–week fixed effects identification comes from within-country-week substitution across labels. To assess dynamics and pre-trends, we also estimate an event-study version that interacts the treatment indicator with week indicators and normalizes the last pre-treatment period to zero.
 
\begin{figure}[!htbp]
  \centering
  \caption{Event-Study Estimates of the UMG Withdrawal Effects on Spotify Top-200 Charts}
  \label{fig:withdrawal_eventstudy}
  \includegraphics[width=0.75\textwidth]{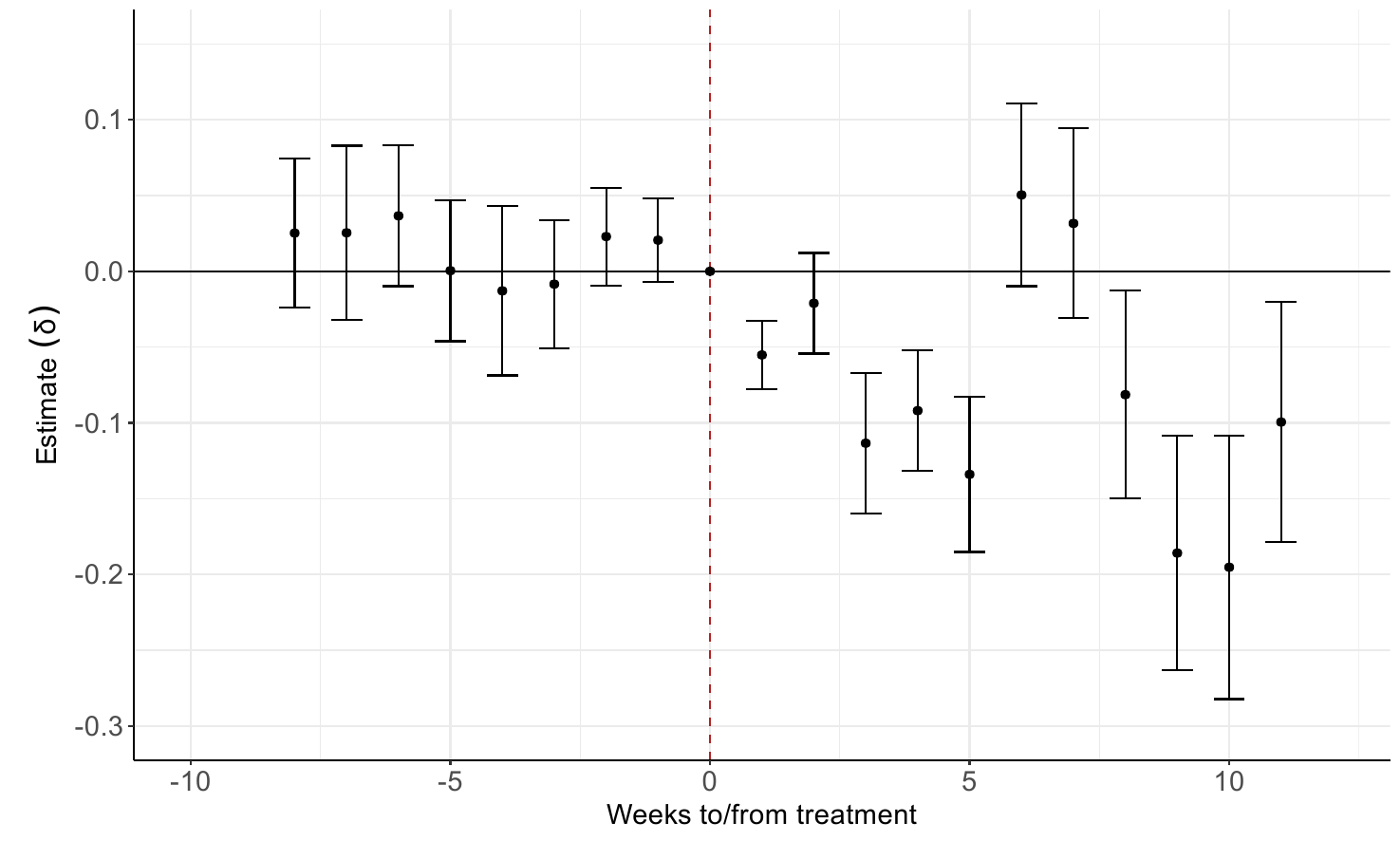}
 
  \vspace{2mm}
  \scriptsize
  \parbox{0.99\textwidth}{\textit{Notes:} The figure reports Poisson pseudo-maximum-likelihood (PPML) event-study estimates from a log-link model where the outcome is the weekly number of unique songs per label and country in Spotify's Top~200 charts. Points show coefficients (log points) for the UMG series relative to other labels, interacted with week indicators and normalized to the last pre-withdrawal week (period 9); whiskers are 95\% confidence intervals. The vertical dashed line marks the first post-withdrawal week. The specification includes label--country fixed effects and country--week fixed effects, and standard errors are clustered at the label--country level.}
\end{figure}
 
As shown in \autoref{fig:withdrawal_eventstudy}, the event-study estimates exhibit no systematic differential pre-trend in UMG’s Top-200 footprint relative to other labels within the same country-week: the pre-treatment coefficients are small and statistically indistinguishable from zero. After the withdrawal, the coefficients turn negative and remain below zero for much of the post window, with several weeks showing statistically significant declines.
 
In the pooled post specification, we estimate an approximately 9\% reduction in UMG’s Top-200 chart footprint per country-week relative to the within-country-week counterfactual defined by the other labels (see results reported in \autoref{tab:top200_ppml_main}). We translate this estimate into a counterfactual ``forgone Top-200 songs'' magnitude and find that it implies roughly 4{,}055 fewer UMG song–country–week Top-200 appearances during the post-withdrawal window.\footnote{We compute forgone Top-200 appearances using the estimated post coefficient from the log-link model. Let $N^{obs}_{post}$ denote the observed total number of UMG song--country--week Top-200 appearances in the post window (counting a song separately each time it appears in a given country-week). Under the multiplicative structure, the no-withdrawal counterfactual is $N^{cf}_{post}=N^{obs}_{post}/\exp(\hat\delta)$, so forgone appearances equal $N^{cf}_{post}-N^{obs}_{post}=N^{obs}_{post}\,(\exp(-\hat\delta)-1)$. With $\hat\delta=-0.0917$, this factor is $\exp(0.0917)-1\approx0.0960$. Using an average of 55.64 UMG songs per country-week across 69 countries and 11 post weeks (759 country-weeks) implies $N^{obs}_{post}\approx42{,}227$ and about $42{,}227\times0.0960\approx4{,}055$ forgone appearances. This is an extensive-margin benchmark and does not correspond to 4{,}055 unique songs.} Averaged over 69 countries and 11 post-withdrawal weeks (759 country-weeks), this corresponds to about 5.3 fewer UMG Top-200 chart entries per country-week. Taken together, these results reinforce the main pattern in the paper: reductions in TikTok exposure disproportionately affect upper-tail, threshold-like “hit” outcomes.

These results should be interpreted with two caveats in mind. First, chart positions are inherently zero-sum: the Top-200 has a fixed number of slots, so a UMG song that drops out of the chart is mechanically replaced by a song from another label. This means that the DiD estimator comparing UMG's chart footprint to that of Sony, Warner, and independent labels double-counts the effect: a single displacement event --- one UMG song dropping out and one non-UMG song entering --- appears as both a loss for the treated group and a gain for the control group, so the estimated gap overstates the pure reduction in UMG chart presence in isolation. Second, and relatedly, this zero-sum structure is a specific manifestation of the broader cross-catalog interference concern discussed in Section~\ref{sec:sutva}: if the withdrawal causes creators or listeners to reallocate activity toward non-UMG content, control labels benefit directly, which further inflates the treated-minus-control gap beyond the mechanical displacement effect. However, as discussed in Section~\ref{sec:sutva} in the context of the main streaming results, this reallocation is not purely a bias to be corrected away. Under Spotify's pro-rata compensation model, rights holders are paid in proportion to their share of total streams rather than on the basis of absolute stream counts, and the same logic applies to chart presence as a market-level outcome: what matters for UMG's competitive position is not only whether its own songs lose streams, but also whether competitors gain. 
The treated-minus-control gap therefore remains informative about 
the change in UMG's relative chart standing---capturing both the 
direct loss of UMG chart entries and the mechanical gain for 
competing labels on a fixed-size chart---though it is best 
interpreted as a measure of relative competitive position rather 
than as a direct mapping to revenue-share consequences. We present these chart results as a descriptive, downstream implication of the withdrawal, and caution that the estimated magnitude conflates the direct effect and the substitution effect, while noting that both margins are genuine components of the economic consequence of the withdrawal for UMG.
 
While we emphasize charts as an economically salient ``winner'' metric for labels, rankings can also operate as an amplification channel in their own right. Charts make ``what others chose'' salient, so a higher position can both reflect demand and shift subsequent attention and choice by providing a visible popularity signal. Stronger social-influence information increases inequality of outcomes, which is consistent with self-reinforcing cascades in hit formation \citep{salganik_experimental_2006}. Field evidence similarly shows that displaying ranked ``top'' options raises subsequent demand for those items \citep{CaiChenFang2009ObservationalLearning}. In our setting, Top-200 inclusion can plausibly amplify listening through chart browsing, algorithmic surfacing, media coverage, and related feedback loops, which means that rankings provide an additional amplification channel through which an exposure shock can depress downstream listening beyond the direct discovery mechanism.

\begin{table}[!ht]
    \centering
    \renewcommand{\arraystretch}{1.0}
    \footnotesize
    \caption{Spotify Top 200 Charts: Average Treatment Effect (PPML)}
    \label{tab:top200_ppml_main}
    
    \begin{threeparttable}
        \begin{tabular}{p{8cm} c}
            \toprule
             & \textbf{PPML} \\
            \midrule
            $UMG_{\ell} \times Post_{t}$ & $-0.0917^{***}$ \\
             & $(0.0216)$ \\
            \addlinespace[0.75em]
            Label--country FE & Yes \\
            Country--week FE & Yes \\
            Observations & 5,520 \\
            Pseudo $R^{2}$ & 0.8272 \\
            \bottomrule
        \end{tabular}
        \begin{tablenotes}[para,flushleft]
            \scriptsize\emph{Notes:} The dependent variable is the weekly number of unique songs from label $\ell$ in country $c$ that appear in Spotify's Top~200 chart ($n\_songs$). Estimates come from a Poisson pseudo-maximum-likelihood (PPML) model with a log link, label--country fixed effects, and country--week fixed effects. Standard errors (in parentheses) are cluster-robust and clustered at the label--country level. Significance levels: $^{***}p<0.001$, $^{**}p<0.01$, $^{*}p<0.05$.
        \end{tablenotes}
    \end{threeparttable}
\end{table}

\newpage

\section{SUTVA and Substitution in TikTok Creations}
\label{sec:sutva_substitution}
 
This analysis provides evidence on how the UMG withdrawal affects user-level content production on TikTok and, in particular, whether creators substitute into non-UMG sounds in a way that could contaminate the song-level treated-versus-control comparison. The motivation is the \emph{stable unit treatment value assumption} (SUTVA), which requires that a unit’s potential outcomes are unaffected by other units’ treatment status. In our setting, SUTVA may be violated if removing UMG sounds induces creators to reallocate posting toward alternative songs, which would mechanically affect outcomes for songs in the control group. The goal of this analysis is to gauge the likely magnitude of such interference and assess whether it is plausibly large enough to matter for interpreting the song-level DiD estimates. Because the song-level design relies on untreated major-label songs as controls, the relevant SUTVA concern is whether creators reallocate posting from UMG to Sony or Warner, and whether any possible reallocation affects the streams of control units.
 
\subsection{Data and Descriptive Statistics}
We source public TikTok user data through Tikapi.io, a proprietary service that enables retrospective access to posting histories and metadata for public accounts. Because there is no comprehensive sampling frame for TikTok usernames, we approximate quasi-randomness using two broad sampling anchors: we first seed the sample with generic, character-based username searches, and we then expand coverage by randomly selecting a large set of sounds that appear in our data and retrieving the most recent public-posting users for each sound, which incorporates both official and original audio. The resulting dataset contains 15{,}454 distinct users who, in the nine-week pre-treatment window, created at least one post that incorporates music.
 
\begin{figure}[!htb]
  \centering
  \caption{Concentration of Creations}
  \label{fig:tk_concentration}
  
  \includegraphics[width=0.6\textwidth]{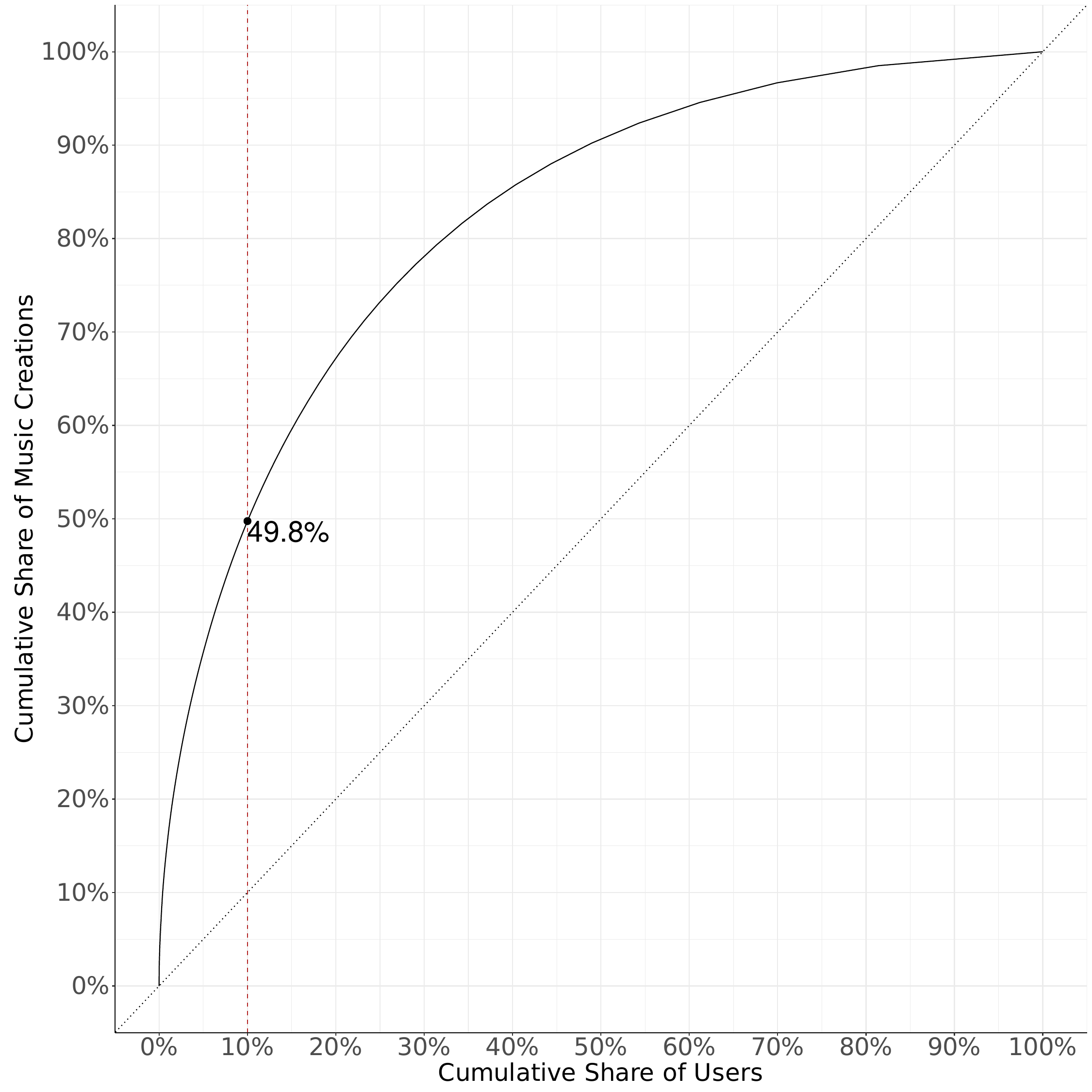}
 
  \vspace{2mm}
  \scriptsize
  \parbox{0.99\textwidth}{\textit{Notes:} Cumulative share of TikTok creations against the cumulative share of users, sorted in descending order of the number of creations. The top 10\% of users account for approximately 49.8\% of the TikTok creations in our sample. }
\end{figure}

In our user sample, we observe 753{,}796 pre-treatment posts, of which 25.64\% contain licensed music (i.e., can be matched to songs using Soundcharts). UMG accounts for the largest share of the matched licensed-music posts, representing 12.00\% of posts (23{,}194 posts). \autoref{tab:descriptives_pre_post} summarizes user–week posting behavior before and during the dispute. A key feature of the data is strong concentration in creator activity, which is typical in social media settings: posting-intensity-weighted means are much larger than unweighted means, which indicates that a relatively small set of highly active users accounts for a disproportionate share of total posts. \autoref{fig:tk_concentration} makes this skewness explicit by plotting the cumulative share of music creations against the cumulative share of users. This concentration motivates reporting post-weighted moments and interpreting effects through a post-level, rather than the average user.

\begin{table}[!htb]
    \centering
    \renewcommand{\arraystretch}{1.0}
    \caption{Descriptive Statistics of User-Week Posting Before and During the Dispute}
    \label{tab:descriptives_pre_post}
    \resizebox{0.85\linewidth}{!}{%
    \begin{threeparttable}
        \begin{tabular}{lcccccccc}
            \toprule
             & \multicolumn{4}{c}{Pre-period} & \multicolumn{4}{c}{Post-period} \\
            \cmidrule(lr){2-5} \cmidrule(lr){6-9}
            Outcome & Wtd Mean & Mean & SD & Max & Wtd Mean & Mean & SD & Max \\
            \midrule
            Total music posts        & 9.60 & 1.39 & 4.39 & 604 & 5.54 & 1.19 & 3.00 & 208 \\
            Non-major posts          & 6.94 & 0.99 & 3.29 & 421 & 4.36 & 0.93 & 2.49 & 176 \\
            UMG posts                & 1.36 & 0.17 & 0.97 & 126 & 0.10 & 0.02 & 0.23 & 33  \\
            Sony/Warner posts        & 1.31 & 0.23 & 0.89 & 89  & 1.09 & 0.23 & 0.92 & 116 \\
            \bottomrule
        \end{tabular}
        \begin{tablenotes}[para,flushleft]
            \scriptsize\emph{Notes:} The table reports descriptive statistics for weekly user-level posting outcomes in the pre- and post-periods. \textit{Wtd Mean} weights each user--week by the user’s pre-period total number of posts, so it reflects the average post rather than the average user. \textit{Mean} is the unweighted arithmetic mean across user--weeks. SD is computed over unweighted user--week observations.
        \end{tablenotes}
    \end{threeparttable}
    }%
\end{table}

 \begin{figure}[!htbp]
    \caption{Weekly User-Generated Audio Creations for the Major Labels in our TikTok User Sample}
    \label{fig:tt_sum_major_creations}
    \includegraphics[width=.9\textwidth]{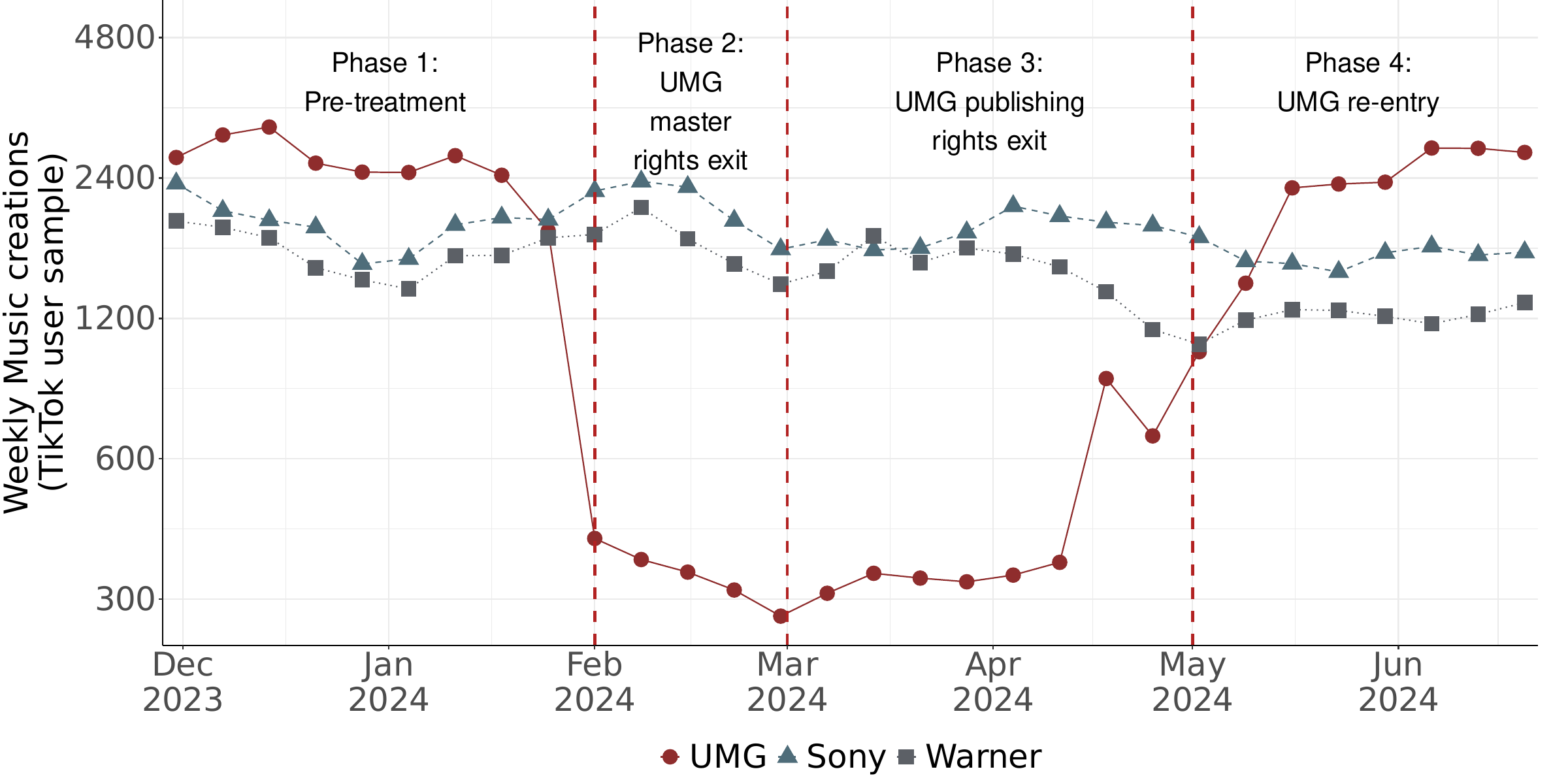}
\end{figure}
 
As shown in \autoref{fig:tt_sum_major_creations}, weekly music posting using UMG-affiliated audio drops sharply at the master-rights withdrawal and then recovers following re-entry. The UMG series does not fall to zero, which likely reflects residual availability and measurement frictions in retrospective public posting data. By contrast, postings using Sony and Warner audio do not exhibit a commensurate upward shift over the dispute window. This pattern suggests that the withdrawal does not primarily reallocate creator activity one-for-one from treated UMG sounds into the major-label control catalogs. We investigate this more formally next.

\subsection{Quantifying Substitution Across Catalogs}
 
To move beyond the descriptive time series and quantify substitution versus contraction, we exploit the fact that users differ in how exposed they are to the withdrawal. Specifically, we relate weekly music posting to each user’s pre-withdrawal reliance on UMG audio, which creates a continuous exposure gradient rather than a binary treated–control split. We begin with an exposure-weighted visualization that serves as a continuous analogue of the standard treated–control mean plot, and then use the same structure in our regression analysis.
 
For each user $u$, let $s_u \in [0,100]$ denote the user’s \emph{pre-withdrawal UMG reliance}, defined as the scaled share of the user’s music posts in Phase~1 that use UMG-affiliated audio (restricting attention to posts that contain music). Let $y_{ut}$ denote the outcome of interest in week $t$, defined as the number of music posts by user $u$ in week $t$ (aggregating across all labels). We then construct two reweighted weekly means:
\[
\bar y^{H}_t \equiv \frac{\sum_u s_u\, y_{ut}}{\sum_u s_u}
\qquad\text{and}\qquad
\bar y^{L}_t \equiv \frac{\sum_u (100-s_u)\, y_{ut}}{\sum_u (100-s_u)}.
\]
We refer to $\bar y^{H}_t$ as the \emph{high-exposure-weighted} series and to $\bar y^{L}_t$ as the \emph{low-exposure-weighted} series. Intuitively, $\bar y^{H}_t$ places more weight on users who relied heavily on UMG audio prior to the dispute, while $\bar y^{L}_t$ places more weight on users who relied little on UMG audio. This construction is a smooth analogue of assigning users to ``high exposure'' versus ``low exposure'' groups based on pre-period reliance, but it avoids introducing an arbitrary threshold and preserves the full exposure gradient. Because the weights are fixed using pre-withdrawal behavior, the series track how posting evolves for users with different baseline exposure to the withdrawal shock, without mechanically changing composition over time.
 
\begin{figure}[!htbp]
  \centering
  \caption{Weekly User-Generated Audio Creations in our TikTok User Sample}
  \label{fig:tk_matched_avgposts}
  \includegraphics[width=0.9\textwidth]{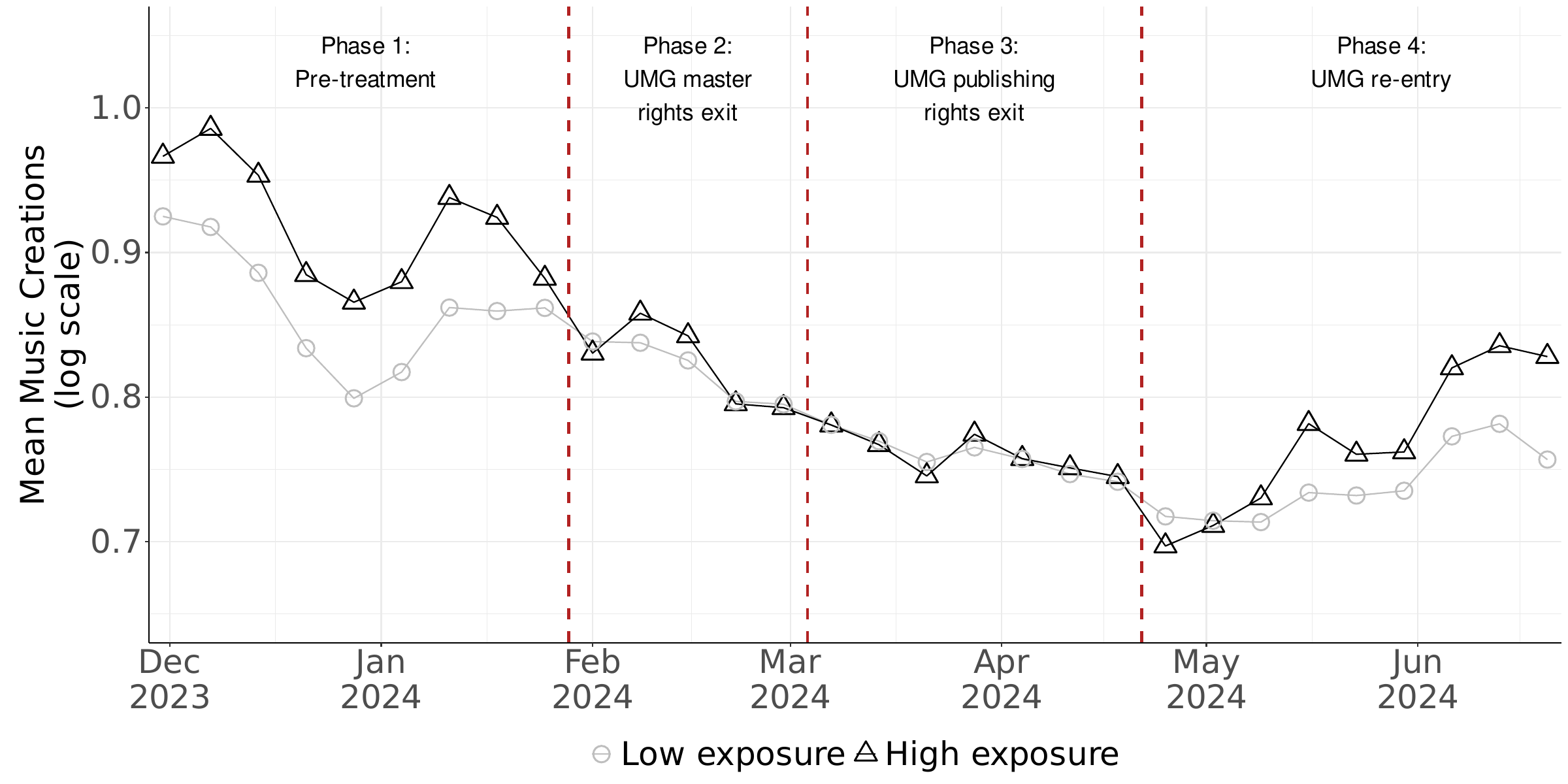}
 
  \vspace{2mm}
  \scriptsize
  \parbox{0.99\textwidth}{\textit{Notes:} The figure plots weekly mean user-generated music creations per user (log scale) in our TikTok user sample, separately for users with high versus low baseline exposure to UMG audio. Exposure is defined by each user’s pre-period share of music posts that use UMG-affiliated audio, with the two series constructed as exposure-weighted averages so that users with higher (lower) UMGShare receive more weight in the High (Low) exposure curve. Vertical dashed lines mark the start of Phase 2 (UMG master-rights exit), Phase 3 (UMG publishing-rights exit), and Phase 4 (UMG re-entry). The outcome counts all music posts regardless of label, so differences across series reflect changes in overall music-posting intensity rather than mechanical relabeling of UMG posts.}
\end{figure}

\autoref{fig:tk_matched_avgposts} plots $\log(\bar y^{H}_t)$ and $\log(\bar y^{L}_t)$ over time. The figure shows a pronounced decline in the high-exposure-weighted series at the withdrawal, followed by a partial recovery after re-entry, while the low-exposure-weighted series is comparatively smooth over the same period. We interpret this pattern as descriptive evidence that the disruption is associated with a contraction in overall music posting among users who were most reliant on UMG audio pre-withdrawal, rather than a one-for-one reallocation of activity toward other catalogs.

Next, we formalize this exposure-gradient pattern in a user--week design that estimates how posting responds during the dispute window as a function of baseline UMG reliance, $s_u$. Because creator activity is highly concentrated (see \autoref{fig:tk_concentration}), we estimate Poisson pseudo-maximum-likelihood (PPML) models. PPML accommodates zero-heavy count outcomes and, with a log link, yields a proportional effect on the conditional mean that is naturally informative about aggregate posting mass rather than the behavior of the average user.
 
Let $Post_t$ indicate the dispute window (Phases~2--3). For each outcome $Y^j_{ut}$, we estimate:
\begin{equation}
\label{eq:ppml_user_catalog_j}
\mathbb{E}\!\left[ Y^{j}_{ut} \mid \cdot \right]
=
\exp\!\left(
    \delta^{j} \left( s_u \times Post_t \right)
    + \mu_u + \gamma_t
\right),
\end{equation}
where $\mu_u$ are user fixed effects and $\gamma_t$ are week fixed effects. We estimate \autoref{eq:ppml_user_catalog_j} for three outcomes: (i) total music posts ($Y^{Tot}_{ut}$), (ii) posts using Sony or Warner audio pooled ($Y^{Maj}_{ut}\equiv Y^{Sony}_{ut}+Y^{Warner}_{ut}$), and (iii) posts using independent/other audio ($Y^{Ind}_{ut}$). The resulting coefficients are reported in \autoref{tab:tt_ppml_main_results}.

\begin{table}[!htb]\centering
    \caption{User-level Poisson Pseudo-Maximum-Likelihood with Continuous Exposure to UMG Audio: Posting Effects During the Dispute Window}
    \label{tab:tt_ppml_main_results}
    \renewcommand{\arraystretch}{1.0}
    \footnotesize
     
    \resizebox{0.85\linewidth}{!}{%
    \begin{threeparttable}
        \begin{tabular}{lccc}
            \toprule
             & Total posts ($\hat\delta$) & Other majors ($\hat\delta^{Maj}$) & Independent ($\hat\delta^{Ind}$) \\
            \midrule
            $s_u \times Post_t$
            & $-0.0046^{***}$ &  $0.0083^{***}$ & $0.0081^{***}$ \\
            & $(0.0008)$      & $(0.0011)$     & $(0.0009)$        \\
            \addlinespace
            User FE & Yes & Yes & Yes \\
            Week FE & Yes & Yes & Yes \\
            Observations (N) & 324{,}534 & 229{,}278 & 314{,}118 \\
            Users (N) & 15{,}454 & 10{,}918 & 14{,}958 \\
            Pseudo R\textsuperscript{2} & 0.463 & 0.298 & 0.437 \\
            \bottomrule
        \end{tabular}
         
        \begin{tablenotes}[para,flushleft]
            \scriptsize\emph{Notes:} Each column reports a Poisson pseudo-maximum-likelihood (PPML) estimate from a user--week panel regression with a log link and user and week fixed effects. The regressor $s_u$ is the user’s pre-period share of music posts that use UMG audio, and $Post_t$ indicates the dispute window (Phases 2--3). Outcomes are weekly counts of music posts: total music posts ($Y^{Tot}_{ut}$), posts using audio from Sony or Warner pooled ($Y^{Maj}_{ut}\equiv Y^{Sony}_{ut}+Y^{Warner}_{ut}$), and posts using independent/other audio ($Y^{Ind}_{ut}$). Standard errors are clustered at the user level. The regressions on the music specific outcomes (columns 2 and 3) can have different estimation samples because many user--weeks have no variation in the specific outcome. The table reports the corresponding observation counts for each column. $^{***}\,p<0.001$.
        \end{tablenotes}
    \end{threeparttable}
    }%
\end{table}

Three patterns in \autoref{tab:tt_ppml_main_results} are informative about the potential for interference. First, the total-posts coefficient is negative, which indicates that higher-exposure users reduce overall music posting during the dispute window. This is inconsistent with one-for-one substitution as the dominant response. Second, the Sony/Warner coefficient is positive, which indicates that some posting is reallocated into the catalogs that form our song-level control group. This is not, by itself, a SUTVA violation at the user level, but it is the most direct identification concern for the song-level DiD because it can increase TikTok exposure, and therefore streams, for control songs relative to their own no-withdrawal counterfactual. Third, the independent coefficient is also positive, which indicates that some of the displaced activity flows into catalogs outside our control pool and therefore mechanically dampens the scope for contamination of the treated-versus-control comparison.
 
 
To translate the user-level PPML estimates into an economically interpretable market-level magnitude, we construct model-based counterfactual predictions for aggregate posting during the dispute window. The goal is to quantify how much music creation is lost or reallocated when users are exposed to the UMG withdrawal. We do this separately for three outcomes: total music posts (\(j=Tot\)), Sony/Warner posts (\(j=Maj\)), and independent posts (\(j=Ind\)). For each outcome \(j\), let \(\hat{\mu}^j_{ut}(s)\) denote the fitted conditional mean from \autoref{eq:ppml_user_catalog_j} evaluated at exposure level \(s\). Using the estimates from \autoref{eq:ppml_user_catalog_j}, we compute two post-period predictions for each user--week. The observed-exposure prediction is \(\hat{\mu}^{j,obs}_{ut}=\hat{\mu}^j_{ut}(s_u)\). The no-UMG counterfactual sets exposure to zero in the post period, \(\hat{\mu}^{j,0}_{ut}=\hat{\mu}^j_{ut}(0)\).

Aggregating these predictions over all user--weeks in the dispute window yields the implied platform-level outcomes
 
\begin{equation}
    \widehat{Y}^{j,obs}_{post}
    =
    \sum_{(u,t) \in Post} \hat{\mu}^{j,obs}_{ut},
    \widehat{Y}^{j,0}_{post}
    =
    \sum_{(u,t) \in Post} \hat{\mu}^{j,0}_{ut}.
\end{equation}

The aggregate effect of UMG exposure for outcome $j$ is then defined as
 
\begin{equation}
    \Delta^j_{post}
    =
    \widehat{Y}^{j,obs}_{post}
    -
    \widehat{Y}^{j,0}_{post},
\end{equation}
 
    with the corresponding proportional effect given by
 
\begin{equation}
    RelativeChange^j_{post}
    =
    \widehat{Y}^{j,obs}_{post} / \widehat{Y}^{j,0}_{post} - 1.
\end{equation}
 
Table \ref{tab:bote_substitution} summarizes the implied platform-level posting responses from the PPML counterfactual aggregation. Relative to the no-UMG counterfactual, total music posting is about 5.2\% lower during the dispute window, which indicates that creators do not substitute one-for-one into alternative sounds and that some posting mass disappears. At the same time, posting shifts toward other catalogs: Sony/Warner posting is about 11.9\% higher than in the counterfactual, and independent/other posting is about 8.0\% higher. These predictions also allow us to quantify the implied reduction in UMG posting during the dispute window. Because UMG activity is not estimated directly in the PPML models, we recover the UMG shortfall residually as the difference between predicted total music posts and the predicted postings for all non-UMG catalogs (Sony/Warner and Independent). Comparing this implied UMG posting under the observed exposure profile to the corresponding no-UMG counterfactual yields an estimate of how much UMG-related creation is displaced during the dispute. Taken together, the pattern is consistent with partial reallocation away from UMG audio into both the major-label control catalogs and the broader non-major repertoire, with the net effect being a contraction in overall music creation rather than pure reshuffling across songs.

\begin{table}[!htb]
    \centering
    \renewcommand{\arraystretch}{1.0}
    \footnotesize
    \caption{Model-Implied Counterfactual Posting Changes}
    \label{tab:bote_substitution}
    \resizebox{0.85\linewidth}{!}{%
        \begin{threeparttable}
            \begin{tabular}{lrrrr}
                \toprule
                Outcome & Post (factual) & Post (counterfactual) & Delta & Relative Change \\
                \midrule
                Total music posts   & 221,103 & 233,160 & -12,057 & -5.17\% \\
                Independent posts   & 173,134 & 160,328 & +12,806 & +8.0\% \\
                Sony/Warner posts   & 43,366  & 38,747  & +4,619  & +11.9\% \\
                \hline 
                {UMG} & {4{,}603} & {34{,}085} & -29{,}482 & -86.5\% \\
                \bottomrule
            \end{tabular}
         
            \begin{tablenotes}[para,flushleft]
                \scriptsize\emph{Notes:} The table reports aggregated model-implied post-period outcomes from the user--week PPML models. ``Post (factual)'' corresponds to predictions evaluated at users' observed pre-period UMG exposure $s_u$, while ``Post (counterfactual)'' sets $s_u = 0$ for all post-period observations. Effects  are computed as absolute values (Delta) and proportional differences (Relative Change) relative to the post-period counterfactual. UMG posting is inferred residually as the difference between total music postings and postings using non-UMG audio.
            \end{tablenotes}
        \end{threeparttable}
    }%
\end{table}
 
To summarize, the counterfactual aggregation exercise shows that substitution into Sony and Warner audio is present but limited. Relative to the no-UMG counterfactual, Sony and Warner posts increase by 4{,}619, compared to an implied reduction of about 29{,}482 UMG posts during the dispute window. This implies that approximately 16\% of the aggregate UMG shortfall is absorbed by Sony and Warner catalogs. 

\subsection{U.S.\ TikTok Outage}
\label{app:tt_ban_results}

This appendix provides additional details on the sample construction and matching procedure for the U.S.\ TikTok outage analysis, and reports the pooled DiD estimate corresponding to the event-study in \autoref{fig:ban_eventstudy}.

The analysis uses a daily song--country panel from Luminate. For each of the 10 countries in the analysis, we begin with the 50{,}000 most streamed tracks and retain song--country observations with strictly positive daily activity on each day of the 28-day event window. This balanced-panel restriction yields 42{,}850 U.S.\ tracks observed on all 28 days, and we apply the same restriction to the control countries.

Relative to the main UMG analysis, the outage sample is concentrated in the upper tail of the popularity distribution. In the United States, the balanced-panel sample has a minimum of 8{,}538 daily streams, a mean of 52{,}131, and a maximum of 5{,}334{,}115.\footnote{To benchmark these levels relative to the weekly global Spotify distribution in the main study, we translate U.S.\ daily streams into approximate global weekly streams. Using Luminate's year-end totals, 1.4 trillion U.S.\ on-demand audio streams out of 4.8 trillion globally \citep{luminate_yearend_2024}, the U.S.\ accounts for about 29\% of global streaming activity. Applying this share and converting to a weekly scale ($\text{daily}\times 7 / 0.29$), these values correspond to roughly 0.21 million, 1.26 million, and 128.6 million global weekly streams, respectively.} Relative to the main-study pre-treatment Spotify deciles, these stream levels fall primarily in the upper part of the distribution.\footnote{Daily stream counts from Luminate cover all major services, while the Soundcharts data in the Web Appendix~\ref{app:decile_cutoffs} use Spotify-only streams. Mapping all-services streams to Spotify would lower the implied Spotify-weekly levels, but given Spotify's large global market share, the majority of these songs would still fall in the upper tail of the Spotify distribution.} As a result, this setting is primarily informative about the part of the distribution where the main analysis finds economically meaningful declines.

We construct the control group using 1:1 minimum-distance matching. Starting from the universe of non-U.S.\ song--country observations, we restrict the potential control pool to song--country pairs that fall in the same pre-treatment per-capita streaming quartile as the treated U.S.\ song. Because the data are daily, pre-treatment streams are noisier than in the weekly UMG analysis and reflect high-frequency shocks such as systematic day-of-week patterns. We therefore match on both the shape of the 14-day pre-treatment trajectory and level moments of the pre-period distribution, including the mean, volatility, and last pre-period level. We also include song age to keep treated and control songs comparable in maturity.\footnote{Specifically, within each quartile-specific control pool we select the control that minimizes a weighted Euclidean distance that combines (i) differences in the within-song standardized 14-day pre-treatment path, (ii) differences in pre-treatment level moments, and (iii) differences in song age. Controls may be reused across treated songs; the resulting matched sample consists of 42{,}850 treated songs and 42{,}850 matched control songs, corresponding to 23{,}915 unique control song--country combinations. After matching, treated and control songs are closely balanced on the key observables. Average daily streams per capita are nearly identical (treated: 153.75, SD = 293.55; control: 153.31, SD = 314.55; SMD $\approx -0.001$), and song age is similarly well aligned (treated: 600.88 weeks, SD = 695.03; control: 562.93 weeks, SD = 623.93; SMD $\approx 0.058$), which indicates that the matched sample is comparable in both baseline demand and maturity.}

\autoref{tab:ban_ppml_main} reports the pooled PPML DiD estimates for a 7-day and a 14-day post window. The coefficients imply an average decline of approximately 5.1\% and 2.8\% in U.S.\ streams relative to the matched cross-country counterfactual during the TikTok disruption window, respectively.

\begin{table}[!htbp]
    \centering
    \renewcommand{\arraystretch}{1.0}
    \footnotesize
    \caption{U.S.\ TikTok Outage: Average Treatment Effect (PPML)}
    \label{tab:ban_ppml_main}
    \begin{threeparttable}
        \begin{tabular}{p{8cm} cc}
            \toprule
             & \textbf{(1)} & \textbf{(2)} \\
             & 7-day post window & 14-day post window \\
            \midrule
            $US_c \times Post_t$ & $-0.0526^{***}$ & $-0.0284^{***}$ \\
             & $(0.0011)$ & $(0.0014)$ \\
            \addlinespace[0.75em]
            Song--country FE & Yes & Yes \\
            Day--country FE & Yes & Yes \\
            Observations & 1,799,700 & 2,399,600 \\
            Pseudo $R^{2}$ & 0.969 & 0.968 \\
            \bottomrule
        \end{tabular}
        \begin{tablenotes}[para,flushleft]
            \scriptsize\emph{Notes:} The dependent variable is daily streams per capita for track $i$ in country $c$ on day $t$. Columns (1) and (2) restrict the post-treatment window to 7 and 14 days after the onset of the U.S.\ TikTok outage, respectively. Estimates come from a Poisson pseudo-maximum-likelihood (PPML) model with a log link, song--country fixed effects, and day $\times$ control-country fixed effects. Standard errors in parentheses are Huber--White sandwich standard errors two-way clustered at the song--country and match-pair level. Significance levels: $^{***}p<0.001$, $^{**}p<0.01$, $^{*}p<0.05$.
        \end{tablenotes}
    \end{threeparttable}
\end{table}
\FloatBarrier

    \FloatBarrier
    \putbib

\end{bibunit}

\begin{thebibliography}{31}
\providecommand{\natexlab}[1]{#1}
\providecommand{\url}[1]{\texttt{#1}}
\providecommand{\urlprefix}{URL }

\bibitem[{Austin(2011)}]{austin_introduction_2011}
Austin PC (2011) An {Introduction} to {Propensity} {Score} {Methods} for
  {Reducing} the {Effects} of {Confounding} in {Observational} {Studies}.
  \emph{Multivariate Behavioral Research} 46(3):399--424.

\bibitem[{Bairathi et~al.(2025)Bairathi, Lambrecht, \protect\BIBand{}
  Rao}]{bairathi_lambrecht_rao_2024}
Bairathi M, Lambrecht A, Rao A (2025) Social {Media}, {Music} {Consumption},
  and {Cross}-{Platform} {Spillover} {Effects}. SSRN Working Paper,
  \urlprefix\url{https://papers.ssrn.com/abstract=4959753}.

\bibitem[{Chen \protect\BIBand{} Roth(2024)}]{ChenRoth2024}
Chen J, Roth J (2024) Logs with zeros? {Some} problems and solutions. \emph{The
  Quarterly Journal of Economics} 139(2):891--936,
  \urlprefix\url{http://dx.doi.org/10.1093/qje/qjad054}.

\bibitem[{Cheng et~al.(2026)Cheng, Ofek, \protect\BIBand{}
  Yoganarasimhan}]{cheng_value_2024}
Cheng MM, Ofek E, Yoganarasimhan H (2026) The value of silence: The effect of
  {UMG}'s licensing dispute with {TikTok} on music demand. \emph{Marketing
  Science} 45(3):493--514.

\bibitem[{Ciani \protect\BIBand{} Fisher(2019)}]{ciani_did_2019}
Ciani E, Fisher P (2019) Dif-in-dif estimators of multiplicative treatment
  effects. \emph{Journal of Econometric Methods} 8(1).

\bibitem[{Donati \protect\BIBand{} Fong(2025)}]{donati2025cost}
Donati D, Fong H (2025) The cost of banning tiktok: Implications for digital
  advertising. \emph{Columbia Business School Research Paper} .

\bibitem[{Fleder \protect\BIBand{} Hosanagar(2009)}]{fleder_blockbuster_2009}
Fleder D, Hosanagar K (2009) Blockbuster {Culture}'s {Next} {Rise} or {Fall}:
  {The} {Impact} of {Recommender} {Systems} on {Sales} {Diversity}.
  \emph{Management Science} 55(5):697--712.

\bibitem[{Herstand(2024)}]{herstand_whos_2024}
Herstand A (2024) Who's {Getting} {Hurt} in the {Universal} {Music}-{TikTok}
  {Standoff}? {Artists} and {Songwriters} ({Guest} {Column}).
  \urlprefix\url{https://variety.com/2024/digital/opinion/universal-music-tiktok-battle-hurts-artists-1235929689/},
  accessed March 21, 2024.

\bibitem[{{IFPI}(2022)}]{IFPI2022engaging}
{IFPI} (2022) Engaging with music.
  \urlprefix\url{https://www.ifpi.org/wp-content/uploads/2022/11/Engaging-with-Music-2022_full-report-1.pdf},
  accessed: July 11, 2024.

\bibitem[{Ingham(2022{\natexlab{a}})}]{ingham_crises_2022}
Ingham T (2022{\natexlab{a}}) Lyor cohen: The music industry is facing one of
  our biggest crises to date.
  \urlprefix\url{https://www.musicbusinessworldwide.com/podcast/lyor-cohen-music-industry-facing-one-of-our-biggest-crises-to-date/},
  accessed March 15, 2024.

\bibitem[{Ingham(2022{\natexlab{b}})}]{ingham_so_2022}
Ingham T (2022{\natexlab{b}}) So. . . how much did tiktok actually pay the
  music industry from its \$4bn in revenues last year?
  \urlprefix\url{https://www.musicbusinessworldwide.com/so-how-much-did-tiktok-actually-pay-the-music-industry-from-its-4bn-in-revenues-last-year/},
  accessed March 15, 2024.

\bibitem[{Ingham(2024)}]{ingham_tiktoks_2024}
Ingham T (2024) {TikTok}'s biggest headache in its fight with {Universal}
  {Music} {Group} is {UMG}'s publishing catalog (and 2 other observations on
  the feud the whole music industry is talking about).
  \urlprefix\url{https://www.musicbusinessworldwide.com/tiktoks-biggest-headache-universal-feud-2/},
  accessed March 22, 2024.

\bibitem[{Johnson et~al.(2014)Johnson, Faraj, \protect\BIBand{}
  Kudaravalli}]{johnson_emergence_2014}
Johnson SL, Faraj S, Kudaravalli S (2014) Emergence of power laws in online
  communities: the role of social mechanisms and preferential attachment.
  \emph{MIS Quarterly} 38(3):795--808.

\bibitem[{Koenig(2023)}]{koenig_technical_2023}
Koenig F (2023) Technical {Change} and {Superstar} {Effects}: {Evidence} from
  the {Rollout} of {Television}. \emph{American Economic Review: Insights}
  5(2):207--223.

\bibitem[{Maheshwari(2023)}]{maheshwari_tiktok_2023}
Maheshwari S (2023) {TikTok} {Is} {Our} {DJ} {Now}. {It}'s {Playing} a {Lot} of
  {Meghan} {Trainor}.
  \urlprefix\url{https://www.nytimes.com/2023/06/24/business/meghan-trainor-made-you-look-tiktok.html},
  accessed March 15, 2024.

\bibitem[{Maheshwari et~al.(2025)Maheshwari, Mickle, \protect\BIBand{}
  Grant}]{maheshwari_tiktok_2025}
Maheshwari S, Mickle T, Grant N (2025) {TikTok} {Returns} to {Apple} and
  {Google} {App} {Stores}.
  \urlprefix\url{https://www.nytimes.com/2025/02/13/technology/tiktok-apple-google-app-stores.html},
  accessed October 9, 2025.

\bibitem[{McConnell(2024)}]{mcconnell_cant_2024}
McConnell B (2024) Can't see the forest for the logs: On the perils of using
  difference-in-differences with a log-dependent variable.
  \urlprefix\url{https://arxiv.org/pdf/2308.00167}, accessed December 8, 2025.

\bibitem[{Rosen(1981)}]{rosen_economics_1981}
Rosen S (1981) The {Economics} of {Superstars}. \emph{American Economic Review}
  71(5):845--858.

\bibitem[{Salganik et~al.(2006)Salganik, Dodds, \protect\BIBand{}
  Watts}]{salganik_experimental_2006}
Salganik MJ, Dodds PS, Watts DJ (2006) Experimental {Study} of {Inequality} and
  {Unpredictability} in an {Artificial} {Cultural} {Market}. \emph{Science}
  311(5762):854--856.

\bibitem[{Santos~Silva \protect\BIBand{}
  Tenreyro(2011)}]{SantosSilvaTenreyro2011}
Santos~Silva JMC, Tenreyro S (2011) Further simulation evidence on the
  performance of the {P}oisson pseudo-maximum likelihood estimator.
  \emph{Economics Letters} 112(2):220--222.

\bibitem[{Silva \protect\BIBand{} Tenreyro(2006)}]{log_gravity}
Silva JMCS, Tenreyro S (2006) The {Log} of {Gravity}. \emph{The Review of
  Economics and Statistics} 88(4):641--658.

\bibitem[{Sisario(2024)}]{sisario_universal_2024}
Sisario B (2024) Universal {Music} {Group} {Pulls} {Songs} {From} {TikTok}.
  \urlprefix\url{https://www.nytimes.com/2024/02/01/arts/music/universal-group-tiktok-music.html},
  accessed February 12, 2024.

\bibitem[{Solon et~al.(2015)Solon, Haider, \protect\BIBand{}
  Wooldridge}]{SolonHaiderWooldridge2015}
Solon G, Haider SJ, Wooldridge JM (2015) What are we weighting for?
  \emph{Journal of Human Resources} 50(2):301--316.

\bibitem[{{Spotify for Artists}(2026)}]{spotify_royalties}
{Spotify for Artists} (2026) Understanding spotify royalties.
  \urlprefix\url{https://support.spotify.com/us/artists/article/understanding-spotify-royalties/},
  accessed May 5, 2026.

\bibitem[{Stokel-Walker(2022)}]{stokel-walker_tiktok_2022}
Stokel-Walker C (2022) {TikTok} {Wants} {Longer} {Videos}--{Whether} {You}
  {Like} {It} or {Not}.
  \urlprefix\url{https://www.wired.co.uk/article/tiktok-wants-longer-videos-like-not},
  accessed March 12, 2024.

\bibitem[{Tomé(2025)}]{tome2025tiktok}
Tomé J (2025) Tiktok ban takes hold: data reveals sharp traffic decline and
  rapid shift to alternatives.
  \urlprefix\url{https://blog.cloudflare.com/tiktok-ban-traffic-decline-alternatives-rednote/},
  accessed May 13, 2026.

\bibitem[{{UMG}(2024)}]{universal_music_group_open_2024}
{UMG} (2024) An open letter to the artist and songwriter community why we must
  call time out on {TikTok}.
  \urlprefix\url{https://www.universalmusic.com/an-open-letter-to-the-artist-and-songwriter-community-why-we-must-call-time-out-on-tiktok/},
  accessed July 7, 2024.

\bibitem[{Witrand(2025)}]{Witrand2025_master_publishing_rights}
Witrand E (2025) Master rights vs. publishing rights: What’s the difference
  and why it matters.
  \urlprefix\url{https://soundcharts.com/blog/master-rights-vs-publishing-rights},
  accessed Jan 3, 2026.

\bibitem[{Wl\"{o}mert et~al.(2024)Wl\"{o}mert, Papies, Clement,
  \protect\BIBand{} Spann}]{wlomert_interplay_2024}
Wl\"{o}mert N, Papies D, Clement M, Spann M (2024) The {Interplay} of
  {User}-{Generated} {Content}, {Content} {Industry} {Revenues}, and {Platform}
  {Regulation}: {Quasi}-{Experimental} {Evidence} from {YouTube}.
  \emph{Marketing Science} 43(1):1--12.

\bibitem[{Wooldridge(2023)}]{wooldridge_simple_2023}
Wooldridge JM (2023) Simple approaches to nonlinear difference-in-differences
  with panel data. \emph{The Econometrics Journal} 26(3):C31--C66.

\bibitem[{Yang et~al.(2024)Yang, Zhang, \protect\BIBand{} Liu}]{yang_ugc_2024}
Yang G, Zhang Y, Liu H (2024) Frontiers: Pirating foes or creative friends?
  effects of user-generated condensed clips on demand for streaming services.
  \emph{Marketing Science} 43(3):469--478.

\end{thebibliography}


\begin{thebibliography}{27}
\providecommand{\natexlab}[1]{#1}
\providecommand{\url}[1]{\texttt{#1}}
\providecommand{\urlprefix}{URL }

\bibitem[{Abadie \protect\BIBand{} Spiess(2022)}]{abadie_spiess_2022}
Abadie A, Spiess J (2022) Robust post-matching inference. \emph{Journal of the
  American Statistical Association} 117(538):983--995,
  \urlprefix\url{http://dx.doi.org/10.1080/01621459.2020.1840383}.

\bibitem[{Aguiar \protect\BIBand{} Waldfogel(2021)}]{aguiar_platforms_2021}
Aguiar L, Waldfogel J (2021) Platforms, {Power}, and {Promotion}: {Evidence}
  from {Spotify} {Playlists}. \emph{The Journal of Industrial Economics}
  69(3):653--691.

\bibitem[{Arkhangelsky et~al.(2021)Arkhangelsky, Athey, Hirshberg, Imbens,
  \protect\BIBand{} Wager}]{arkhangelsky2021synthetic}
Arkhangelsky D, Athey S, Hirshberg DA, Imbens GW, Wager S (2021) Synthetic
  difference-in-differences. \emph{American Economic Review}
  111(12):4088--4118.

\bibitem[{Autor et~al.(2020)Autor, Dorn, Katz, Patterson, \protect\BIBand{}
  Van~Reenen}]{autor_fall_2020}
Autor D, Dorn D, Katz LF, Patterson C, Van~Reenen J (2020) The {Fall} of the
  {Labor} {Share} and the {Rise} of {Superstar} {Firms}. \emph{The Quarterly
  Journal of Economics} 135(2):645--709.

\bibitem[{Bairathi et~al.(2025)Bairathi, Lambrecht, \protect\BIBand{}
  Rao}]{bairathi_lambrecht_rao_2024}
Bairathi M, Lambrecht A, Rao A (2025) Social {Media}, {Music} {Consumption},
  and {Cross}-{Platform} {Spillover} {Effects}. SSRN Working Paper,
  \urlprefix\url{https://papers.ssrn.com/abstract=4959753}.

\bibitem[{Cai et~al.(2009)Cai, Chen, \protect\BIBand{}
  Fang}]{CaiChenFang2009ObservationalLearning}
Cai H, Chen Y, Fang H (2009) Observational learning: Evidence from a randomized
  natural field experiment. \emph{American Economic Review} 99(3):864--882.

\bibitem[{Callaway \protect\BIBand{}
  Sant’Anna(2021)}]{callawayDifferenceinDifferencesMultipleTime2021}
Callaway B, Sant’Anna PHC (2021) Difference-in-differences with multiple time
  periods. \emph{Journal of Econometrics} 225(2):200--230,
  \urlprefix\url{http://dx.doi.org/10.1016/j.jeconom.2020.12.001}.

\bibitem[{Cheng et~al.(2026)Cheng, Ofek, \protect\BIBand{}
  Yoganarasimhan}]{cheng_value_2024}
Cheng MM, Ofek E, Yoganarasimhan H (2026) The value of silence: The effect of
  {UMG}'s licensing dispute with {TikTok} on music demand. \emph{Marketing
  Science} 45(3):493--514.

\bibitem[{Ciani \protect\BIBand{} Fisher(2019)}]{ciani_did_2019}
Ciani E, Fisher P (2019) Dif-in-dif estimators of multiplicative treatment
  effects. \emph{Journal of Econometric Methods} 8(1).

\bibitem[{Hiller(2016)}]{scott_hiller_sales_2016}
Hiller SR (2016) Sales displacement and streaming music: {Evidence} from
  {YouTube}. \emph{Information Economics and Policy} 34:16--26.

\bibitem[{{IFPI}(2025)}]{IFPI2025gmr}
{IFPI} (2025) Global music report.
  \urlprefix\url{https://www.ifpi.org/wp-content/uploads/2024/03/GMR2025_SOTI.pdf},
  accessed: January 22, 2026.

\bibitem[{Kretschmer \protect\BIBand{} Peukert(2020)}]{kretschmer_video_2020}
Kretschmer T, Peukert C (2020) Video {Killed} the {Radio} {Star}? {Online}
  {Music} {Videos} and {Recorded} {Music} {Sales}. \emph{Information Systems
  Research} 31(3):776--800.

\bibitem[{{Luminate}(2025)}]{luminate_yearend_2024}
{Luminate} (2025) Luminate releases 2024 year-end report.
  \urlprefix\url{https://luminatedata.com/reports/yearend-music-industry-report-2024/},
  accessed January 13, 2026.

\bibitem[{MacKinnon et~al.(2023)MacKinnon, Nielsen, \protect\BIBand{}
  Webb}]{mackinnon_cluster_2023}
MacKinnon JG, Nielsen MO, Webb MD (2023) Cluster-robust inference: A guide to
  empirical practice. \emph{Journal of Econometrics} 232(2):272--299,
  \urlprefix\url{http://dx.doi.org/https://doi.org/10.1016/j.jeconom.2022.04.001}.

\bibitem[{Manning \protect\BIBand{} Mullahy(2001)}]{ManningMullahy2001}
Manning WG, Mullahy J (2001) Estimating log models: To transform or not to
  transform? \emph{Journal of Health Economics} 20(4):461--494.

\bibitem[{McConnell(2024)}]{mcconnell_cant_2024}
McConnell B (2024) Can't see the forest for the logs: On the perils of using
  difference-in-differences with a log-dependent variable.
  \urlprefix\url{https://arxiv.org/pdf/2308.00167}, accessed December 8, 2025.

\bibitem[{Roth \protect\BIBand{} Sant’Anna(2023)}]{Roth_SantAnna_2023}
Roth J, Sant’Anna PHC (2023) When is parallel trends sensitive to functional
  form? \emph{Econometrica} 91(2):737–747.

\bibitem[{Salganik et~al.(2006)Salganik, Dodds, \protect\BIBand{}
  Watts}]{salganik_experimental_2006}
Salganik MJ, Dodds PS, Watts DJ (2006) Experimental {Study} of {Inequality} and
  {Unpredictability} in an {Artificial} {Cultural} {Market}. \emph{Science}
  311(5762):854--856.

\bibitem[{Santos~Silva \protect\BIBand{}
  Tenreyro(2011)}]{SantosSilvaTenreyro2011}
Santos~Silva JMC, Tenreyro S (2011) Further simulation evidence on the
  performance of the {P}oisson pseudo-maximum likelihood estimator.
  \emph{Economics Letters} 112(2):220--222.

\bibitem[{Silva \protect\BIBand{} Tenreyro(2006)}]{log_gravity}
Silva JMCS, Tenreyro S (2006) The {Log} of {Gravity}. \emph{The Review of
  Economics and Statistics} 88(4):641--658.

\bibitem[{Solon et~al.(2015)Solon, Haider, \protect\BIBand{}
  Wooldridge}]{SolonHaiderWooldridge2015}
Solon G, Haider SJ, Wooldridge JM (2015) What are we weighting for?
  \emph{Journal of Human Resources} 50(2):301--316.

\bibitem[{{Spotify}(2023)}]{spotify_streamon_2023}
{Spotify} (2023) Spotify reveals more opportunities and features for creators
  during stream on.
  \url{https://newsroom.spotify.com/2023-03-08/stream-on-announcements-new-features-updates-2023-event/},
  accessed: 2026-01-07.

\bibitem[{White(1982)}]{White1982}
White H (1982) Maximum likelihood estimation of misspecified models.
  \emph{Econometrica} 50(1):1--25.

\bibitem[{Winkler et~al.(2026)Winkler, Wl{\"o}mert, \protect\BIBand{}
  Liaukonyt{\.e}}]{winkler_separating_2026}
Winkler D, Wl{\"o}mert N, Liaukonyt{\.e} J (2026) Express: Separating the
  artist from the art: Social media boycotts, platform sanctions, and music
  consumption. \emph{Journal of Marketing Research} .

\bibitem[{Wl\"{o}mert et~al.(2024)Wl\"{o}mert, Papies, Clement,
  \protect\BIBand{} Spann}]{wlomert_interplay_2024}
Wl\"{o}mert N, Papies D, Clement M, Spann M (2024) The {Interplay} of
  {User}-{Generated} {Content}, {Content} {Industry} {Revenues}, and {Platform}
  {Regulation}: {Quasi}-{Experimental} {Evidence} from {YouTube}.
  \emph{Marketing Science} 43(1):1--12.

\bibitem[{Wlömert et~al.(2025)Wlömert, Papies, \protect\BIBand{} van
  Heerde}]{wlomert_vanheerde_papies}
Wlömert N, Papies D, van Heerde HJ (2025) Driving {Music} {Demand} in the
  {Age} of {Streaming}: {Understanding} the {Heterogeneity} in {Curated}
  {Playlist} {Effectiveness}. \emph{Journal of Marketing} 00222429251395986.

\bibitem[{Yang et~al.(2024)Yang, Zhang, \protect\BIBand{} Liu}]{yang_ugc_2024}
Yang G, Zhang Y, Liu H (2024) Frontiers: Pirating foes or creative friends?
  effects of user-generated condensed clips on demand for streaming services.
  \emph{Marketing Science} 43(3):469--478.

\end{thebibliography}
\end{document}